\UseRawInputEncoding

\documentclass[11pt]{article}
\usepackage{}
\usepackage{amssymb}
\usepackage{amsfonts}
\usepackage{mathrsfs}
\usepackage{graphicx}
\usepackage{amsmath}
\usepackage{color}
\usepackage{amsthm}
\usepackage{epstopdf}
\usepackage{subfigure}
\usepackage{pifont,bm}
\usepackage{tikz}
\usepackage{enumerate}
\usepackage{mathrsfs}
\usepackage{cite}
\usepackage{hyperref}
\usepackage{booktabs}

\usepackage{diagbox}
\usepackage{makecell}

\theoremstyle{definition}

\makeatletter
\def\@biblabel#1{[#1]}
\makeatother

\makeatletter \@addtoreset{equation}{section}

\begin{document}

\begin{titlepage}
\title{\bf{Data-driven vector localized waves and parameters discovery for Manakov system using deep learning approach
\footnote{Corresponding authors.\protect\\
\hspace*{3ex} E-mail addresses: ychen@sei.ecnu.edu.cn (Y. Chen)}
}}
\author{Juncai Pu$^{a}$, Yong Chen$^{a,b,*}$\\
\small \emph{$^{a}$School of Mathematical Sciences, Shanghai Key Laboratory of Pure Mathematifcs and} \\
\small \emph{Mathematical Practice, East China Normal University, Shanghai, 200241, China} \\
\small \emph{$^{b}$College of Mathematics and Systems Science, Shandong University}\\
\small \emph{of Science and Technology, Qingdao, 266590, China} \\ \date{}}
\thispagestyle{empty}
\end{titlepage}
\maketitle

\vspace{-0.5cm}
\begin{center}
\rule{15cm}{1pt}\vspace{0.3cm}

\parbox{15cm}{\small
{\bf Abstract}\\
\hspace{0.5cm}
An improved physics-informed neural network (IPINN) algorithm with four output functions and four physics constraints, which possesses neuron-wise locally adaptive activation function and slope recovery term, is appropriately proposed to obtain the data-driven vector localized waves, including vector solitons, breathers and rogue waves (RWs) for the Manakov system with initial and boundary conditions, as well as data-driven parameters discovery for Manakov system with unknown parameters. The data-driven vector RWs which also contain interaction waves of RWs and bright-dark solitons, interaction waves of RWs and breathers, as well as RWs evolved from bright-dark solitons are learned to verify the capability of the IPINN algorithm in training complex localized wave. In the process of parameter discovery, routine IPINN can not accurately train unknown parameters whether using clean data or noisy data. Thus we introduce parameter regularization strategy with adjustable weight coefficients into IPINN to effectively and accurately train prediction parameters, then find that once setting the appropriate weight coefficients, the training effect is better as using noisy data. Numerical results show that IPINN with parameter regularization shows superior noise immunity in parameters discovery problem.
}

\vspace{0.5cm}
\parbox{15cm}{\small{

\vspace{0.3cm} \emph{Key words:Data-driven vector localized waves, Vector rogue waves, Parameters discovery, Manakov system, Improved PINN}  \\

} }
\end{center}
\vspace{0.3cm} \rule{15cm}{1pt} \vspace{0.2cm}

\section{Introduction}
Localized waves, containing soliton, breather and rogue wave (RW), have become a significant research direction in the field of nonlinear science in recent decades \cite{Khaykovich2002,Akhmediev1987,Ma1979,AkhmedievN2009}. Compared with the breathers which are localized in space or time (Akhmediev or Ma breathers \cite{Akhmediev1987,Ma1979}), the RWs are localized both in space and time with the heights of the RWs are two or more times those of the surrounding waves, and RW is the name given by oceanographers to isolated large amplitude waves which appears suddenly and disappear without a trace \cite{Kharif2003}. In 1964, Draper first observed the existence of RWs and called them ``Freak ocean waves" \cite{Draper1966}, and found RWs widely occur both in deep ocean and in shallow water \cite{Peregrine1976,Akhmediev2009,Akhmediev2010}. Hitherto, the research of RWs has gone far beyond the oceanographic background, containing other spatial-temporal continuous systems, such as water tank \cite{Chabchoub2011}, ultra-cold bosonic gases \cite{Charalampidis2018}, microwave transport \cite{Hohmann2010}, capillary waves \cite{Shats2010}, atmosphere \cite{Stenflo2009} and plasma \cite{Moslem2011}. In any of these aforementioned disciplines, new discoveries which indicate that RWs may be rather universal enrich their concept and lead to progress towards a comprehensive understanding of a phenomenon which still remains largely unexplored to obtain state-of-the-art theories on the subject.

The so-called Peregrine soliton provided a formal mathematical description of RWs, this solitary wave is the solution of the (1+1)-dimensional scalar nonlinear Schr\"{o}dinger equation (NLS), which is localized in both coordinate systems and describes a unique RW event \cite{Peregrine1983}. This solution is also unique in the mathematical sense, due to it is expressed in terms of rational functions of coordinates, which is different from most other known soliton solutions of the NLS, and the determinant representation of the $N$-order RW of the NLS has been directly obtained directly by a series of row operations on matrices appeared in the $N$-fold Darboux transformation \cite{Wang2017}. In a variety of complex systems, such as optical fibers \cite{kaup1993}, financial systems \cite{Yan2011} and Bose-Einstein condensates \cite{Bludov2010}, several amplitudes rather than a single one need to be considered. The resulting coupled systems can describe extreme waves more accurately than the scalar NLS model \cite{Bludov2010,Manakov1974}. Furthermore, for scalar systems, the velocity of the background field has no practical effect on the mode structure of RWs, since the corresponding solutions can be correlated by Galileo transform, but the relative velocity between different component fields for the coupled systems has real physical effects, and can not be eliminated by any simple transformation \cite{Ling2014}. In recent years, some important results have been obtained in the study of solitons and RWs for local and nonlocal multicomponent coupled nonlinear systems, such as the two-component derivative nonlinear schr\"{o}dinger equation \cite{GuoL2019} and nonlocal $M$-component NLS \cite{Rao2020}, but the research of exact and numerical solutions for local and nonlocal multicomponent coupled nonlinear systems is relatively few compared with scalar nonlinear systems in general. Therefore, the extended researches on vector RWs for multi-component coupled systems are nontrivial and meaningful.

With the revolution of hardware equipment and software technology, the explosive growth of available data and the great improvement of computer operation speed promote the application of machine learning and big data analysis technology in practice \cite{Mjolsness2001,Bishop2006}. Furthermore, the computational complexity of neural network (NN) was greatly reduced by introducing back-propagation NN algorithm and the gradient vanishing problem was solved, deep learning was born and has attracted extensive attention \cite{Rumelhart1986,LeCun2015}. Since that, deep learning detonated a large number of landing applications, such as face recognition \cite{Sun2018}, medical imaging \cite{Alipanahi2015}, video surveillance \cite{Shao2020}, speech recognition \cite{Nassif2019}, language understanding \cite{Collobert2011} and mathematical physics \cite{Raissi2019}. Among these applications mentioned above, a new physics-informed neural network (PINN) which controlled by mathematical physical systems based on the deep learning patterns of multi-layer NNs, have been proposed and proved to be particularly suitable for dealing with both the forward problems and highly ill-posed inverse problems by obtaining the approximate solutions of governing equations and discovering parameters involved in the governing equation are inferred from the training data, and found the PINN approach only needs less data sets to complete high-dimensional network tasks well \cite{Raissi2019}. Recently, PINN has played an important role in many physical applications \cite{KarniadakisGE2021}. Afterwards, Jagtap and collaborators proposed two different kinds of adaptive activation functions, namely global adaptive activation functions and locally adaptive activation functions, to approximate smooth and discontinuous functions as well as solutions of linear and nonlinear partial differential equations by introducing a scalable parameters in the activation function and adding a slope recovery term based on activation slope to the loss function of locally adaptive activation functions, and demonstrated the locally adaptive activation functions further improve the training speed, performance and speed up the training process of NNs \cite{Jagtap2020,JagtapA2020}.

Recently, recovering the data-drive solutions and revealing the dynamic behaviors of nonlinear partial differential equations have attracted extensive attention and sparked interest in their research by applying the PINN \cite{Raissi2019}. Due to the abundant sample space and good properties of integrable systems, it has become a research hotspots by applying deep learning based on PINN to the field of integrable systems, and significant numerical calculation results and some interesting unknown phenomena will be found. Chen research group is committed to the research of integrable deep learning algorithm, and obtains a series of high-quality data-driven solutions for classical integrable systems by using PINN and improved PINN (IPINN) algorithms. Specifically, data-driven solutions with abundant dynamic behaviors for some classical nonlinear evolution partial differential equations have been obtained by utilizing the PINN framework \cite{LiJ2020}. Especially, Pu et al. recovered the solitons, breathers and rogue wave solutions of the NLS with the aid of the PINN model \cite{Pu2021}. The data-driven rogue periodic wave of nonlinear partial differential equation has been learned by applying multi-layer PINN for the first time \cite{Peng2021}, and the PINN is applied to high-dimensional system to solve the ($N$+1)-dimensional initial boundary value problem with 2$N$+1 hyperplane boundaries \cite{Miao2021}. It is worth mentioning that a two-stage PINN method which is tailored to the nature of equations by introducing features of physical systems into NNs, is used to simulate abundant localized wave solutions of integrable equations \cite{Lin2021}. Furthermore, an PINN approach with neuron-wise locally adaptive activation function was presented to derive rational soliton solutions and rogue wave solutions of the derivative nonlinear Schr\"{o}dinger equation in complex space, and numerical results demonstrated the improved approach has faster convergence and better simulation effect than classical PINN method \cite{PuJ2021,PuJC2021}. In addition, other scholars have also done important works on data-driven solutions and parameter discovery of other nonlinear systems, such as higher-order NLS and defocusing NLS with the time-dependent potential \cite{Fang2021,Wang2021}. However, these aforementioned works utilize various PINNs with up to two outputs and two coupled physical constraints, but
we need the NNs with at least four outputs and four coupled physical constraints to simulate their corresponding data-driven localized waves for multi-component coupled nonlinear systems, this is also the difficulty we need to solve when we reconstruct a new NN to solve the data-driven localized waves of multi-component coupled system. To our best knowledge, although two parallel PINN with two inputs and two outputs are proposed to study the dynamic process and model parameters of the vector optical solitons for the coupled NLS in birefringent fibers \cite{WuGZ2021}, but the vector solitons, breathers and RWs for the Manakov system which also be called two-component coupled NLS have not been studied by applying the PINN with four outputs and four physical constraints and combining with integrable system theory \cite{Manakov1974}. Therefore, we design and propose an single IPINN model with four outputs and four constraint equations for the study of Manakov systems with the corresponding initial boundary value conditions for the first time to recover the vector localized waves and learn unknown parameters of Manakov system in this paper.

Recently, we proposed PINN with two inputs and three outputs to solve data-driven forward-inverse problems for Yajima-Oikawa system by using parameter regularization strategy \cite{PuYO2021}. Next, we focus on the following Manakov system with unknown parameters $\lambda_1$ and $\lambda_2$, the expression is as follows
\begin{align}\label{E1}
\begin{split}
\begin{cases}
\mathrm{i}q_{1t}+\lambda_1q_{1xx}+\lambda_2(|q_1|^2+|q_2|^2)q_1=0,\,x\in[L_0,L_1],\, t\in[T_0,T_1],\\
\mathrm{i}q_{2t}+\lambda_1q_{2xx}+\lambda_2(|q_1|^2+|q_2|^2)q_2=0,\,x\in[L_0,L_1],\, t\in[T_0,T_1],\\
q_1(x,T_0)=q_1^0(x),\,q_2(x,T_0)=q_2^0(x),\,x\in[L_0,L_1],\\
q_1(L_0,t)=q_1^{\mathrm{lb}}(t),\,q_1(L_1,t)=q_1^{\mathrm{ub}}(t),\,t\in[T_0,T_1],\\
q_2(L_0,t)=q_2^{\mathrm{lb}}(t),\,q_2(L_1,t)=q_2^{\mathrm{ub}}(t),\,t\in[T_0,T_1],\\
\end{cases}
\end{split}
\end{align}
where $``\mathrm{i}"$ satisfied $\mathrm{i}^2=-1$ is an imaginary number, the subscripts denote the partial derivatives of the complex fields $q_1(x,t)$ and $q_2(x,t)$ with respect to the space $x$ and time $t$, while the $L_0$ and $L_1$ represent the lower and upper boundaries of $x$ respectively. Similarly, $T_0$ and $T_1$ represent the initial and final times of $t$ respectively. Moreover, the $q_1^0(x)$ ($q_2^0(x)$) represents initial value of the $q_1(x,t)$ ($q_2(x,t)$) at $t=T_0$, the $q_1^{\mathrm{lb}}(t)$ and $q_1^{\mathrm{ub}}(t)$ ($q_2^{\mathrm{lb}}(t)$ and $q_2^{\mathrm{ub}}(t)$) are the lower and upper boundaries of the $q_1(x,t)$ ($q_2(x,t)$) corresponding to $x=L_0$ and $x=L_1$ respectively. Eq. \eqref{E1} applies to a Kerr medium with the electrostrictive mechanism of nonlinearity \cite{Kaplan1983}, as well as to randomly birefringent fiber optic transmission links \cite{Wang1999}.

In recent years, a variety of explicit vector solitons, breathers and RWs of the Manakov system have been obtained. By utilizing the RHP approach, Yang derived the exact determinant form of vector $N$-soliton for Manakov system with $3\times3$ Lax pair in detail \cite{Yang2010}. Priya et al. presented explicit forms of general breather, Akhmediev breather, Ma soliton, and RWs of the Manakov system by means of the Hirota bilinear method \cite{Priya2013}. The exact RW solutions, breathers, and rogue-bright-dark solutions for the Manakov system have been constructed by the Darboux transformation \cite{Guo2011}. A novel multiparametric vector freak solutions which feature both exponential and rational dependence on coordinates for the Manakov system has been analytically constructed and discussed, and the family of exact solutions includes known vector Peregrine (rational) solutions has also been derived \cite{Baronio2012}. Zhao and Liu obtained the dark RWs analytically for the first time in the Manakov system, and found that two RWs can appear in the temporal-spatial distribution \cite{Zhao2012}. According to the generalized Darboux transformation, Ling et al. studied the dynamics of high-order RWs in the Manakov system, and pointed out that four fundamental RWs can emerge from second-order vector RWs in the coupled system, in contrast to the high-order ones in single-component systems \cite{Ling2014}. Wang et al. investigated the novel higher-order localized waves account for the Manakov system are investigated by using the generalized Darboux transformation, and exhibited that two dark-bright solitons together with a second-order RW of fundamental or triangular pattern and two breathers together with a second-order RW of fundamental or triangular pattern coexist in the second-order localized wave for the coupled system \cite{Wang2014}. Rao et al. employed the Kadomtsev-Petviashvili hierarchy reduction method for deriving the general vector RW of the $M$-coupled NLS systems which contain the Manakov system in a compact form \cite{Rao2019}. From the previous references, we find out that the exact vector RWs of the Manakov system is mainly obtained via Darboux transformation, so it is difficult to accurately solve the vector RWs of the multi-component coupled nonlinear systems, and the means are extremely scarce, especially for some non integrable coupled nonlinear systems. Therefore, we consider how to apply the deep learning NN architecture to solve multi-component coupled nonlinear systems, specifically how to recover various vector licalized waves when only the initial-boundary value conditions of the multi-component coupled nonlinear systems are known. In the following, we focus on the data-driven vector localized waves which contain vector solitons, breather and RWs, as well as parameters discovery for the Manakov system with initial-boundary value conditions by utilizing IPINN pattern.

The rest of this paper is organized as follows. In section 2, we introduce briefly discussions of the IPINN method with locally adaptive activation function for the Manakov system, where also discuss about training data, loss function, optimization method and the training environment. Moreover, the algorithm flow schematic and algorithm steps for the Manakov system based on IPINN model are exhibited in detail. In Section 3, the data-driven solitons and breather with vivid plots and dynamic behavior analyses of the Manakov system have been exhibited via IPINN model. Section 4 provides the data-driven vector RWs of the Manakov system by utilizing the IPINN approach, and related plots and dynamic analyses are revealed in detail. Section 5 presents experimental results with different trade-off norm penalty term coefficients during learning data-driven parameter discovery of Manakov system. Conclusions and discussions are given out in last section.

\section{Methodology}

In general, we consider the general (1 + 1)-dimensional coupled nonlinear time-dependent systems with unknown parameters $\lambda_1$ and $\lambda_2$ in complex space, its general form is as shown below
\begin{align}\label{E2}
\begin{split}
&\mathrm{i}q_{1t}+\mathcal{N}[q_1,q_2;\lambda_1,\lambda_2]=0,\\
&\mathrm{i}q_{2t}+\mathcal{N}'[q_1,q_2;\lambda_1,\lambda_2]=0,
\end{split}
\end{align}
where $q_1$ and $q_2$ are complex-valued solutions of $x$ and $t$ to be determined later, $\mathcal{N}[\cdot,\cdot;\lambda_1,\lambda_2]$ and $\mathcal{N}'[\cdot,\cdot;\lambda_1,\lambda_2]$ are nonlinear differential operators in space. Due to the complexity of the structure of the complex-valued solutions $q_1(x,t)$ and $q_2(x,t)$ in Eq. \eqref{E2}, we decompose $q_1(x,t)$ ($q_2(x,t)$) into the real part $u(x,t)$ ($m(x,t)$) and the imaginary part $v(x,t)$ ($n(x,t)$) by employing real-valued functions $u(x,t)$ and $v(x,t)$ ($m(x,t)$ and $n(x,t)$), that is $q_1(x,t)=u(x,t)+\mathrm{i}v(x,t)$ and $q_2(x,t)=m(x,t)+\mathrm{i}n(x,t)$. After substituting it into Eq. \eqref{E2}, then letting the real and imaginary parts be equal to 0, we have
\begin{align}\label{E3}
\begin{split}
&-v_t+\mathcal{N}_u[u,v,m,n;\lambda_1,\lambda_2]=0,\quad u_t+\mathcal{N}_v[u,v,m,n;\lambda_1,\lambda_2]=0,\\
&-n_t+\mathcal{N}'_m[u,v,m,n;\lambda_1,\lambda_2]=0,\,\,m_t+\mathcal{N}'_n[u,v,m,n;\lambda_1,\lambda_2]=0,
\end{split}
\end{align}
accordingly, the $\mathcal{N}_u$, $\mathcal{N}_v$, $\mathcal{N}'_m$ and $\mathcal{N}'_n$ are nonlinear differential operators in space. Then the physics-informed neural networks $f_u(x,t)$, $f_v(x,t)$, $f_m(x,t)$ and $f_n(x,t)$ can be defined as
\begin{align}\label{E4}
\begin{split}
&f_u:=-v_t+\mathcal{N}_u[u,v,m,n;\lambda_1,\lambda_2],\,\quad f_v:=u_t+\mathcal{N}_v[u,v,m,n;\lambda_1,\lambda_2],\\
&f_m:=-n_t+\mathcal{N}'_m[u,v,m,n;\lambda_1,\lambda_2],\,f_n:=m_t+\mathcal{N}'_n[u,v,m,n;\lambda_1,\lambda_2],
\end{split}
\end{align}

From Ref. \cite{PuJ2021}, one can know about the classical PINN method could not accurately reconstruct some solutions with complex forms in some complicated nonlinear systems, while the PINN approach with neuron-wise locally adaptive activation function and slope recovery term can improve the convergence speed and stability of the loss function in the training process. Therefore, considering the training accuracy, performance requirements and the structural complexity of multi-component coupled nonlinear systems, based on the aforementioned PINN method, we extend and propose an IPINN model for investigating the data-driven licalized waves of the Manakov system effectively in this paper. It changes the slope of the activation function adaptively, resulting in non-vanishing gradients and faster training of the network.

We establish an IPINN of depth $D$ corresponding to the NN with an input layer, $D-1$ hidden-layers and an output layer. In the $d$th hidden-layer, $N_d$ number of neurons are present. Each hidden-layer of the IPINN receives an output $\textbf{x}^{d-1}\in\mathbb{R}^{N_{d-1}}$ from the previous layer, where an affine transformation can be written as follows form
\begin{align}\label{E-buchong}
\mathcal{L}_d(\textbf{x}^{d-1})\triangleq\textbf{W}^d\textbf{x}^{d-1}+\textbf{b}^d,
\end{align}
where the network weights $\textbf{W}^{d}\in\mathbb{R}^{N_d\times N_{d-1}}$ and bias term $\textbf{b}^d\in\mathbb{R}^{N_d}$ associated with the $d$th layer. Specifically, we define such neuron-wise locally adaptive activation function as
\begin{align}\nonumber
\sigma\left(na^d_i\left(\mathcal{L}_d\left(\textbf{x}^{d-1}\right)\right)_i\right),\,d=1,2,\cdots,D-1,\,i=1,2,\cdots,N_d,
\end{align}
where $\sigma$ is the activation function, and $n>1$ is a scaling factor and $\{a^d_i\}$ are additional $\sum\limits_{d=1}^{D-1}N_d$ parameters to be optimized. Note that, there is a critical scaling factor $n_{c}$, and the optimization algorithm will become sensitive when $n\geqslant n_c$ in each problem set. The neuron activation function acts as a vector activation function in each hidden layer, and each neuron has its own slope of activation function.

The IPINN with neuron-wise locally adaptive activation function can be represented as
\begin{align}\label{E5}
q(\textbf{x};\bar{\Theta})=\big(\left(\mathcal{L}_D\right)_{i'}\circ\sigma\circ na^{D-1}_{i}\left(\mathcal{L}_{D-1}\right)_{i}\circ\cdots\circ\sigma\circ na^1_i\left(\mathcal{L}_1\right)_i\big)(\textbf{x}),\,i'=1,2,3,4,
\end{align}
where $\textbf{x}$ and $q(\textbf{x};\bar{\Theta})$ represent the two inputs and four outputs in the IPINN, respectively. The set of trainable parameters $\bar{\Theta}\in\bar{\mathcal{P}}$ consists of $\big\{\textbf{W}^d,\textbf{b}^d\big\}_{d=1}^{D}$, unknown parameters ($\lambda_1$ and $\lambda_2$) to be learned and $\big\{a_i^d\big\}_{d=1}^{D-1},\forall i=1,2,\cdots,N_d$, $\bar{\mathcal{P}}$ is the parameter space. In this paper, the initialization of scalable parameters are carried out in the case of $na_i^d=1,\forall n\geqslant1$.

The resulting optimization algorithm will attempt to find the optimized parameters including the weights, biases and additional coefficients in the activation to minimize the new loss function defined as
\begin{align}\label{E6}
\mathscr{L}(\bar{\Theta})=Loss=Loss_{q_1}+Loss_{q_2}+Loss_{f_1}+Loss_{f_2}+Loss_a,
\end{align}
where $Loss_{q_1}, Loss_{q_2}, Loss_{f_1}$ and $Loss_{f_2}$ are defined as following
\begin{align}\label{E7}
\begin{split}
&Loss_{q_1}=\frac{1}{N_q}\left[\sum^{N_q}_{j=1}\big|\hat{u}(x^j,t^j)-u^j\big|^2+\sum^{N_q}_{j=1}\big|\hat{v}(x^j,t^j)-v^j\big|^2\right],\\
&Loss_{q_2}=\frac{1}{N_q}\left[\sum^{N_q}_{j=1}\big|\hat{m}(x^j,t^j)-m^j\big|^2+\sum^{N_q}_{j=1}\big|\hat{n}(x^j,t^j)-n^j\big|^2\right],
\end{split}
\end{align}
and
\begin{align}\label{E8}
\begin{split}
Loss_{f_1}=\frac{1}{N_f}\left[\sum^{N_f}_{l=1}\big|f_u(x_f^l,t_f^l)\big|^2+\sum^{N_f}_{l=1}\big|f_v(x_f^l,t_f^l)\big|^2\right],\\
Loss_{f_2}=\frac{1}{N_f}\left[\sum^{N_f}_{l=1}\big|f_m(x_f^l,t_f^l)\big|^2+\sum^{N_f}_{l=1}\big|f_n(x_f^l,t_f^l)\big|^2\right],
\end{split}
\end{align}
where $\{x^j,t^j,u^j,v^j,m^j,n^j\}^{N_q}_{j=1}$ denote the initial-boundary value inputs data on Eqs. \eqref{E3} and \eqref{E4}. Here $\hat{u}(x^j,t^j), \hat{v}(x^j,t^j), \hat{m}(x^j,t^j)$ and $\hat{n}(x^j,t^j)$ represent the optimal training outputs data through the IPINN. Furthermore, $\{x_f^l,t_f^l\}^{N_{f}}_{l=1}$ represent the collocation points on networks $f_u(x,t)$, $f_v(x,t)$, $f_m(x,t)$ and $f_n(x,t)$. The last slope recovery term $Loss_a$ in the loss function \eqref{E6} is defined as
\begin{align}\label{E9}
Loss_a=\frac{1}{\frac{N_a}{D-1}\sum\limits_{d=1}^{D-1}\mathrm{exp}\Bigg(\frac{\sum\limits_{i=1}^{N_d}a_i^d}{N_d}\Bigg)},
\end{align}
where $1/N_a$ is hyper-parameter for slope recovery term $Loss_a$, and we all take $N_a=100$ for dominating the loss function and ensuring that the final loss value is not too large in this paper. Here, term $Loss_a$ forces the NN to increase the activation slope value quickly, which ensures the non-vanishing of the gradient of the loss function and improves the network's training speed. Consequently, $Loss_{q_1}$ and $Loss_{q_2}$ correspond to the loss on the initial and boundary data, the $Loss_{f_1}$ and $Loss_{f_2}$ penalize the Manakov system not being satisfied on the collocation points, and the $Loss_a$ changes the topology of $Loss$ function and improves the convergence speed and network optimization ability. Moreover, in order to better measure the training error, we introduce $\mathbb{L}_2$ norm error, which is defined as follows
\begin{align}\nonumber
\mathrm{Error}=\frac{\sqrt{\sum\limits_{k=1}^{N}\big|q^{\mathrm{exact}}(\textbf{x}_{k})-q^{\mathrm{predict}}(\textbf{x}_{k};\bar{\Theta})\big|^2}}{\sqrt{\sum\limits_{k=1}^{N}\big|q^{\mathrm{exact}}(\textbf{x}_{k})\big|^2}},
\end{align}
where $q^{\mathrm{predict}}(\textbf{x}_{k};\bar{\Theta})$ and $q^{\mathrm{exact}}(\textbf{x}_{k})$ represent the model training prediction solution and exact analytical solution at point $\textbf{x}_{k}=(x_k,t_k)$, respectively.

In order to understand the IPINN approach more clearly, the IPINN algorithm flow chart of the Manakov system is shown in following Fig. \ref{F1}, where one can see the NN along with the supplementary physics-informed part, and the loss function is evaluated using the contribution from the NN part as well as the residual from the governing equation given by the physics-informed part. Then, one can seek the optimal values of weights $\textbf{W}$, biases $\textbf{b}$, parameters $\lambda_1$ and $\lambda_2$ as well as scalable parameter $a^d_i$ in order to minimize the loss function below certain tolerance $\varepsilon$ until a prescribed maximum number of iterations. From Fig. \ref{F1}, since the Manakov system contains two components $q_1(x,t)$ and $q_2(x,t)$, one can see that the ``NN" part has four output functions $\{u,v,m,n\}$, and there are four nonlinear equation constraints in the ``PDE" part, that is, in terms of the nonlinear coupled system with more components, the number of output functions and nonlinear equation constraints of the IPINN will increase multiply. Furthermore, in order to further understand the IPINN, we also showcase the corresponding procedure steps of the IPINN with adaptive activation function and slope recovery term in the following table.
\begin{figure}[htbp]
\centering
\begin{minipage}[t]{0.99\textwidth}
\centering
\includegraphics[height=9cm,width=15cm]{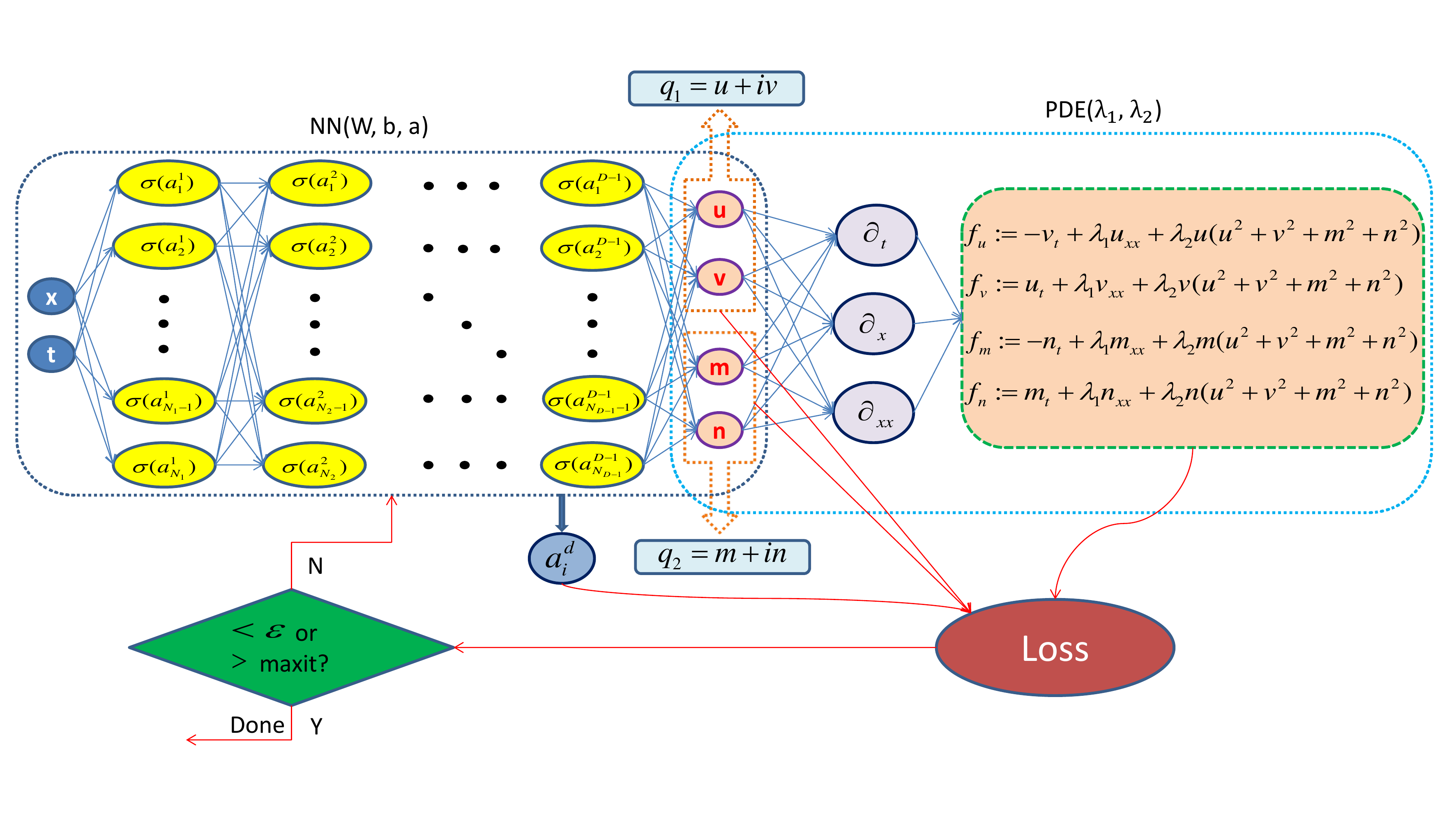}
\end{minipage}
\centering
\caption{(Color online) Schematic of IPINN for the Manakov system with unknown parameters $\lambda_1$ and $\lambda_2$. The left NN is the universal approximation network while the right one induced by the governing equation is the physics-informed network. The two NNs share hyper-parameters and they both contribute to the loss function.}
\label{F1}
\end{figure}

\begin{table}[htbp]
  \label{Tab:bookRWCal}
  \centering
  \begin{tabular}{p{15cm}}
  \toprule[2pt]
  Algorithm: IPINN algorithm with adaptive activation function and slope recovery term.  \\
  \midrule[2pt]
  \quad \textbf{Step 1}: Specification of training set in computational domain:\\
  \quad \emph{Training data}: $\{x^j,t^j,u^j,v^j,m^j,n^j\}^{N_q}_{j=1}$, \emph{Residual training points}: $\{x_f^l,t_f^l\}^{N_{f}}_{l=1}.$\\
  \quad \textbf{Step 2}: Construct NN $q(\textbf{x};\bar{\Theta})$ with random initialization of parameters $\bar{\Theta}$.\\
  \quad \textbf{Step 3}: Construct the residual NN $\{f_u,f_v,f_m,f_q\}$ by substituting surrogate $q(\textbf{x};\bar{\Theta})$ into the governing equations using automatic differentiation and other arithmetic operations.\\
  \quad \textbf{Step 4}: Specification of the loss function $\mathscr{L}(\bar{\Theta})$ that includes the slope recovery term.\\
  \quad \textbf{Step 5}: Find the best parameters $\bar{\Theta}^*$ using a suitable optimization method for minimizing the loss function $\mathscr{L}(\bar{\Theta})$ as\\
  \qquad\qquad\qquad\qquad\qquad\qquad\qquad\quad $\bar{\Theta}^*=\mathop{\mathrm{arg\,min}}\limits_{\bar{\Theta}\in\bar{\mathcal{P}}}\mathscr{L}(\bar{\Theta})$.\\
  \bottomrule[2pt]
  \end{tabular}
\end{table}
In this method, all loss functions are simply optimized by employing the Adam and L-BFGS algorithm, in which Adam optimization algorithm is a variant of the traditional stochastic gradient descent, whereas L-BFGS optimization algorithm is a full-batch gradient descent optimization algorithm based on a quasi-Newton method.\cite{Kingma2014,Liu1989}. Especially, the scalable parameters in the adaptive activation function are initialized generally as $n=10,a_i^d=0.1$, unless otherwise specified. In addition, we select relatively simple multi-layer perceptrons (i.e., feedforward NNs) with the Xavier initialization and the hyperbolic tangent ($\tanh$) as activation function. All the codes in this article is based on Python 3.7 and Tensorflow 1.15, and all numerical experiments reported here are run on a DELL Precision 7920 Tower computer with 2.10 GHz 8-core Xeon Silver 4110 processor, 64 GB memory and 11 GB Nvidia GeForce GTX 1080 Ti video card.

\section{Data-driven vector solitons and breathers of the Manakov system}
In this section, we will focus on the data-driven vector solitons and breather for the Manakov system with $\lambda_1=1$ and $\lambda_2=2$ by means of 9 hidden layers deep IPINN with 40 neurons per layer, thus the physics-informed parts Eq. \eqref{E4} of the IPINN for Manakov sysem \eqref{E1} become the following formula
\begin{align}\label{E-pi}
\begin{split}
&f_u:=-v_t+u_{xx}+2u(u^2+v^2+m^2+n^2),\\
&f_v:=u_t+v_{xx}+2v(u^2+v^2+m^2+n^2),\\
&f_m:=-n_t+m_{xx}+2m(u^2+v^2+m^2+n^2),\\
&f_n:=m_t+n_{xx}+2n(u^2+v^2+m^2+n^2).
\end{split}
\end{align}
The vector $N$-soliton and breather of the Manakov system with $\lambda_1=1$ and $\lambda_2=2$ have been derived by the Riemann-Hilbert method \cite{Yang2010}, the general vector form of $N$-soliton solution can be expressed as
\begin{align}\label{E10}
\begin{bmatrix} q_1(x,t) \\ q_2(x,t) \\\end{bmatrix}=2\mathrm{i}\sum_{j,k=1}^N\bigg(\begin{array}{c} \alpha_j \\ \beta_j \\ \end{array}\bigg) \mathrm{e}^{\theta_j-\theta^*_k}(M^{-1})_{jk},
\end{align}
where $\theta_l=-\mathrm{i}\zeta_lx-2\mathrm{i}\zeta_l^2t$, and $\alpha_l$, $\beta_l$ and $\zeta_l$ are arbitrary constants in complex space $(l=1,2,\cdots,N)$. The symbol $``*"$ represents complex conjugate, and $M$ is $N\times N$ matrix whose elements are given by
\begin{align}\label{E11}
M_{jk}=\frac{1}{\zeta_j^*-\zeta_k}\Big[\mathrm{e}^{-(\theta_j^*+\theta_k)}+(\alpha_j^*\alpha_k+\beta_j^*\beta_k)\mathrm{e}^{\theta_j^*+\theta_k}\Big].
\end{align}

Next, we will exhibit the vector one-soliton, two-soliton and breather for the Manakov system in more detail.

$\bullet$ \textbf{Vector One-Solitons}

In order to get the vector one-soliton solution, we set $N=1$ in the above formulae \eqref{E10}, and take $\alpha_1=1/4,\,\beta_1=1/4$ and $\zeta_1=1+2\mathrm{i}$, then the vector one-soliton solution in the Manakov system can be rewritten as
\begin{align}\label{E12}
\begin{split}
&q_{1,\mathrm{os}}=\frac{16\mathrm{e}^{2\mathrm{i}(6t-x)}}{8\mathrm{e}^{-4x-16t}+\mathrm{e}^{4x+16t}},\\
&q_{2,\mathrm{os}}=\frac{16\mathrm{e}^{2\mathrm{i}(6t-x)}}{8\mathrm{e}^{-4x-16t}+\mathrm{e}^{4x+16t}}.
\end{split}
\end{align}

$\bullet$ \textbf{Vector Two-Solitons}

When $N=2$ and $\mathrm{Re}\zeta_1\neq\mathrm{Re}\zeta_2$, solution \eqref{E10} describe the collision of two one-soliton, to illustrate, we take $\zeta_1=0.1+0.7\mathrm{i},\,\zeta_2=-0.1+0.4\mathrm{i},\,\alpha_1=\alpha_2=1,\,\beta_1=1/4,\,\beta_2=0$ and the corresponding solution are derived as
\begin{small}
\begin{align}\label{E13}
\begin{split}
&q_{1,\mathrm{ts}}=\frac{\Omega}{3584\cos\big(\frac{33}{25}t-\frac25x\big)-2000\mathrm{e}^{-\frac{22}{25}t-\frac35x}-333\mathrm{e}^{\frac{6}{25}t+\frac{11}{5}x}-208\mathrm{e}^{-\frac{6}{25}t-\frac{11}{5}x}-2125\mathrm{e}^{\frac{22}{25}+\frac35x}},\\
&q_{2,\mathrm{ts}}=\frac{\Delta}{3584\cos\big(\frac{33}{25}t-\frac25x\big)-2000\mathrm{e}^{-\frac{22}{25}t-\frac35x}-333\mathrm{e}^{\frac{6}{25}t+\frac{11}{5}x}-208\mathrm{e}^{-\frac{6}{25}t-\frac{11}{5}x}-2125\mathrm{e}^{\frac{22}{25}+\frac35x}},
\end{split}
\end{align}
\end{small}
with
\begin{small}
\begin{align}\nonumber
\begin{split}
&\Omega=\frac85\mathrm{i}\big(1036\mathrm{i}{\rm e}^{{\frac{48}{25}}\mathrm{i}t-\frac15\mathrm{i}x-\frac{8}{25}t+\frac45x}+1036\mathrm{i}{\rm e}^{{\frac{48}{25}}\mathrm{i}t-\frac15\mathrm{i}x+{\frac{8}{25}}t-\frac45x}-464\mathrm{i}{{\rm e}^{\frac35\mathrm{i}t+\frac15\mathrm{i}x-{\frac{14}{25}}t-\frac75x}}-\\
&\qquad339\mathrm{i}{{\rm e}^{\frac35\mathrm{i}t+\frac15\mathrm{i}x+{\frac{14}{25}}t+\frac75x}}-448{{\rm e}^{{\frac{48}{25}}\mathrm{i}t-\frac15\mathrm{i}x+{\frac{8}{25}}t-\frac45x}}+448{{\rm e}^{\frac35\mathrm{i}t+\frac15\mathrm{i}x-{\frac{14}{25}}t-\frac75x}}+\\
&\qquad448{{\rm e}^{{\frac{48}{25}}\mathrm{i}t-\frac15\mathrm{i}x-{\frac{8}{25}}t+\frac45x}}-448{{\rm e}^{\frac35\mathrm{i}t+\frac15\mathrm{i}x+{\frac{14}{25}}t+\frac75x}}\big),\\
&\Delta={\frac{56}{5}}\mathrm{i}\big(125\mathrm{i}{{\rm e}^{{\frac{48}{25}}\mathrm{i}t-\frac15\mathrm{i}x-{\frac{8}{25}}t+\frac45x}}+37\mathrm{i}{{\rm e}^{{\frac{48}{25}}\mathrm{i}t-\frac15\mathrm{i}x+{\frac{8}{25}}t-\frac45x}}-88\mathrm{i}{{\rm e}^{\frac35\mathrm{i}t+\frac15\mathrm{i}x+{\frac{14}{25}}t+\frac75x}}-\\
&\qquad16{\rm e}^{{\frac{48}{25}}\mathrm{i}t-\frac15\mathrm{i}x+{\frac{8}{25}}t-\frac45x}-16{\rm e}^{\frac35\mathrm{i}t+\frac15\mathrm{i}x+{\frac{14}{25}}t+\frac75x}\big).
\end{split}
\end{align}
\end{small}

$\bullet$ \textbf{Vector Breathers}

Specially, if $N=2$ and $\mathrm{Re}\zeta_1=\mathrm{Re}\zeta_2$, one can obtain the vector breather solution for the Manakov system, by substituting $\zeta_1=0.7\mathrm{i},\,\zeta_2=0.4\mathrm{i},\,\alpha_1=\alpha_2=1,\,\beta_1=1/4,\,\beta_2=0$ into Eq. \eqref{E10}, we have
\begin{small}
\begin{align}\label{E14}
\begin{split}
&q_{1,\mathrm{bs}}=\frac{88}{5}\frac{84{{\rm e}^{{\frac{49}{25}}\mathrm{i}t+\frac45x}}+84{{\rm e}^{{\frac{49}{25}}\mathrm{i}t-\frac45x}}-48{{\rm e}^{-\frac75x+{\frac{16}{25}}\mathrm{i}t}}-37{{\rm e}^{\frac75x+{\frac{16}{25}}\mathrm{i}t}}}{-3584\cos\big({\frac{33}{25}}t\big) +1936{\rm e}^{-\frac35x}+265{\rm e}^{{\frac{11}{5}}x}+144{\rm e}^{-{\frac{11}{5}}x}+2057{\rm e}^{\frac35x}},\\
&q_{2,\mathrm{bs}}=\frac{616}{5}\frac{11{{\rm e}^{{\frac{49}{25}}\mathrm{i}t+\frac45x}}+3{{\rm e}^{{\frac{49}{25}}\mathrm{i}t-\frac45x}}-8{{\rm e}^{\frac75x+{\frac{16}{25}}\mathrm{i}t}}}{-3584\cos\big({\frac{33}{25}}t\big) +1936{{\rm e}^{-\frac35x}}+265{{\rm e}^{{\frac{11}{5}}x}}+144{{\rm e}^{-{\frac{11}{5}}x}}+2057{{\rm e}^{\frac35x}}}.
\end{split}
\end{align}
\end{small}

Furthermore, we can also obtain higher-order vector solitons and breathers by taking different $N$ from Eq. \eqref{E10}. Accordingly, the form and dynamic behavior of higher-order vector solitons and breathers are more complex, with more parameters and stronger adjustability. In this paper, we only consider data-driven vector one-soliton, two-soliton and breather, and the higher-order case is similar.

\subsection{Data-driven vector one-solitons}
In this subsection, we will focus on the data-driven vector one-solitons with the initial conditions $q_{r}^0(x)$ and Dirichlet boundary conditions $q_{r}^{\mathrm{lb}}(t)$ and $q_{r}^{\mathrm{ub}}(t),\,r=1,2$. Here one can take $[L_0,L_1]$ and $[T_0,T_1]$ in Eq. \eqref{E1} as $[-3.0,3.0]$ and $[-0.2,0.2]$, respectively. Then, we obtain the corresponding initial conditions as shown below
\begin{align}\label{E15}
\begin{split}
&q_{r}^0(x)=q_{r,\mathrm{os}}(x,-0.2),\,x\in[-3.0,3.0],
\end{split}
\end{align}
and the Dirichlet boundary conditions
\begin{align}\label{E16}
q_{r}^{\mathrm{lb}}(t)=q_{r,\mathrm{os}}(-3.0,t),\,q_{r}^{\mathrm{ub}}(t)=q_{r,\mathrm{os}}(3.0,t),\,t\in[-0.2,0.2],\,r=1,2.
\end{align}

We employ the traditional finite difference scheme on even grids in Matlab to simulate vector one-solitons which contains the initial datas \eqref{E15} and boundary datas \eqref{E16} to acquire the original training data. More specifically, we divide spatial region $[-3.0,3.0]$ into 2500 points and temporal region $[-0.2,0.2]$ into 1000 points, the vector one-solitons \eqref{E12} is discretized into 1000 snapshots accordingly. We generate a smaller training dataset containing initial-boundary data by randomly extracting $N_q=2500$ from original dataset and $N_f = 40000$ collocation points which are generated by the Latin Hypercube Sampling method (LHS) \cite{Stein1987}. After giving a dataset of initial and boundary points, the latent vector one-solitons $q_r(x,t)(r=1,2)$ have been successfully learned by tuning all learnable parameters of the 9 layers IPINN with 40 neurons per layer and regulating the loss function, in which we utilize the 50000 steps Adam firstly and subsequently use 15134 steps L-BFGS optimizations for minimizing the loss function \eqref{E6}. The relative $\mathbb{L}_2$ errors of the IPINN model are 3.813803$\rm e^{-2}$ for $q_1(x,t)$ and 3.754237$\rm e^{-2}$ for $q_2(x,t)$, the total number of iterations is 65134.

\begin{figure}[htbp]
\centering
\subfigure[]{
\begin{minipage}[t]{0.48\textwidth}
\centering
\includegraphics[height=6.5cm,width=6cm]{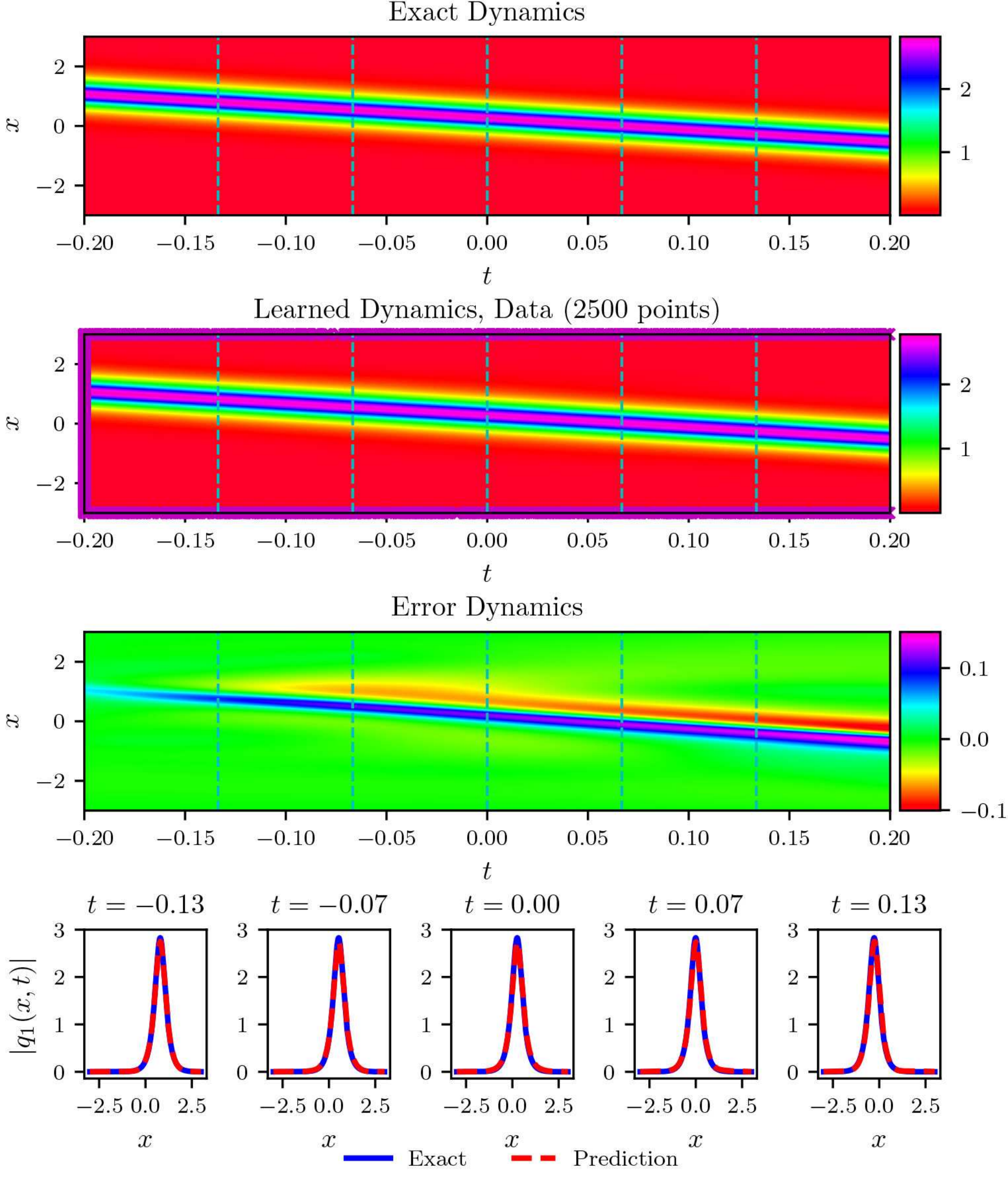}
\end{minipage}
}%
\subfigure[]{
\begin{minipage}[t]{0.48\textwidth}
\centering
\includegraphics[height=6.5cm,width=5cm]{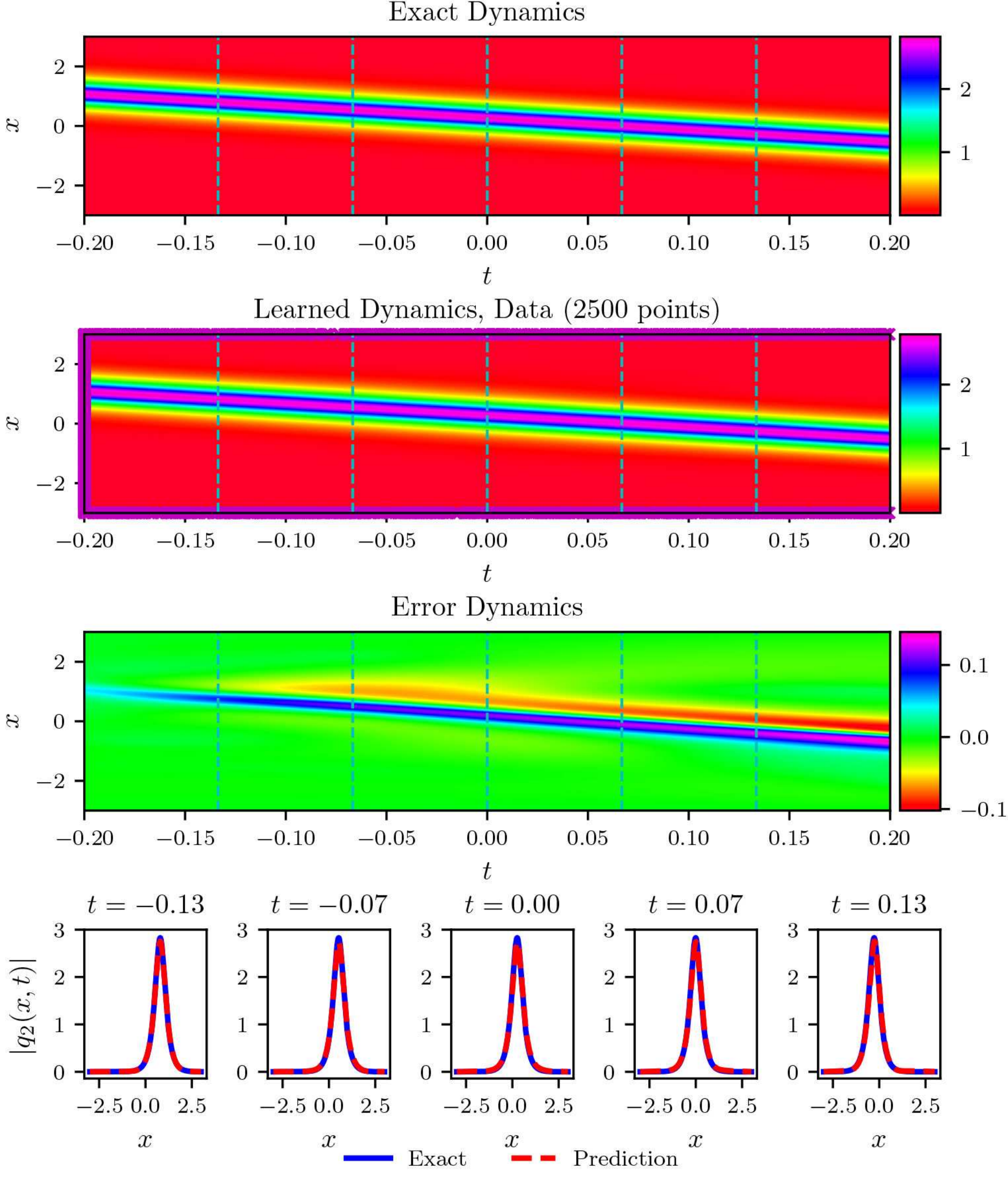}
\end{minipage}%
}%
\centering
\caption{(Color online) The vector one-solitons $q_r(x,t)$ $(r=1,2)$ resulted from the IPINN with the randomly chosen initial and boundary points $N_q=2500$ which have been shown by using mediumorchid $``\times"$ in learned dynamics , and $N_f = 40000$ collocation points in the corresponding spatiotemporal region. The exact, learned and error dynamics density plots for the vector one-solitons $q_r(x,t)$ with five distinct tested times $t=-0.13, -0.07, 0.00, 0.07$ and 0.13 (darkturquoise dashed lines), and the sectional drawings which contain the learned and explicit vector one-solitons $q_r(x,t)$ at the aforementioned five distinct times: (a) The density plots and sectional drawings for the one-soliton $q_1(x,t)$; (b) The density plots and sectional drawings for the one-soliton $q_2(x,t)$.}
\label{F2}
\end{figure}

\begin{figure}[htbp]
\centering

\begin{minipage}[t]{0.99\textwidth}
\centering
\includegraphics[height=6cm,width=14cm]{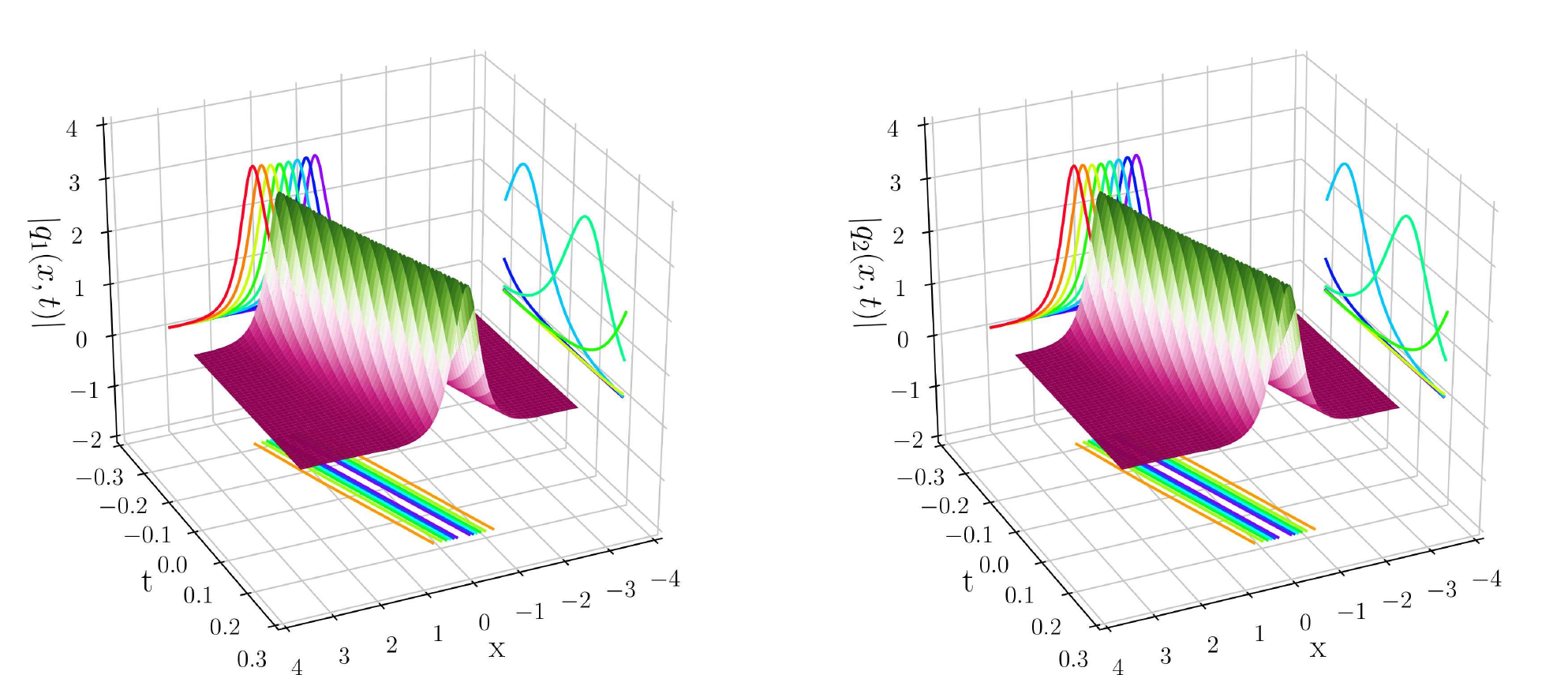}
\end{minipage}
\centering
\caption{(Color online) The three-dimensional plots with contour map on three planes of the predicted vector one-solitons $q_r(x,t)$ $(r=1,2)$ based on the IPINN: (Left side panel) The 3D plot for the $q_1(x,t)$; (Right side panel) The 3D plot for the $q_2(x,t)$.}
\label{F3}
\end{figure}

\begin{figure}[htbp]
\centering
\subfigure[]{
\begin{minipage}[t]{0.48\textwidth}
\centering
\includegraphics[height=5cm,width=7cm]{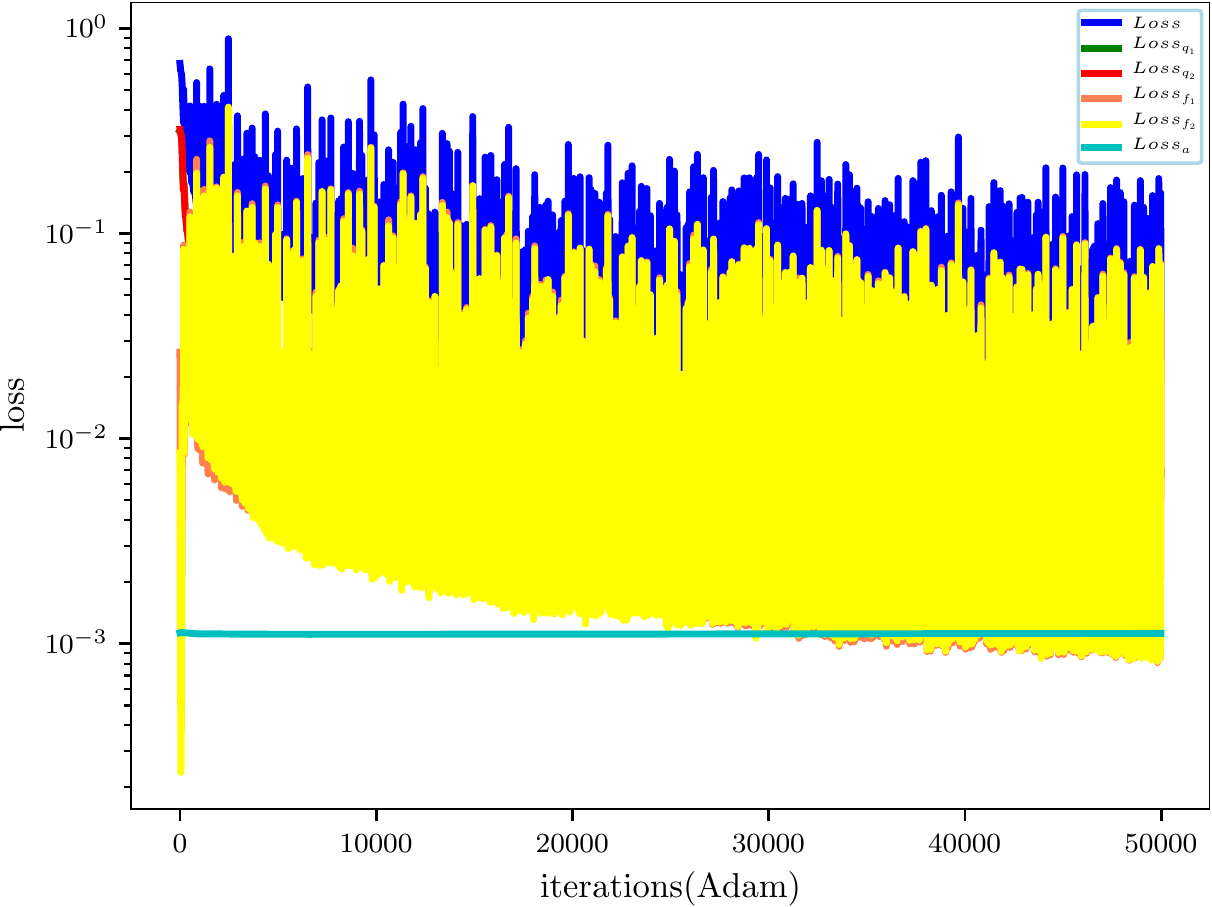}
\end{minipage}
}%
\subfigure[]{
\begin{minipage}[t]{0.48\textwidth}
\centering
\includegraphics[height=5cm,width=7cm]{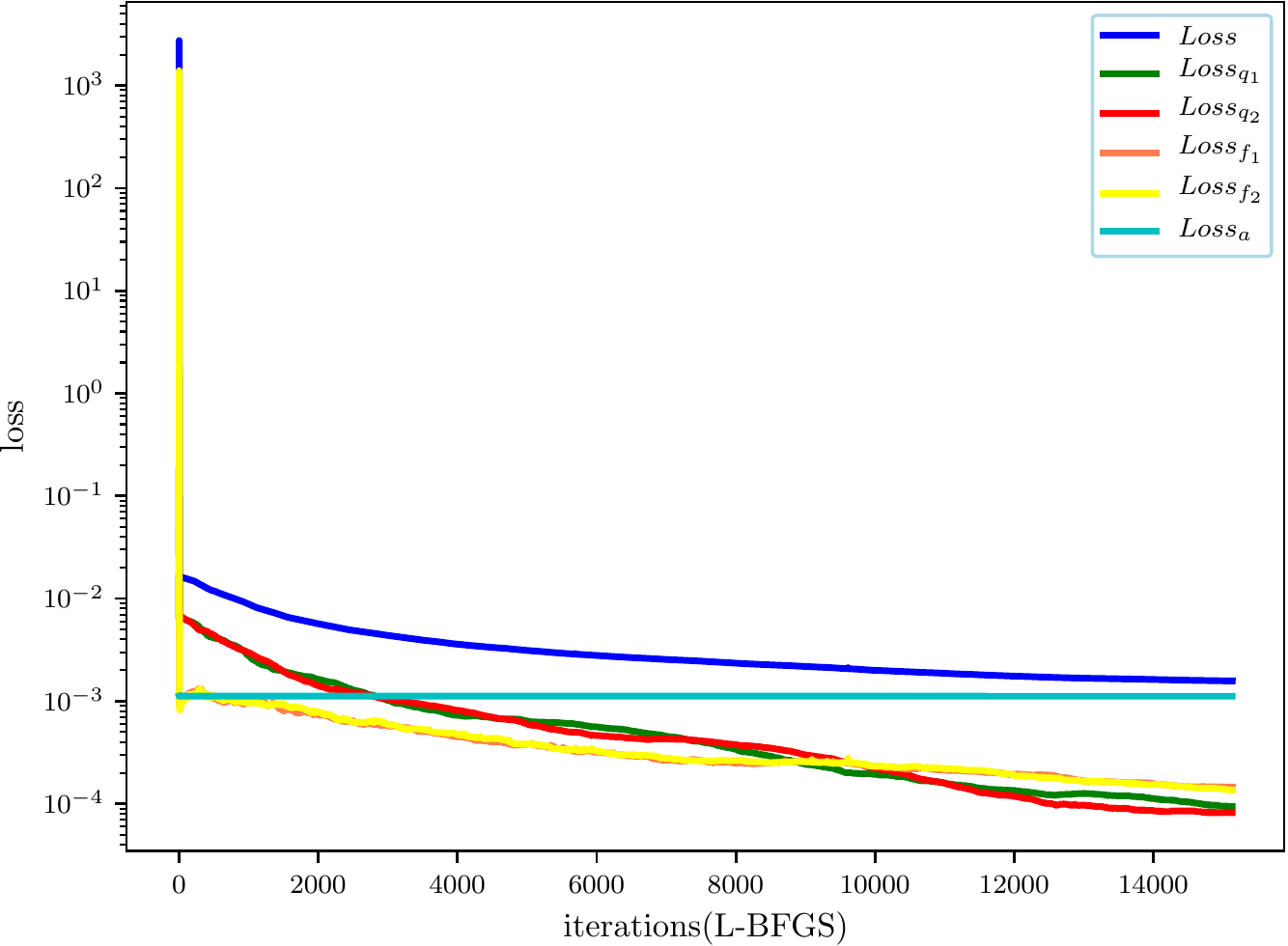}
\end{minipage}%
}%
\centering
\caption{(Color online) The loss function curve figures of the vector one-solitons $q_r(x,t)$ $(r=1,2)$ arising from the IPINN with the 50000 steps Adam and 15134 steps L-BFGS optimizations: (a) The loss function curve for the 50000 Adam optimization iterations; (b) The loss function curve for the 15134 L-BFGS optimization iterations.}
\label{F4}
\end{figure}

Figs. \ref{F2} - \ref{F4} display the deep learning results of vector one-solitons based on the IPINN related to the initial boundary value problem \eqref{E15} and \eqref{E16} of the Manakov system \eqref{E1}. Specifically, the density plots with the corresponding peak scale for diverse dynamics which contain exact dynamics, learned dynamics and error dynamics have been exhibited in detail, and the sectional drawings of one soliton solutions $q_r(x,t)$ $(r=1,2)$ at different moments arising from the IPINN in Fig. \ref{F2}. From the bottom panels of Fig. \ref{F2} (a) and (b), one can indicate vector one-solitons propagate from right to left along the $x$-axis, and show that the amplitudes of solitons are constant with the development of time $t$. Fig. \ref{F3} exhibit the three-dimensional plots and corresponding contour maps of the predicted vector one-soliton solutions $q_r(x,t)$ stemmed from the IPINN. The loss function curve figures account for the vector one-solitons $q_r(x,t)$ have been given out in Fig. \ref{F4}, the left panel (a) of Fig. \ref{F4} demonstrates that Adam is an optimizer with oscillatory loss function curves, the left panel (b) in Fig. \ref{F4} illustrates L-BFGS is an optimization algorithm with linear loss function curves. Due to the $Loss_a$ possesses unique topology structure \eqref{E9}, the $Loss_a$ curves decrease slowly around 0.001 and have a strong stability in Fig. \ref{F4}. From Fig. \ref{F4} (a), we will find that the loss function curves of $Loss_{q_1}$, $Loss_{q_2}$, $Loss_{f_1}$ and $Loss_{f_2}$ overlap, which is caused by the consistency of the initial-boundary value conditions of $q_1(x,t)$ and $q_2(x,t)$. By the way, according to Eqs. \eqref{E10} - \eqref{E12}, one can observe that we can take unequal values of $\alpha_1$ and $\beta_1$ in Eq. \eqref{E10} to adjust the amplitude and initial boundary value conditions for the vector one-solitons. Of course, for simplicity, we take the case of equal amplitude in this subsection.

\subsection{Data-driven vector two-solitons}

In what follows, we will consider the initial-boundary value problem of the Manakov system for obtaining the data-driven vector two-solitons by applying the multilayer IPINN. Similarly, taking $[L_0,L_1]$ and $[T_0,T_1]$ in Eq. \eqref{E1} as $[-6.0,6.0]$ and $[-3.0,3.0]$ respectively, we derive the corresponding initial condition $q_{r}^0(x)$ and Dirichlet boundary conditions $q_{r}^{\mathrm{lb}}(t)$ and $q_{r}^{\mathrm{ub}}(t),\,r=1,2$ as shown in the following formulas
\begin{align}\label{E17}
\begin{split}
&q_{r}^0(x)=q_{r,\mathrm{ts}}(x,-3.0),\,x\in[-6.0,6.0],
\end{split}
\end{align}
and
\begin{align}\label{E18}
q_{r}^{\mathrm{lb}}(t)=q_{r,\mathrm{ts}}(-6.0,t),\,q_{r}^{\mathrm{ub}}(t)=q_{r,\mathrm{ts}}(6.0,t),\,t\in[-3.0,3.0],\,r=1,2.
\end{align}

By means of Matlab, we discretize the Eq. \eqref{E13} by utilizing the traditional finite difference scheme on even grids, and obtain the original training data which only contains initial data \eqref{E17} and boundary data \eqref{E18} by dividing the spatial region $[-6.0,6.0]$ into 2000 points and the temporal region $[-3.0,3.0]$ into 1500 points, the remaining data will be used to obtain training errors by comparing with predicted vector two-solitons. After that, we generate a smaller training dataset containing initial-boundary data by randomly extracting $N_q=2000$ from original training dataset and $N_f=30000$ collocation points produced via LHS in the corresponding spatiotemporal region. Then, the latent vector two-solitons $q_r(x,t)$ $(r=1,2)$ have been successfully learned by imposing a 9-hidden-layer IPINN with 40 neurons per layer, and the related loss functions are optimized through 20000 Adam iterations and 26239 L-BFGS iterations. The relative $\mathbb{L}_2$ errors of the IPINN model are 2.678111$\rm e^{-2}$ for $q_1(x,t)$ and 2.356420$\rm e^{-2}$ for $q_2(x,t)$, the total number of iterations is 46239.

\begin{figure}[htbp]
\centering
\subfigure[]{
\begin{minipage}[t]{0.48\textwidth}
\centering
\includegraphics[height=6.5cm,width=6cm]{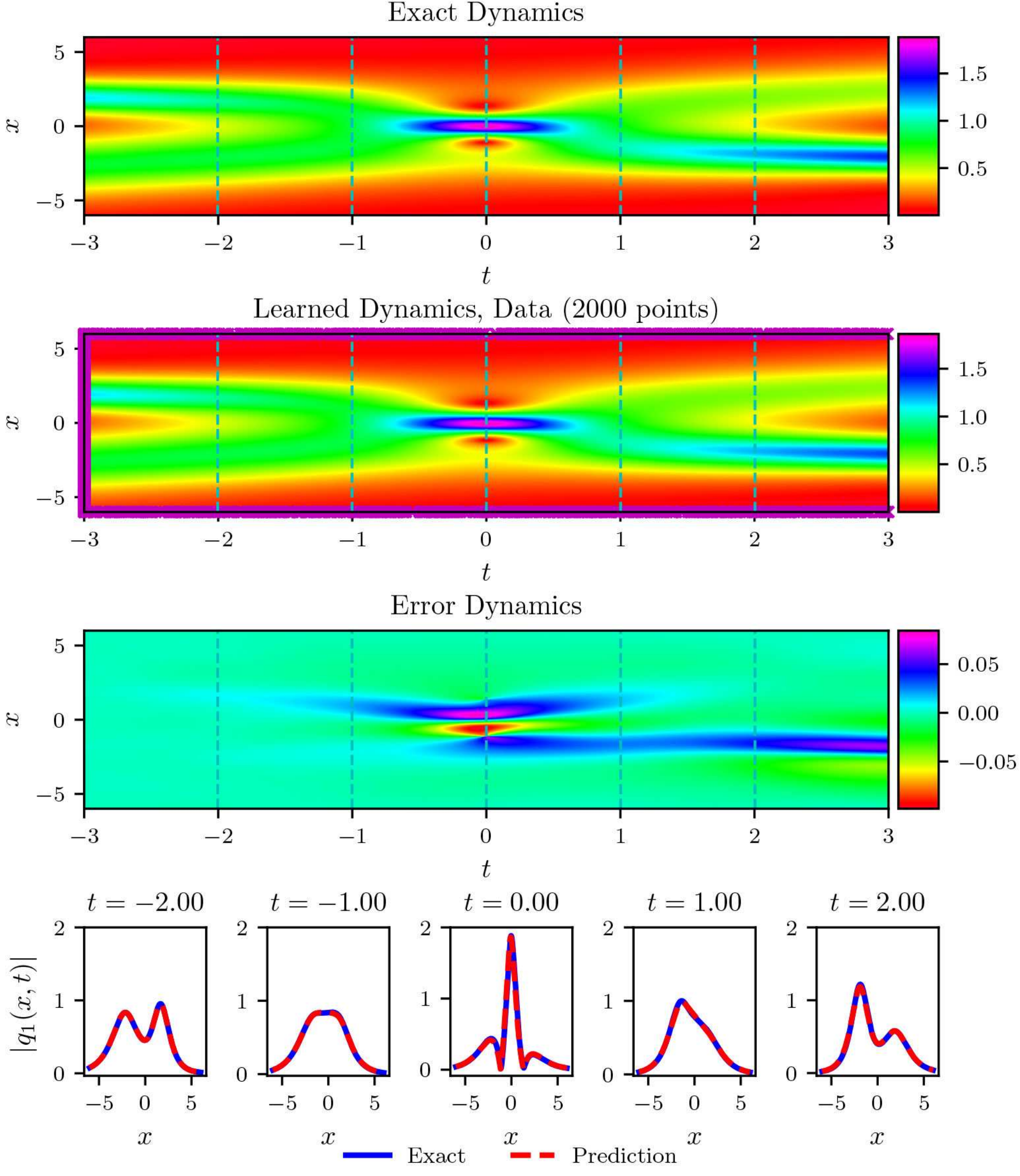}
\end{minipage}
}%
\subfigure[]{
\begin{minipage}[t]{0.48\textwidth}
\centering
\includegraphics[height=6.5cm,width=5cm]{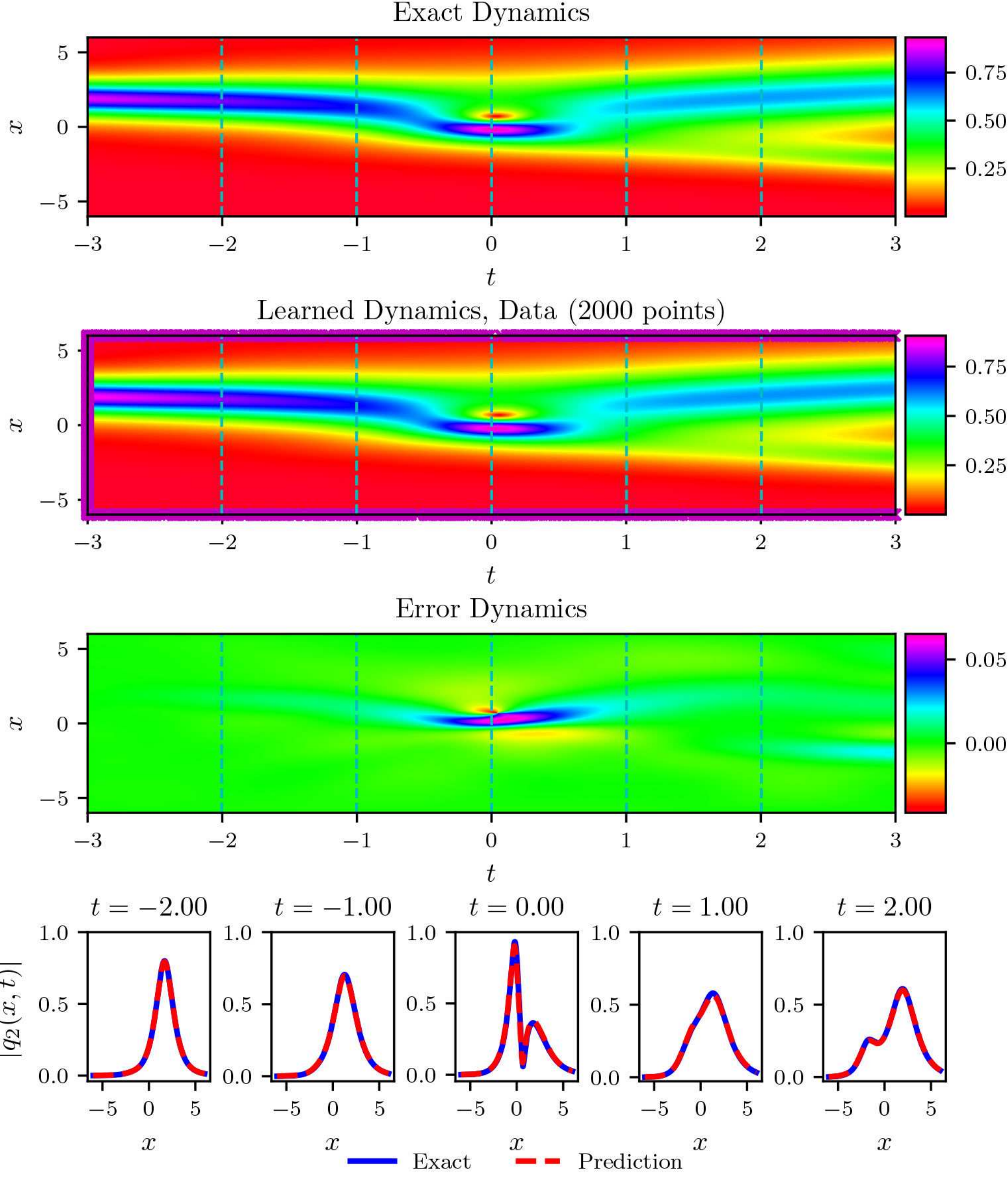}
\end{minipage}%
}%
\centering
\caption{(Color online) The vector two-solitons $q_r(x,t)$ $(r=1,2)$ resulted from the IPINN with the randomly chosen initial and boundary points $N_q=2000$ which have been shown by using mediumorchid $``\times"$ in learned dynamics , and $N_f = 30000$ collocation points in the corresponding spatiotemporal region. The exact, learned and error dynamics density plots for the vector two-solitons $q_r(x,t)$ with five distinct tested times $t=-2.00, -1.00, 0.00, 1.00$ and 2.00 (darkturquoise dashed lines), and the sectional drawings which contain the learned and explicit vector two-solitons $q_r(x,t)$ at the aforementioned five distinct times: (a) The density plots and sectional drawings for the two-soliton $q_1(x,t)$; (b) The density plots and sectional drawings for the two-soliton $q_2(x,t)$.}
\label{F5}
\end{figure}

\begin{figure}[htbp]
\centering

\begin{minipage}[t]{0.99\textwidth}
\centering
\includegraphics[height=6cm,width=14cm]{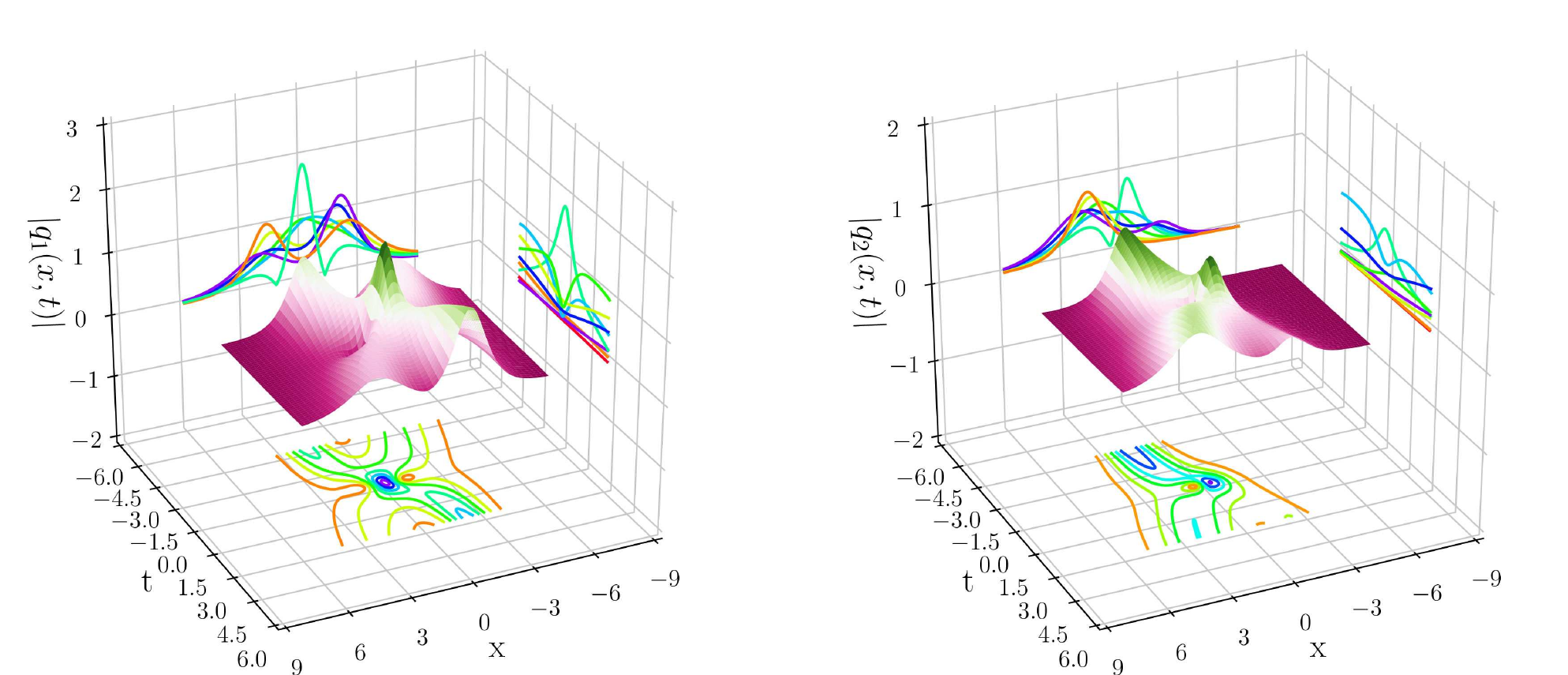}
\end{minipage}
\centering
\caption{(Color online) The three-dimensional plots with contour map on three planes of the predicted vector two-solitons $q_r(x,t)$ $(r=1,2)$ based on the IPINN: (Left side panel) The 3D plot for the $q_1(x,t)$; (Right side panel) The 3D plot for the $q_2(x,t)$.}
\label{F6}
\end{figure}

\begin{figure}[htbp]
\centering
\subfigure[]{
\begin{minipage}[t]{0.48\textwidth}
\centering
\includegraphics[height=5cm,width=7cm]{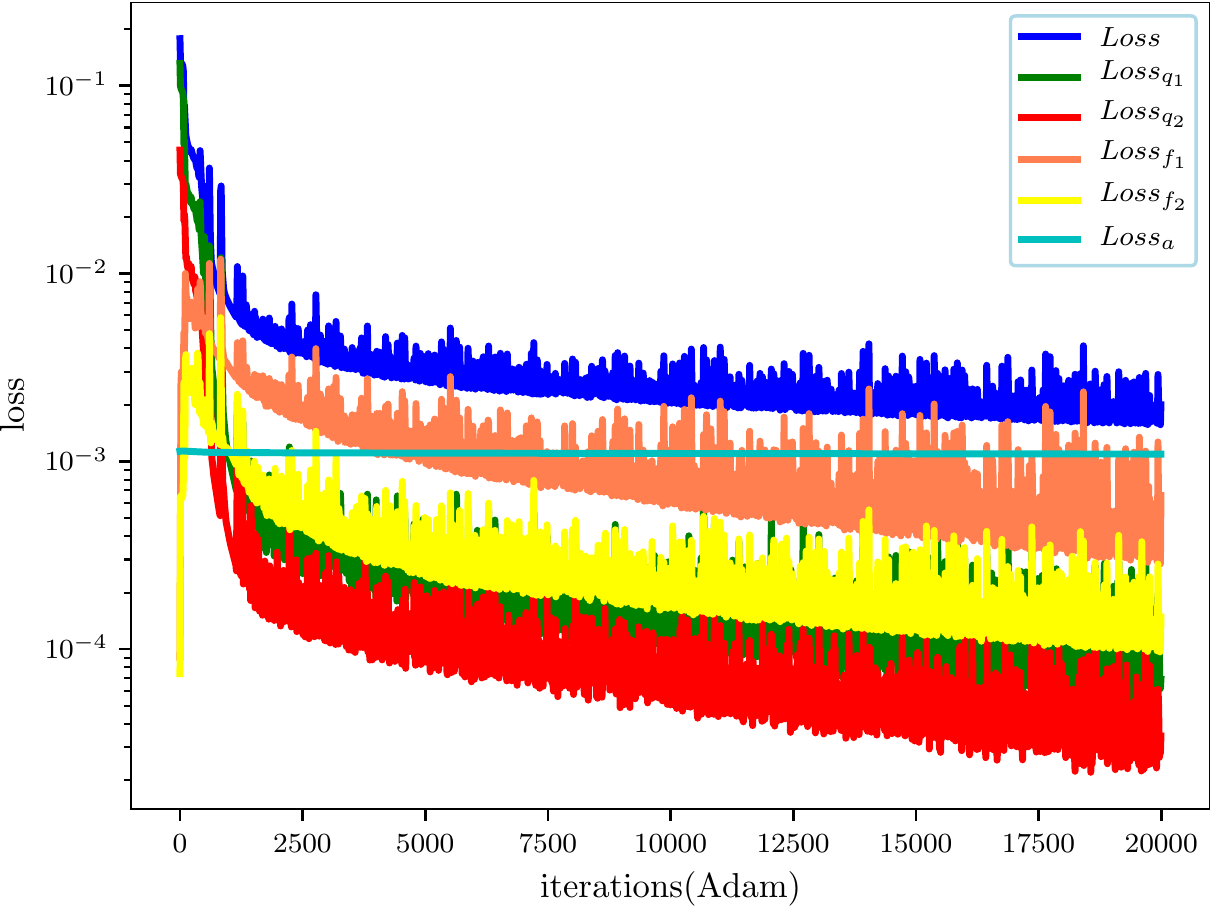}
\end{minipage}
}%
\subfigure[]{
\begin{minipage}[t]{0.48\textwidth}
\centering
\includegraphics[height=5cm,width=7cm]{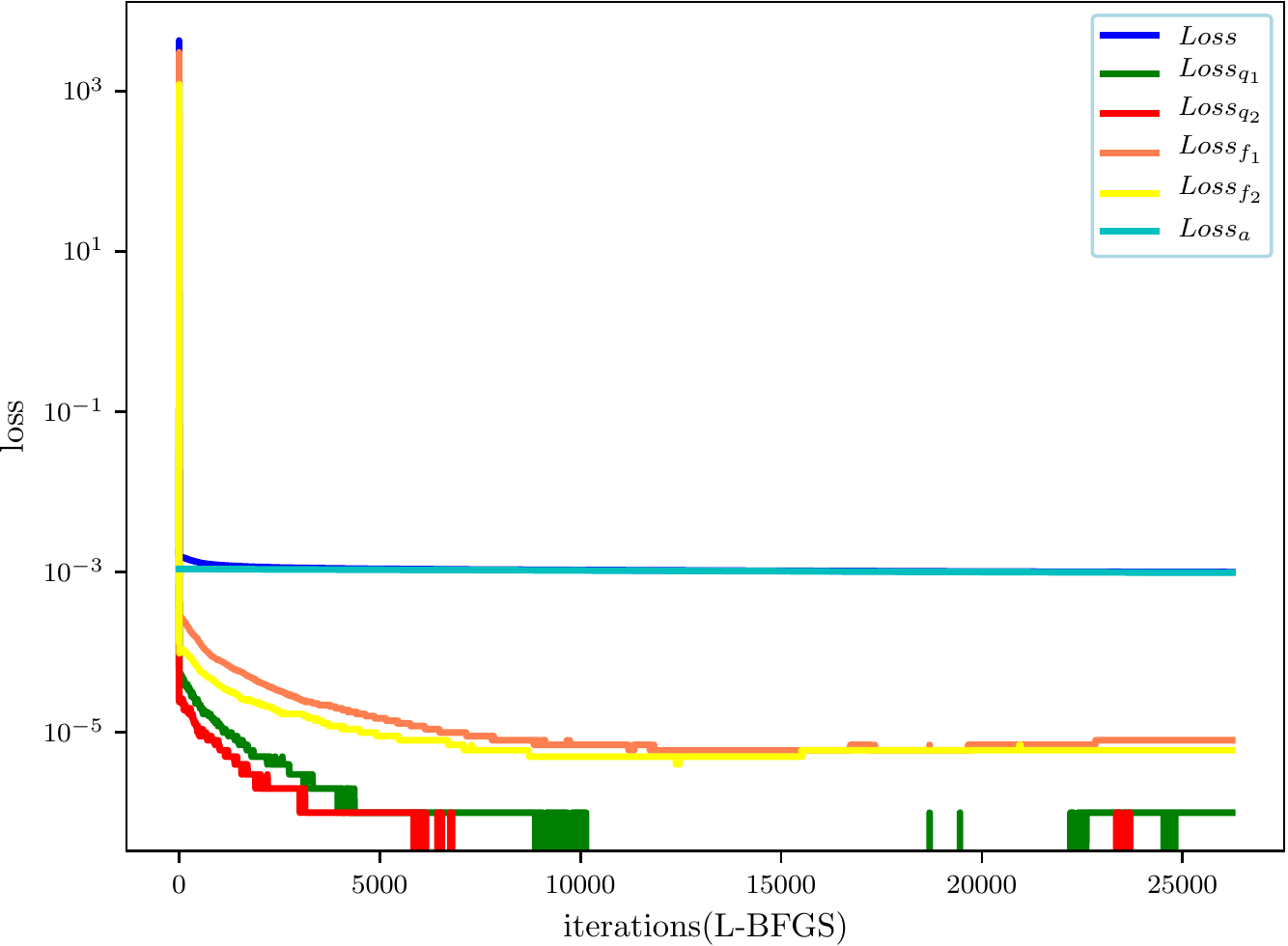}
\end{minipage}%
}%
\centering
\caption{(Color online) The loss function curve figures of the vector two-solitons $q_r(x,t)$ $(r=1,2)$ arising from the IPINN with the 20000 steps Adam and 26239 steps L-BFGS optimizations: (a) The loss function curve for the 20000 Adam optimization iterations; (b) The loss function curve for the 26239 L-BFGS optimization iterations.}
\label{F7}
\end{figure}

Figs. \ref{F5} - \ref{F7} display the training results of the vector two-solitons $q_r(x,t)$ $(r=1,2)$ based on the IPINN related to the initial-boundary value problem \eqref{E17} and \eqref{E18} of the Manakov system \eqref{E1}. Fig. \ref{F5} depicts various dynamic density plots and sectional drawing at different moments, in which the panel (a) corresponds to two-soliton solution $q_1(x,t)$ and panel (b) corresponds to two-soliton solution $q_2(x,t)$ for the Manakov system \eqref{E1}. As we can see from the bottom panels in Fig. \ref{F5} (a) and (b), the $q_1(x,t)$ soliton before collision has two components, but the $q_2(x,t)$ soliton before collision has only a component. However, after collision its another component of the $q_2(x,t)$ soliton appears when this soliton has re-emerged on the right side by combining learned dynamics density plot in Fig. \ref{F5} and 3D plots in Fig. \ref{F6}, it also means that the two-soliton solution $q_2(x,t)$ satisfies a soliton fission process. Especially, the power of each Manakov system soliton remains the before and after collision. Fig. \ref{F6} exhibits the three-dimensional plots with projected contours on three planes for the learning vector two-solitons $q_r(x,t)$, which corresponds to the two density plots of learned dynamics in Fig. \ref{F5}. Fig. \ref{F7} showcases curve plots of the loss function after 20000 Adam optimization iterations and 26239 L-BFGS optimization iterations in IPINN framework. Different from the vector one-soliton solutions in section 3.1, the loss function curves of Adam optimization for the vector two-soliton solutions are more hierarchical, which is caused by the different initial-boundary value conditions of $q_1(x,t)$ and $q_2(x,t)$. Obviously, from the bottom panels (a) and (b) of Fig. \ref{F5}, one can find that the amplitude of $q_1(x,t)$ is always greater than that of $q_2(x,t)$ at a fixed moment, this also causes the loss function values $Loss_{q_1}$ and $Loss_{f_1}$ to be greater than $Loss_{q_2}$ and $Loss_{f_2}$ respectively during the Adam optimization iteration. On the other hand, the loss function curves $Loss_{q_1}$ and $Loss_{q_2}$ of L-BFGS optimization are missing after a certain number of iterations, that is because the loss function value in this part is less than $1\rm e^{-6}$, which is beyond the statistical range of $``\%\mathrm{f}"$ in Python 3.7.

\subsection{Data-driven vector breathers}

It is well known that the breather solution is a special two-soliton solution, so it is also called the soliton bound state solution. Next, we will consider the data-driven vector breather solutions for the Manakov system by utilizing the initial-boundary value conditions and the 9-layer IPINN with 40 neurons per layer. Similarly, considering the initial conditions $q_{r}^0(x)$ $(r=1,2)$ and Dirichlet boundary conditions $q_{r}^{\mathrm{lb}}(t)$ and $q_{r}^{\mathrm{ub}}(t)$ of Eq. \eqref{E1} arising from the exact breathers Eq. \eqref{E14}, we take $[L_0,L_1]$ and $[T_0,T_1]$ in Eq. \eqref{E1} as $[-6.0,6.0]$ and $[-3.0,3.0]$, respectively. Therefore, the corresponding initial conditions can be written as bellows
\begin{align}\label{E19}
\begin{split}
&q_{r}^0(x)=q_{r,\mathrm{bs}}(x,-3.0),\,x\in[-6.0,6.0],
\end{split}
\end{align}
and the Dirichlet boundary conditions become
\begin{align}\label{E20}
q_{r}^{\mathrm{lb}}(t)=q_{r,\mathrm{bs}}(-6.0,t),\,q_{r}^{\mathrm{ub}}(t)=q_{r,\mathrm{bs}}(6.0,t),\,t\in[-3.0,3.0],\,r=1,2.
\end{align}

Similarly, discretizing Eq. \eqref{E14} with the aid of the traditional finite difference scheme on even grids in Matlab, and we obtain the original training data which contain initial data \eqref{E19} and boundary data \eqref{E20} by dividing the spatial region $[-6.0,6.0]$ into 2000 points and the temporal region $[-3.0,3.0]$ into 1000 points. Then, one can generate a smaller training dataset that containing initial-boundary data by randomly extracting $N_q=2000$ from original dataset and $N_f=30000$ collocation points which are produced by the LHS. After that, the latent vector breathers $q_r(x,t)$ have been successfully learned by tuning all learnable parameters of the IPINN and regulating the loss function. The model of IPINN achieved relative $\mathbb{L}_2$ error of 1.631581$\rm e^{-2}$ for the breather solution $q_1(x,t)$ and relative $\mathbb{L}_2$ error of 1.442727$\rm e^{-2}$ for the breather solution $q_2(x,t)$, and the total number of iterations which contain 20000 Adam optimizations and 22374 L-BFGS optimizations is 42374.

\begin{figure}[htbp]
\centering
\subfigure[]{
\begin{minipage}[t]{0.48\textwidth}
\centering
\includegraphics[height=6.5cm,width=6cm]{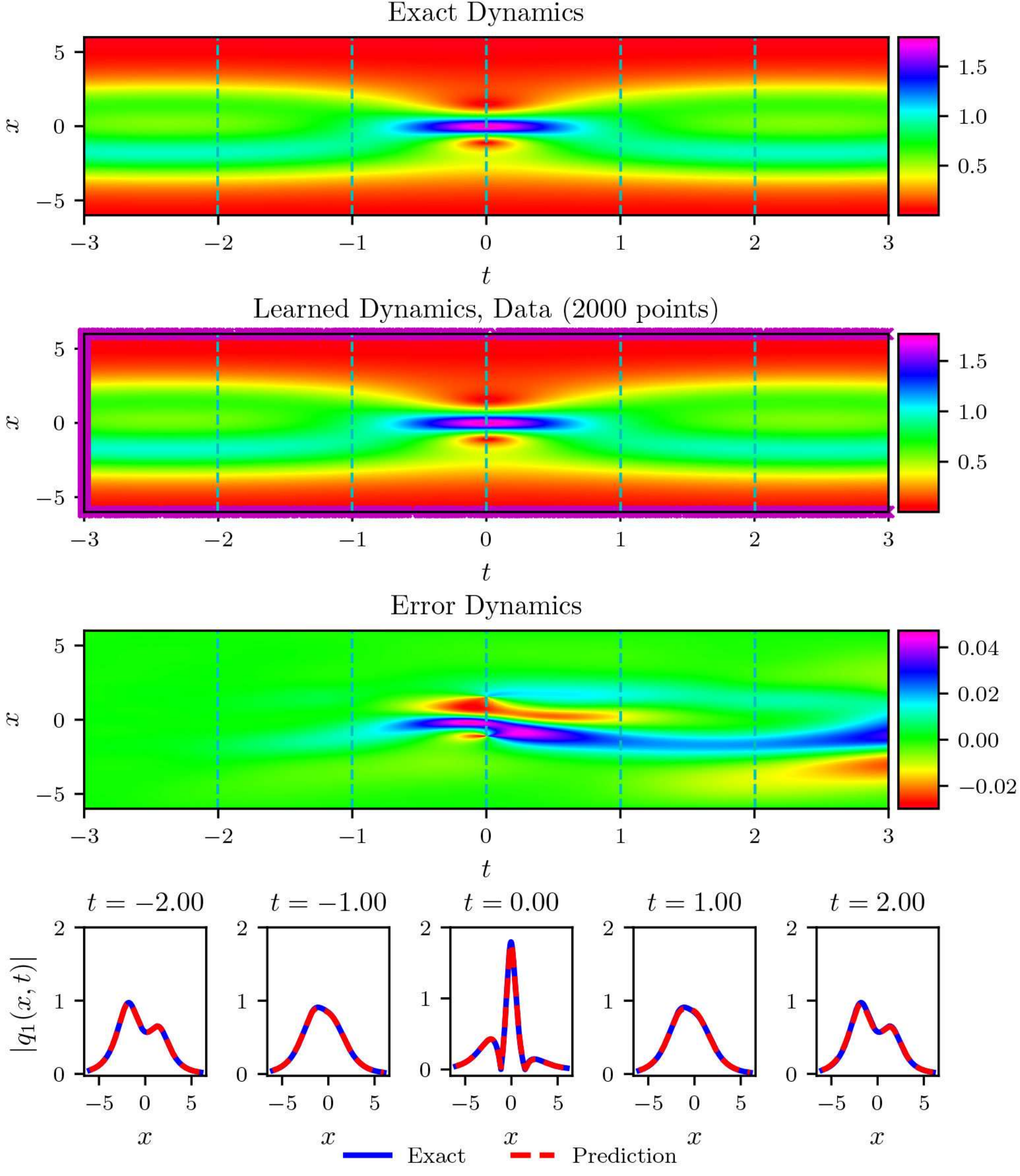}
\end{minipage}
}%
\subfigure[]{
\begin{minipage}[t]{0.48\textwidth}
\centering
\includegraphics[height=6.5cm,width=5cm]{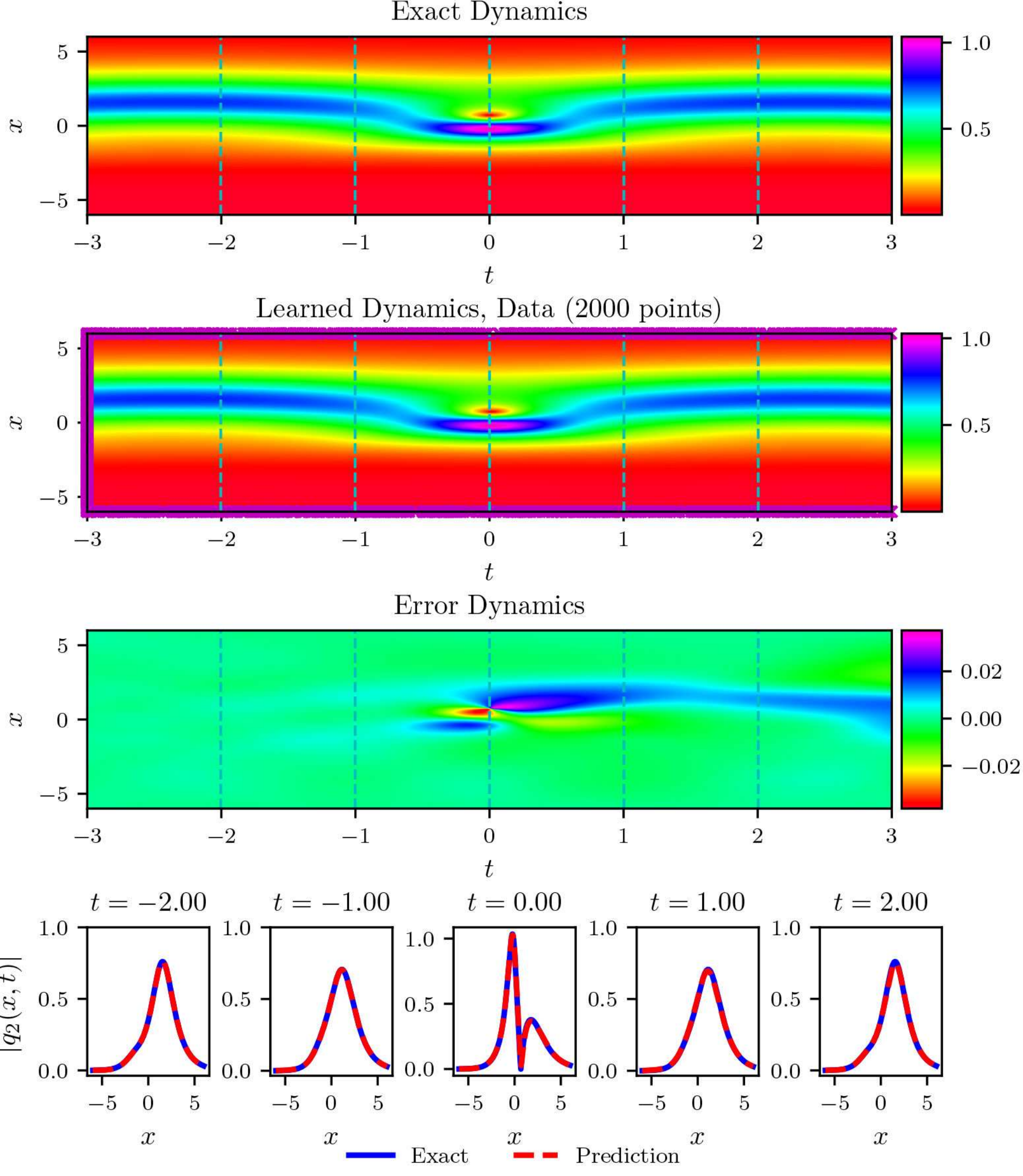}
\end{minipage}%
}%
\centering
\caption{(Color online) The vector breathers $q_r(x,t)$ $(r=1,2)$ resulted from the IPINN with the randomly chosen initial and boundary points $N_q=2000$ which have been shown by using mediumorchid $``\times"$ in learned dynamics , and $N_f = 30000$ collocation points in the corresponding spatiotemporal region. The exact, learned and error dynamics density plots for the vector breathers $q_r(x,t)$ with five distinct tested times $t=-2.00, -1.00, 0.00, 1.00$ and 2.00 (darkturquoise dashed lines), and the sectional drawings which contain the learned and explicit vector breathers $q_r(x,t)$ at the aforementioned five distinct times: (a) The density plots and sectional drawings for the breather $q_1(x,t)$; (b) The density plots and sectional drawings for the breather $q_2(x,t)$.}
\label{F8}
\end{figure}

\begin{figure}[htbp]
\centering

\begin{minipage}[t]{0.99\textwidth}
\centering
\includegraphics[height=6cm,width=14cm]{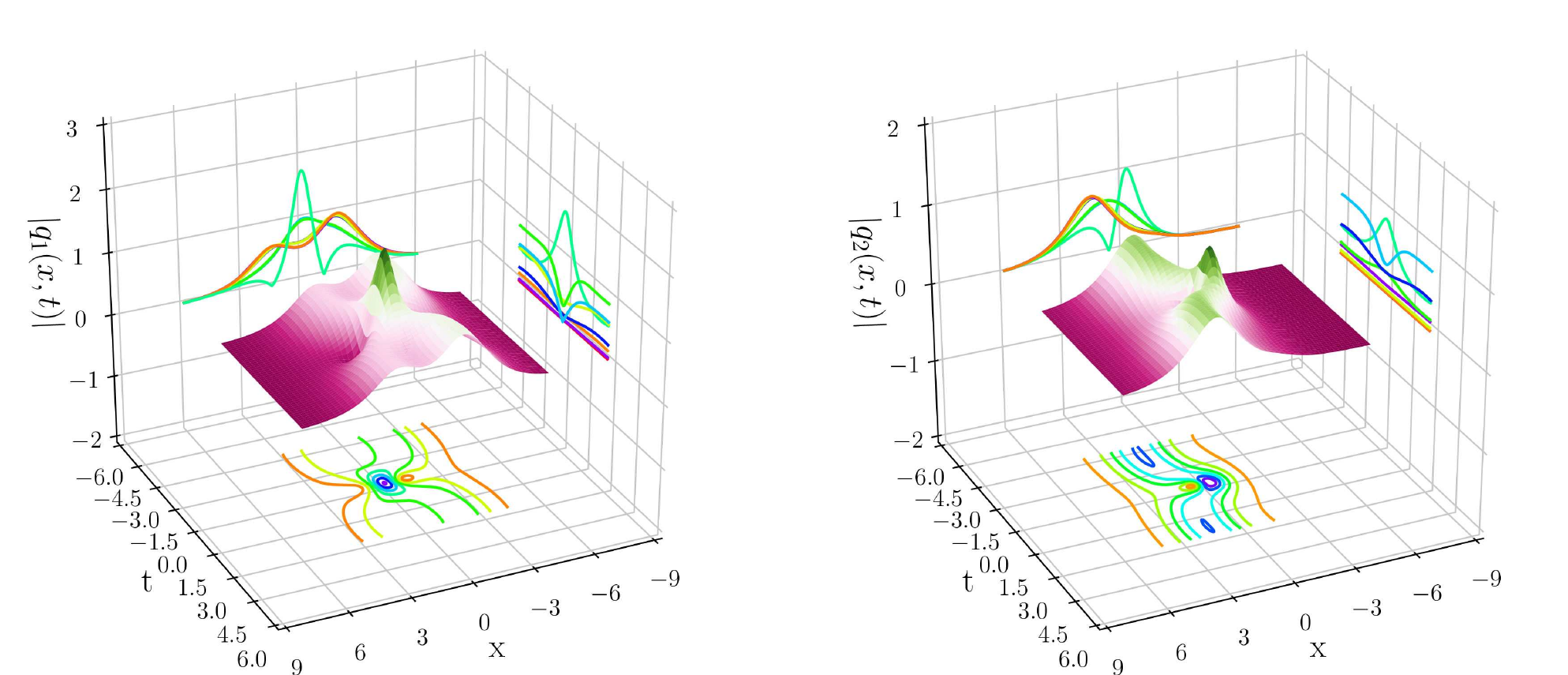}
\end{minipage}
\centering
\caption{(Color online) The three-dimensional plots with contour map on three planes of the predicted breather solutions $q_r(x,t)$ $(r=1,2)$ based on the IPINN: (Left side panel) The 3D plot for the $q_1(x,t)$; (Right side panel) The 3D plot for the $q_2(x,t)$.}
\label{F9}
\end{figure}

\begin{figure}[htbp]
\centering
\subfigure[]{
\begin{minipage}[t]{0.48\textwidth}
\centering
\includegraphics[height=5cm,width=7cm]{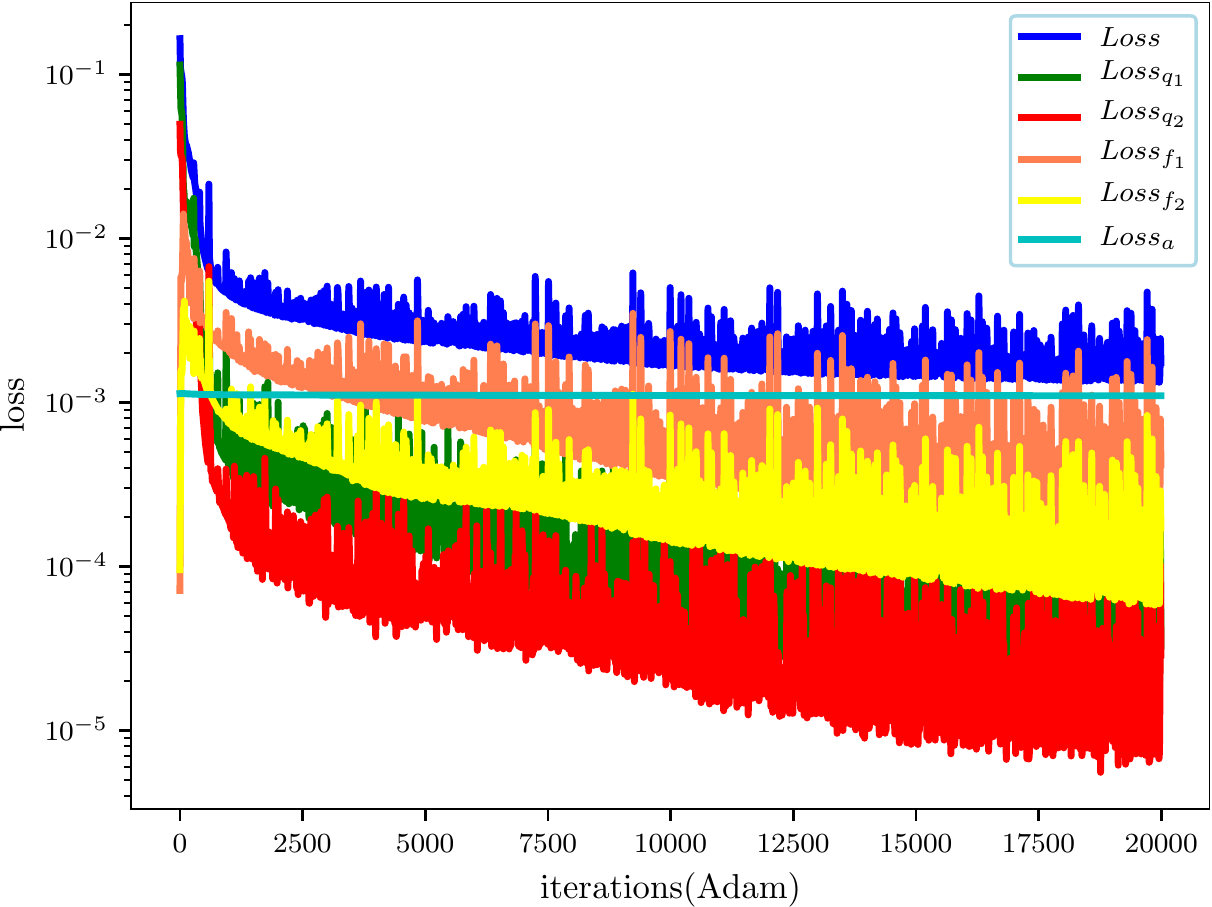}
\end{minipage}
}%
\subfigure[]{
\begin{minipage}[t]{0.48\textwidth}
\centering
\includegraphics[height=5cm,width=7cm]{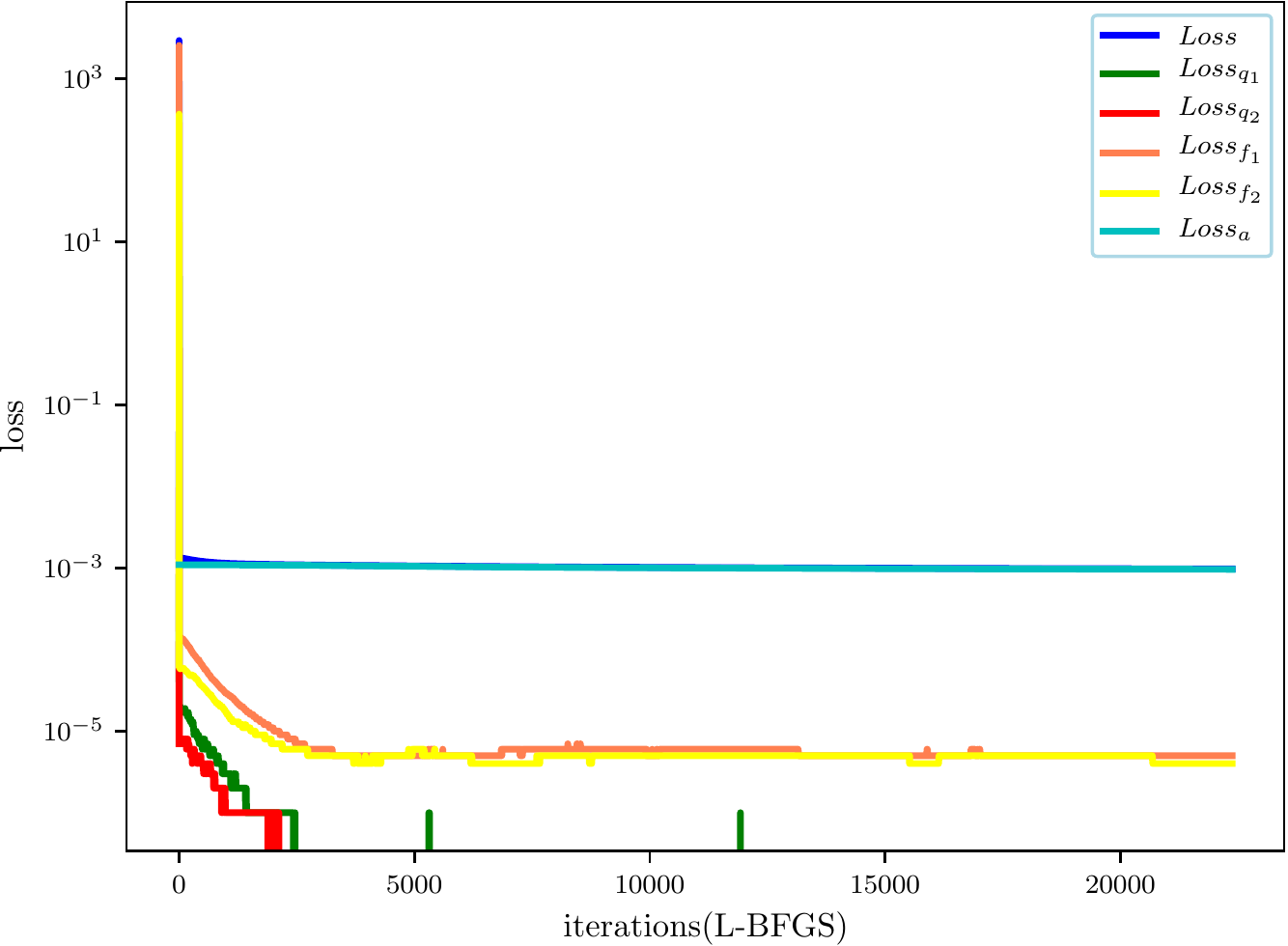}
\end{minipage}%
}%
\centering
\caption{(Color online) The loss function curve figures of the vector breathers $q_r(x,t)$ $(r=1,2)$ arising from the IPINN with the 20000 steps Adam and 22374 steps L-BFGS optimizations: (a) The loss function curve for the 20000 Adam optimization iterations; (b) The loss function curve for the 22374 L-BFGS optimization iterations.}
\label{F10}
\end{figure}

Figs. \ref{F8} - \ref{F10} display the deep learning results of the vector breathers $q_r(x,t)$ $(r=1,2)$ based on the IPINN related to the initial-boundary value problem \eqref{E19} and \eqref{E20} of the Manakov system \eqref{E1}. The dynamics behaviors which contain exact, learned, error dynamics of the vector breathers are shown in Fig. \ref{F8}. Compared with the dynamics behaviors of vector two-solitons in the bottom panels of Fig. \ref{F5}, one can notice that the shapes of the vector breathers are the same at $t=-1$ $(t=-2)$ and $t=1$ $(t=2)$ in the bottom panels of Fig. \ref{F8}, which are different from the vector two solitons. From the learned dynamics density plots in Fig. \ref{F8}, it can be seen that this is a localized periodic oscillation solution, whose shape is similar to respiratory fluctuation, so it is also called ``breather''. Fig. \ref{F9} exhibits three-dimensional plots of the predicted vector breathers $q_r(x,t)$ for the Eq. \eqref{E1} result from the IPINN. The graph in Fig. \ref{F10} (a) is similar to that the case of vector two-solitons in Fig. \ref{F7} (a), and there are more missing parts of the $Loss_{q_1}$ and $Loss_{q_2}$ in Fig. \ref{F10} (b), which also indicate that the optimization for the vector breathers is better than the vector two-solitons arising from the L-BFGS optimization iterations.

In this section, based on the initial-boundary value problem of the Manakov system, we discuss the data-driven vector one-solitons, two-solitons and breathers via the IPINN model. However, for more data-driven vector $N$-solitons and $N$-breathers, such as the vector two-breathers which derived from the vector four-solitons, we can also recover them through the IPINN framework. Of course, it will require us to take more initial-boundary value data points and residual configuration points, establish deeper and wider NNs, and demand stronger CPU and GPU of computer. Tab. \ref{Tab-SB} exhibits relative $\mathbb{L}_2$ norm errors of three different types of localized waves by means of IPINN.

\begin{table}[htbp]
  \caption{Relative $\mathbb{L}_2$ errors of three different localized waves types in IPINN model}
  \label{Tab-SB}
  \centering
  \scalebox{0.8}{
  \begin{tabular}{l|c|c|c}
  \toprule
  \diagbox{\textbf{\textbf{Component}}}{\textbf{Wave Types}} & Vector one-solitons & Vector two-solitons & Vector breathers\\
  \hline
  Component $q_1$   & 3.813803$\rm e$-02 & 2.678111$\rm e$-02 & 1.631581$\rm e$-02 \\
  \hline
  Component $q_2$   & 3.754237$\rm e$-02 & 2.356420$\rm e$-02 & 1.442727$\rm e$-02 \\
  \bottomrule
  \end{tabular}}
\end{table}

\section{Data-driven vector rogue waves of the Manakov system}
In various complex systems, the research of vector RWs has more important physical significance, and the multi-component model has more appropriate practical significance. In this section, we will focus on the data-driven vector RWs of the Manakov system \eqref{E1} with $\lambda_1=1$ and $\lambda_2=2$ based on the IPINN. That is the physics-informed parts are consistent with Eq. \eqref{E-pi} in section 3. The abundant interaction solutions between RWs and various other types of waves and $N$-order RWs with parameter adjustment have been obtained by means of Darboux transformation. Next, in order to compare the errors of vector RWs learned by using the IPINN, we give out the exact solution expressions of several common vector RWs.

$\bullet$ \textbf{Vector Rogue Waves $\rm\uppercase\expandafter{\romannumeral1}$}

One can obtain the explicit vector RWs of the Manakov system with $\lambda_1=1$ and $\lambda_2=2$ as following
\begin{small}
\begin{align}\label{E21}
\begin{split}
&q_{1,\mathrm{rw1}}(x,t)=-\frac{3}{10}\cdot{\frac{\Omega_1}{900t\sqrt{3}+1500\sqrt{3}x-972{t}^{2}-810tx-675{x}^{2}-3125}},\\
&q_{2,\mathrm{rw1}}(x,t)=\frac{3}{10}\cdot{\frac{\Omega_2}{900t\sqrt{3}+1500\sqrt{3}x-972{t}^{2}-810tx-675{x}^{2}-3125}},\\
&\Omega_1=\big(972\mathrm{i}\sqrt{3}{t}^{2}+810\mathrm{i}\sqrt{3}tx+675\mathrm{i}\sqrt{3}{x}^{2}-4050\mathrm{i}t-3375\mathrm{i}x+450t\sqrt{3}-\\
&\qquad\quad2625\sqrt{3}x+972{t}^{2}+810tx+675{x}^{2}+5000\big) {{\rm e}^{{\frac{36}{25}}\mathrm{i}t}},\\
&\Omega_2=\big(972\mathrm{i}\sqrt{3}{t}^{2}+810\mathrm{i}\sqrt{3}tx+675\mathrm{i}\sqrt{3}{x}^{2}-3375\mathrm{i}x+3600t\sqrt{3}+2625\sqrt{3}x-\\
&\qquad\quad972{t}^{2}-810tx-675{x}^{2}-5000\big){{\rm e}^{{\frac{3}{25}}\mathrm{i}(-5x+9t)}},
\end{split}
\end{align}
\end{small}
the vector RWs of the Manakov system have been derived by Guo and Ling \cite{Guo2011}, this kind of RW solutions are similar to the first-order RW solutions for the NLS \cite{Akhmediev2009}.

$\bullet$ \textbf{Vector Rogue Waves $\rm\uppercase\expandafter{\romannumeral2}$}

Furthermore, in addition to the above vector RWs $\rm\uppercase\expandafter{\romannumeral1}$, Guo and Ling also obtained a more complex vector RWs $\rm\uppercase\expandafter{\romannumeral2}$ \cite{Guo2011}, the form of the explicit vector RWs $\rm\uppercase\expandafter{\romannumeral2}$ is as follows
\begin{align}\label{E22}
\begin{split}
&q_{1,\mathrm{rw2}}(x,t)=-\frac{3}{10}{\frac {\Delta_1}{\Xi}},\\
&q_{2,\mathrm{rw2}}(x,t)=\frac{3}{10}{\frac { \Delta_2}{\Xi}},\\
\end{split}
\end{align}
where
\begin{small}
\begin{align}\nonumber
\begin{split}
&\Delta_1=\big(-3125000+524880\mathrm{i}\sqrt{3}{t}^{3}x+656100\mathrm{i}\sqrt{3}{t}^{2}{x}^{2}+364500\mathrm{i}\sqrt{3}t{x}^{3}+5062500\mathrm{i}t\sqrt{3}x+\\
&\qquad6682500{t}^{2}-2187000\sqrt{3}{t}^{2}x-1215000\sqrt{3}t{x}^{2}+562500t\sqrt{3}+1012500tx+5906250{x}^{2}\\
&\qquad-1875000\sqrt{3}x-15187500\mathrm{i}t-3936600\mathrm{i}{t}^{3}-1518750\mathrm{i}{x}^{3}+524880{t}^{3}x+656100{t}^{2}{x}^{2}+\\
&\qquad364500t{x}^{3}-1181250\sqrt{3}{x}^{3}-145800\sqrt{3}{t}^{3}+314928\mathrm{i}\sqrt{3}{t}^{4}+151875\mathrm{i}\sqrt{3}{x}^{4}+4252500\mathrm{i}\sqrt{3}{t}^{2}\\
&\qquad+843750\mathrm{i}\sqrt{3}{x}^{2}-6561000\mathrm{i}{t}^{2}x-3645000\mathrm{i}t{x}^{2}+314928{t}^{4}+151875{x}^{4}\big){{\rm e}^{{\frac{36}{25}}\mathrm{i}t}},\\
&\Delta_2=\big(3125000+524880\mathrm{i}\sqrt{3}{t}^{3}x+656100\mathrm{i}\sqrt{3}{t}^{2}{x}^{2}+364500\mathrm{i}\sqrt{3}t{x}^{3}-1312200\mathrm{i}{t}^{3}-13972500{t}^{2}\\
&\qquad+4374000\sqrt{3}{t}^{2}x+3037500\sqrt{3}t{x}^{2}+2812500t\sqrt{3}-13162500tx-5906250{x}^{2}+1875000\sqrt{3}x\\
&\qquad-3037500\mathrm{i}t\sqrt{3}x-1518750\mathrm{i}{x}^{3}-524880{t}^{3}x-656100{t}^{2}{x}^{2}-364500t{x}^{3}+1181250\sqrt{3}{x}^{3}+\\
&\qquad2770200\sqrt{3}{t}^{3}+15187500\mathrm{i}t+314928\mathrm{i}\sqrt{3}{t}^{4}+151875\mathrm{i}\sqrt{3}{x}^{4}+843750\mathrm{i}\sqrt{3}{x}^{2}-4374000\mathrm{i}{t}^{2}x-\\
&\qquad1822500\mathrm{i}t{x}^{2}-314928{t}^{4}-151875{x}^{4}-607500\mathrm{i}\sqrt{3}{t}^{2}\big) {{\rm e}^{{\frac{3}{25}}\mathrm{i}( -5x+9t)}},\\
&\Xi=1020600\sqrt{3}{t}^{3}+2187000\sqrt{3}{t}^{2}x+1215000\,\sqrt{3}t{x}^{2}+675000\sqrt{3}{x}^{3}-314928{t}^{4}-\\
&\qquad524880{t}^{3}x-656100{t}^{2}{x}^{2}-364500t{x}^{3}-151875{x}^{4}+1125000t\sqrt{3}+1875000\sqrt{3}x\\
&\qquad-6682500{t}^{2}-4050000tx-3375000{x}^{2}-1562500.\\
\end{split}
\end{align}
\end{small}

The vector RWs $\rm\uppercase\expandafter{\romannumeral2}$ are different behaviors to RWs of NLS \cite{Akhmediev2009}, and different form Ref. \cite{Yan2011} by a symmetry analysis, which is nothing but the RWs of NLS an different altitude.

$\bullet$ \textbf{Vector Rogue Waves $\rm\uppercase\expandafter{\romannumeral3}$}

In order to obtain the interaction solutions between dark-bright solitons and RWs, authors also derived this interesting solutions by Darboux transformation method in Ref. \cite{Guo2011}, as shown below

\begin{align}\label{E23}
\begin{split}
&q_{1,\mathrm{rw3}}(x,t)={\frac{\Lambda}{\Pi}},\\
&q_{2,\mathrm{rw3}}(x,t)={\frac{4\big(\mathrm{i}x-1-\mathrm{i}+4t-x\big){{\rm e}^{-\frac23t+\frac13x+2\mathrm{i}t+\mathrm{i}x}}}{\Pi}},\\
\end{split}
\end{align}
where
\begin{small}
\begin{align}\nonumber
\begin{split}
&\Lambda={{\rm e}^{\mathrm{i}t+\mathrm{i}x}}\big(8\mathrm{i}{{\rm e}^{\frac43t-\frac23x}}t-16{{\rm e}^{\frac43t-\frac23x}}{t}^{2}+8{{\rm e}^{\frac43t-\frac23x}}tx-2{{\rm e}^{\frac43t-\frac23x}}{x}^{2}-\\
&\qquad4\mathrm{i}{{\rm e}^{\frac43t-\frac23x}}+12{{\rm e}^{\frac43t-\frac23x}}t-2{{\rm e}^{\frac43t-\frac23x}}x-{{\rm e}^{\frac43t-\frac23x}}+2{{\rm e}^{-\frac83t+\frac43x}}\big),\\
&\Pi=16{{\rm e}^{\frac43t-\frac23x}}{t}^{2}-8{{\rm e}^{\frac43t-\frac23x}}tx+2{{\rm e}^{\frac43t-\frac23x}}{x}^{2}-12{{\rm e}^{\frac43t-\frac23x}}t+2{{\rm e}^{\frac43t-\frac23x}}x+3{{\rm e}^{\frac43t-\frac23x}}+2{{\rm e}^{-\frac83t+\frac43x}}.
\end{split}
\end{align}
\end{small}

This kind of solutions behaves like the nonlinear superposition for the RWs and dark-bright soliton. This vector RWs \eqref{E23} reveal the mechanism and dynamic behavior of RWs produced in dark-bright soliton.

$\bullet$ \textbf{Vector Rogue Waves $\rm\uppercase\expandafter{\romannumeral4}$}

For the vector RWs $\rm\uppercase\expandafter{\romannumeral4}$, we will exhibit the dark-bright solitons together with single Peregrine solitons, which have been arrived at by means of the Darboux transformation theory \cite{Wang2014}, it can be expressed by the following formula

\begin{align}\label{E24}
\begin{split}
&q_{1,\mathrm{rw4}}(x,t)={\frac{{{\rm e}^{2\mathrm{i}t}}\big(1600\mathrm{i}t-1600{t}^{2}-400{x}^{2}+{{\rm e}^{-2x}}+300 \big) }{1600{t}^{2}+400{x}^{2}+{{\rm e}^{-2x}}+100}},\\
&q_{2,\mathrm{rw4}}(x,t)={\frac{\big(20-20\mathrm{i}\big)\big(2x+4\mathrm{i}t+1\big){{\rm e}^{-x+3\mathrm{i}t}}}{1600{t}^{2}+400{x}^{2}+{{\rm e}^{-2x}}+100}}.\\
\end{split}
\end{align}

Baronio obtained this solution by employing the Darboux dressing technique for the first time \cite{Baronio2012}. Later, Wang et al. also obtained this solution through the generalized Darboux transformation \cite{Wang2014}. These results provide evidence of an attractive interaction between the dark-bright waves and RWs, and the observed behavior may also be interpreted as a mechanism of generation of one RW out of a slowly moving boomeronic soliton.

$\bullet$ \textbf{Vector Rogue Waves $\rm\uppercase\expandafter{\romannumeral5}$}

Moreover, once the vector RWs $\rm\uppercase\expandafter{\romannumeral4}$ are obtained, the vector RWs $\rm\uppercase\expandafter{\romannumeral5}$ are also derived correspondingly, as shown below
\begin{align}\label{E25}
\begin{split}
&q_{1,\mathrm{rw5}}(x,t)=\frac{\Upsilon_1}{\sqrt{2}{{\rm e}^{-2\sqrt{2}x}}+3200{t}^{2}+400{x}^{2}+50},\\
&q_{2,\mathrm{rw5}}(x,t)=\frac{\Upsilon_2}{\sqrt{2}{{\rm e}^{-2\sqrt{2}x}}+3200{t}^{2}+400{x}^{2}+50}.\\
\end{split}
\end{align}
where
\begin{small}
\begin{align}\nonumber
\begin{split}
&\Upsilon_1=40\mathrm{i}\sqrt[4]{2}{{\rm e}^{-\sqrt{2}x+6\mathrm{i}t}}x-80\cdot{2}^{\frac34}{{\rm e}^{-\sqrt{2}x+6\mathrm{i}t}}t-40\sqrt[4]{2}{{\rm e}^{-\sqrt{2}x+6\mathrm{i}t}}x-80\cdot{2}^{\frac34}\mathrm{i}{{\rm e}^{-\sqrt{2}x+6\mathrm{i}t}}t+10\cdot{2}^{\frac34}\mathrm{i}\\
&\qquad{{\rm e}^{-\sqrt{2}x+6\mathrm{i}t}}-10\cdot{2}^{\frac34}{{\rm e}^{-\sqrt{2}x+6\mathrm{i}t}}+\sqrt{2}{{\rm e}^{4\mathrm{i}t-2\sqrt{2}x}}+1600\mathrm{i}{{\rm e}^{4\mathrm{i}t}}t-3200{{\rm e}^{4\mathrm{i}t}}{t}^{2}-400{{\rm e}^{4\mathrm{i}t}}{x}^{2}+150{{\rm e}^{4\mathrm{i}t}},\\
&\Upsilon_2=-40\mathrm{i}\sqrt[4]{2}{{\rm e}^{-\sqrt{2}x+6\mathrm{i}t}}x+80\cdot{2}^{\frac34}{{\rm e}^{-\sqrt{2}x+6\mathrm{i}t}}t+40\sqrt[4]{2}{{\rm e}^{-\sqrt{2}x+6\mathrm{i}t}}x+80\cdot{2}^{\frac34}\mathrm{i}{{\rm e}^{-\sqrt{2}x+6\mathrm{i}t}}t-10\cdot{2}^{\frac34}\mathrm{i}\\
&\qquad{{\rm e}^{-\sqrt{2}x+6\mathrm{i}t}}+10\cdot{2}^{\frac34}{{\rm e}^{-\sqrt{2}x+6\mathrm{i}t}}+\sqrt{2}{{\rm e}^{4\mathrm{i}t-2\sqrt{2}x}}+1600\mathrm{i}{{\rm e}^{4\mathrm{i}t}}t-3200{{\rm e}^{4\mathrm{i}t}}{t}^{2}-400{{\rm e}^{4\mathrm{i}t}}{x}^{2}+150{{\rm e}^{4\mathrm{i}t}}.
\end{split}
\end{align}
\end{small}

This is an interaction solutions obtained by the nonlinear superposition of breathers and RWs, which are more complex than the vector RWs $\rm\uppercase\expandafter{\romannumeral4}$ \cite{Baronio2012,Wang2014}. Furthermore, due to the introduction of periodic oscillatory breather, the study of this solution plays an important role in the stability analysis of optical fiber transmission.

\subsection{Data-driven vector rogue waves \rm\uppercase\expandafter{\romannumeral1}}
In this section, in order to recover the data-driven vector RWs $\rm\uppercase\expandafter{\romannumeral1}$ of the Manakov system with $\lambda_1=1$ and $\lambda_2=2$, we will commit to introducing the initial boundary value conditions of the Manakov system to the 9-layer IPINN with 40 neurons per layer. Selecting $[L_0,L_1]$ and $[T_0,T_1]$ in Eq. \eqref{E1} as $[-3.0,6.0]$ and $[-2.0,2.0]$ respectively, we can write the corresponding initial value conditions as follows
\begin{align}\label{E26}
\begin{split}
&q_{r}^0(x)=q_{r,\mathrm{rw1}}(x,-2.0),\,x\in[-3.0,6.0],
\end{split}
\end{align}
and the Dirichlet boundary conditions become
\begin{align}\label{E27}
q_{r}^{\mathrm{lb}}(t)=q_{r,\mathrm{rw1}}(-3.0,t),\,q_{r}^{\mathrm{ub}}(t)=q_{r,\mathrm{rw1}}(6.0,t),\,t\in[-2.0,2.0],\,r=1,2.
\end{align}

In order to obtain the original training data set of the above initial boundary value conditions \eqref{E26} and \eqref{E27}, we discretize the exact vector RWs $\rm\uppercase\expandafter{\romannumeral1}$ \eqref{E21} based on the finite difference method by dividing the spatial region $[-3.0,6.0]$ into 2000 points and the temporal region $[-2.0,2.0]$ into 1000 points in Matlab. Furthermore, in addition to the data set composed of the aforementioned initial boundary value conditions, the residual data set is used to calculate the $\mathbb{L}_2$ norm error by comparing with the predicted vector RWs $\rm\uppercase\expandafter{\romannumeral1}$. After that, a smaller training dataset that containing initial-boundary data will be generated by randomly extracting $N_q=2000$ from original dataset and $N_f=30000$ collocation points which are produced by the LHS. According to 20000 Adam iterations and 25117 L-BFGS iterations, the latent vector RWs $\rm\uppercase\expandafter{\romannumeral1}$ $q_r(x,t)$ have been successfully learned by employing the IPINN, and the network achieved relative $\mathbb{L}_2$ error of 7.505391$\rm e^{-3}$ for the RW $q_1(x,t)$ and relative $\mathbb{L}_2$ error of 7.988676$\rm e^{-3}$ for the RW $q_2(x,t)$, and the total number of iterations is 45117.

\begin{figure}[htbp]
\centering
\subfigure[]{
\begin{minipage}[t]{0.48\textwidth}
\centering
\includegraphics[height=6.5cm,width=6cm]{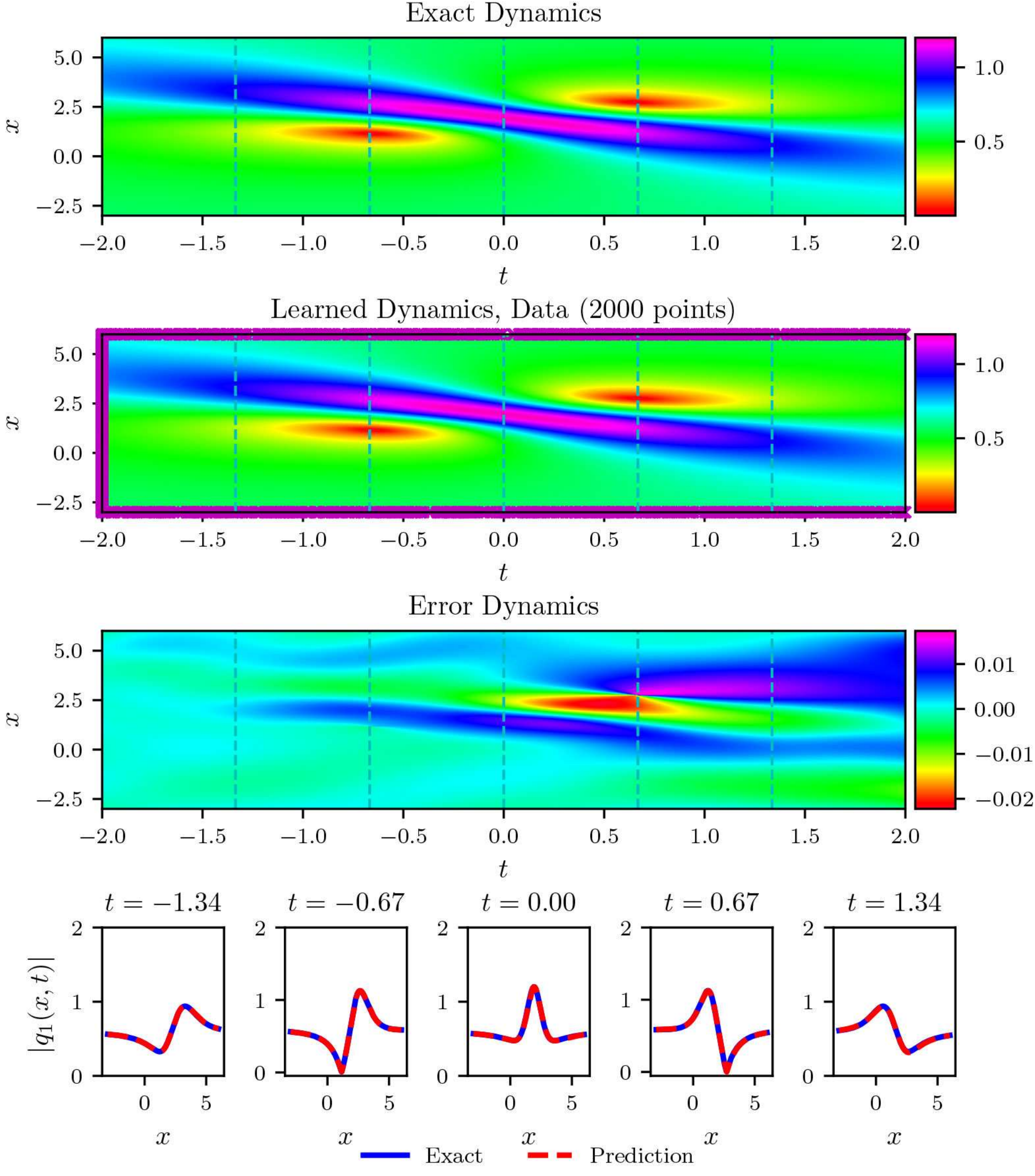}
\end{minipage}
}%
\subfigure[]{
\begin{minipage}[t]{0.48\textwidth}
\centering
\includegraphics[height=6.5cm,width=5cm]{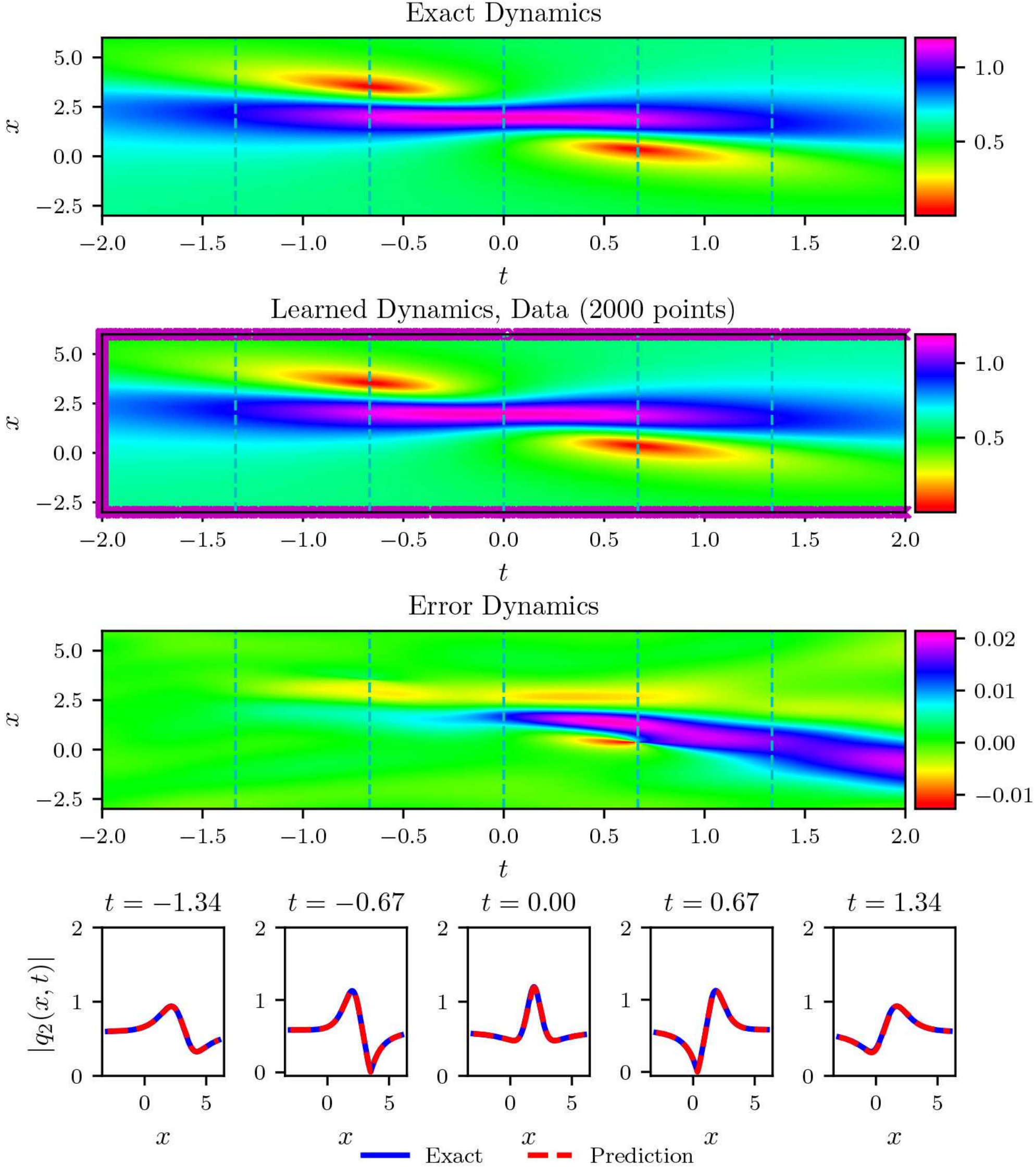}
\end{minipage}%
}%
\centering
\caption{(Color online) The vector RWs $\rm\uppercase\expandafter{\romannumeral1}$ $q_r(x,t)$ $(r=1,2)$ resulted from the IPINN with the randomly chosen initial and boundary points $N_q=2000$ which have been shown by using mediumorchid $``\times"$ in learned dynamics , and $N_f = 30000$ collocation points in the corresponding spatiotemporal region. The exact, learned and error dynamics density plots for the vector RWs $\rm\uppercase\expandafter{\romannumeral1}$ $q_1(x,t)$ with five distinct tested times $t=-1.34, -0.67, 0.00, 0.67$ and 1.34 (darkturquoise dashed lines), and the sectional drawings which contain the learned and explicit vector RWs $\rm\uppercase\expandafter{\romannumeral1}$ $q_r(x,t)$ at the aforementioned five distinct times: (a) The density plots and sectional drawings for the RW $\rm\uppercase\expandafter{\romannumeral1}$ $q_1(x,t)$; (b) The density plots and sectional drawings for the RW $\rm\uppercase\expandafter{\romannumeral1}$ $q_2(x,t)$.}
\label{F11}
\end{figure}

\begin{figure}[htbp]
\centering

\begin{minipage}[t]{0.99\textwidth}
\centering
\includegraphics[height=6cm,width=14cm]{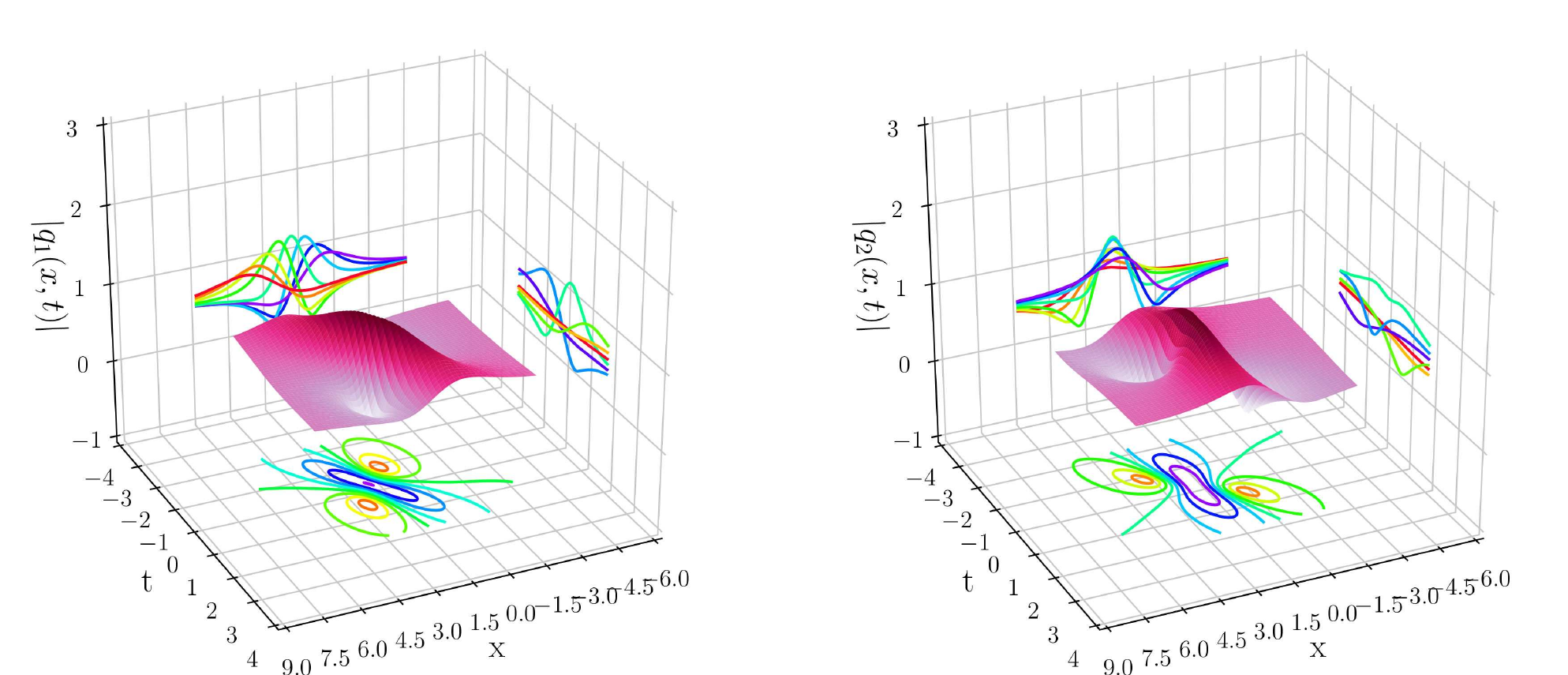}
\end{minipage}
\centering
\caption{(Color online) The three-dimensional plots with contour map on three planes of the predicted vector RWs $\rm\uppercase\expandafter{\romannumeral1}$ $q_r(x,t)$ $(r=1,2)$ based on the IPINN: (Left side panel) The 3D plot for the $q_1(x,t)$; (Right side panel) The 3D plot for the $q_2(x,t)$.}
\label{F12}
\end{figure}

\begin{figure}[htbp]
\centering
\subfigure[]{
\begin{minipage}[t]{0.48\textwidth}
\centering
\includegraphics[height=5cm,width=7cm]{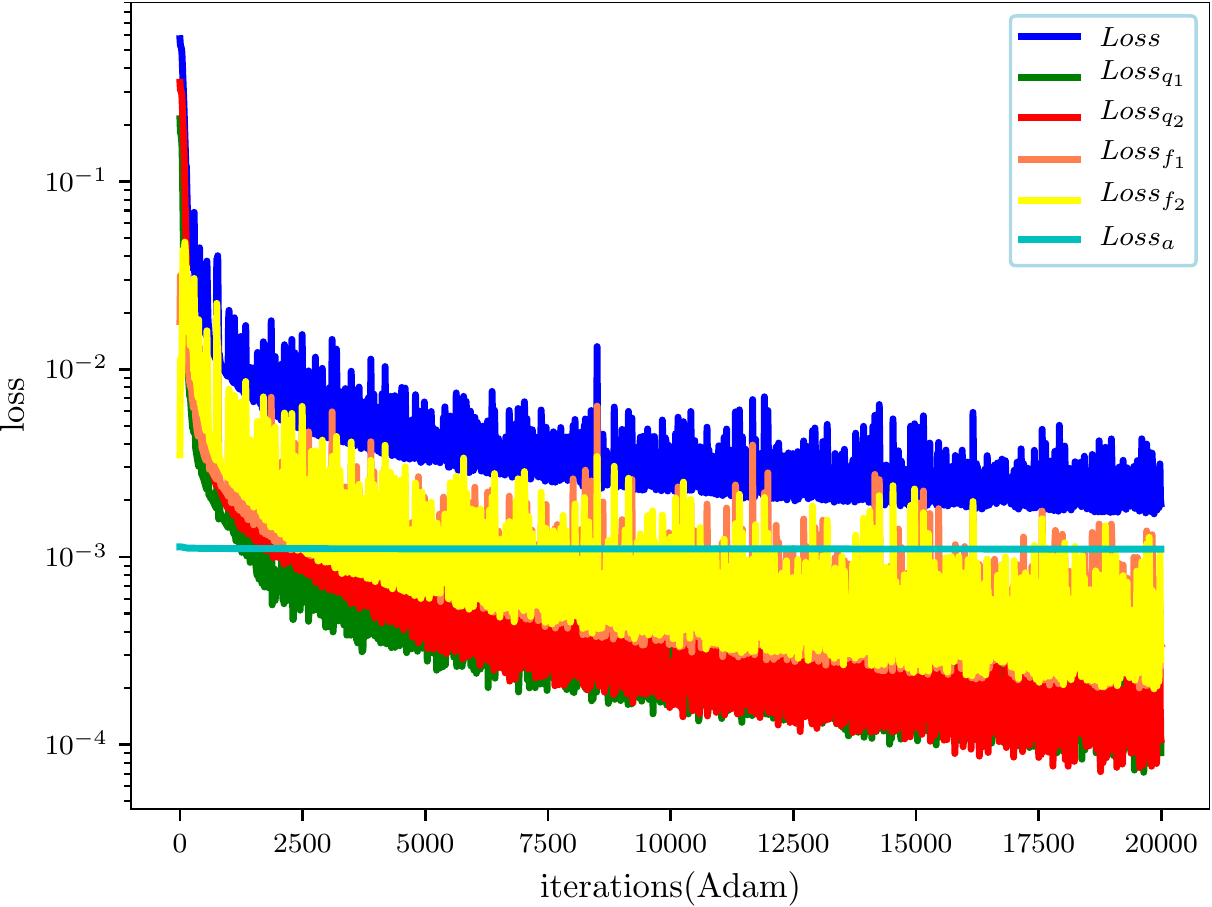}
\end{minipage}
}%
\subfigure[]{
\begin{minipage}[t]{0.48\textwidth}
\centering
\includegraphics[height=5cm,width=7cm]{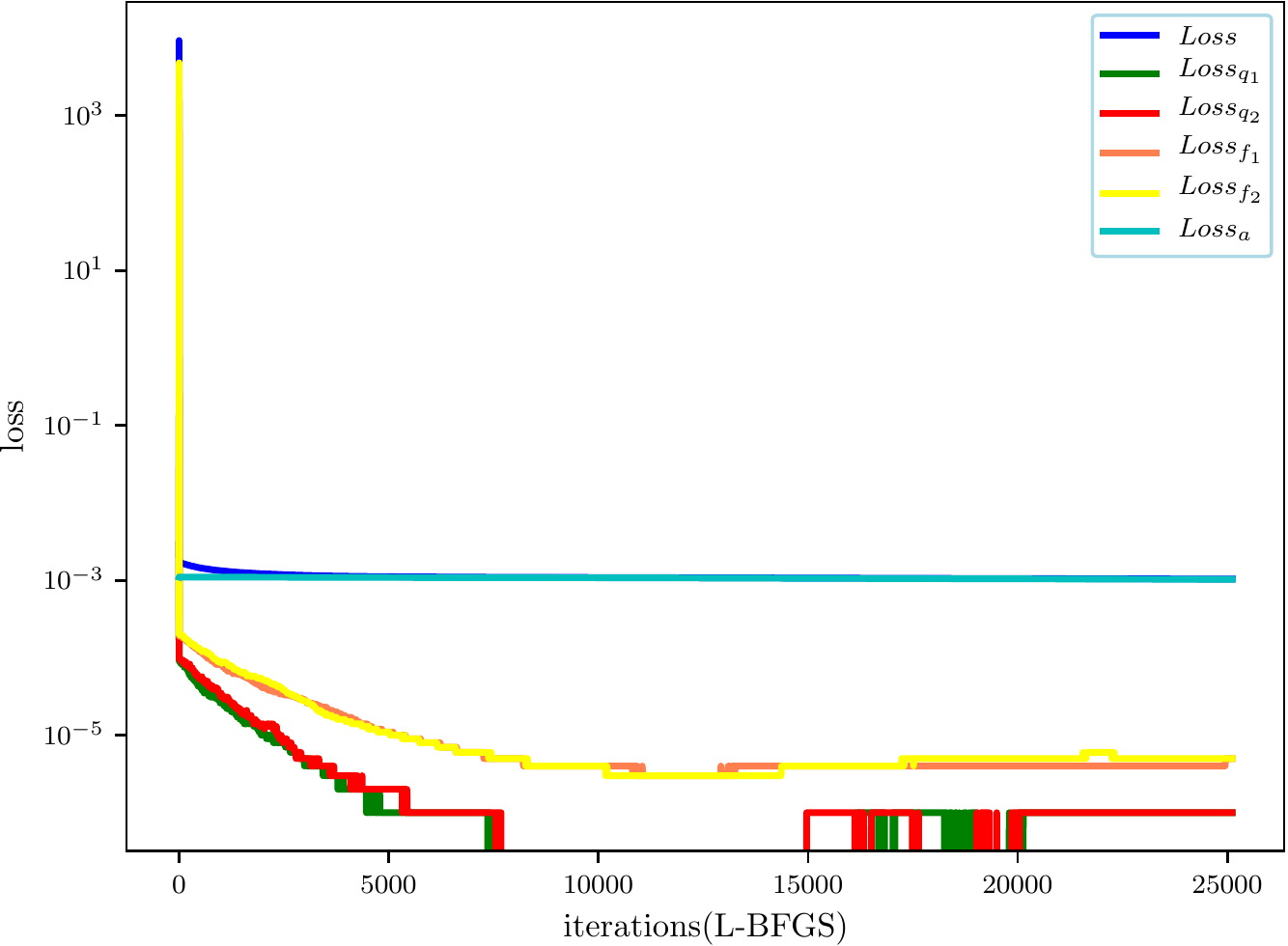}
\end{minipage}%
}%
\centering
\caption{(Color online) The loss function curve figures of the vector RWs $\rm\uppercase\expandafter{\romannumeral1}$ $q_r(x,t)$ $(r=1,2)$ arising from the IPINN with the 20000 steps Adam and 25117 steps L-BFGS optimizations: (a) The loss function curve for the 20000 Adam optimization iterations; (b) The loss function curve for the 25117 L-BFGS optimization iterations.}
\label{F13}
\end{figure}

Figs. \ref{F11} - \ref{F13} present the deep learning results of the vector RWs $\rm\uppercase\expandafter{\romannumeral1}$ $q_r(x,t)$ $(r=1,2)$ based on the IPINN for the Manakov system with the initial-boundary value problem \eqref{E26} and \eqref{E27}. Fig. \ref{F11} displays the exact, learned and error dynamics of the vector RWs $\rm\uppercase\expandafter{\romannumeral1}$, and exhibits the sectional drawings which contain the learned and explicit RWs at five different moments. From the density plots of learned dynamics and profiles which reveal amplitude and error of exact and prediction RWs in Fig. \ref{F11}, we can observe that the shapes are similar and amplitudes are identical for the vector RWs $q_1(x,t)$ and $q_2(x,t)$, but the positions and orientations of the two-component RWs are different. The 3D plots of the predicted vector RWs $\rm\uppercase\expandafter{\romannumeral1}$ $q_r(x,t)$ have been shown through the IPINN in Fig. \ref{F12}. Since the predicted vector RWs $\rm\uppercase\expandafter{\romannumeral1}$ $q_r(x,t)$ are similar in shape, the loss function curves of $Loss_{q_1}$ and $Loss_{q_2}$ ($Loss_{f_1}$ and $Loss_{f_2}$) are overlapped in Fig. \ref{F13} (a). Due to the loss functions of $Loss_{q_1}$ and $Loss_{q_2}$ are small enough, some parts of the loss function curves are missing. However, when the number of iterations is 15000 to 20000 in Fig. \ref{F13} (b), the loss function curve appears frequently from the missing state, which indicates that there are many oscillations in the loss function optimization at this stage by utilizing the L-BFGS optimizer, benefit from the introduction of loss function $Loss_a$, so that the total loss function $Loss$ decreases smoothly.

\subsection{Data-driven vector rogue waves \rm\uppercase\expandafter{\romannumeral2}}
For recovering the data-driven vector RWs \rm\uppercase\expandafter{\romannumeral2} of the Manakov system with $\lambda_1=1$ and $\lambda_2=2$ in the 9-layer IPINN with 40 neurons per layer, we consider the initial value conditions
\begin{align}\label{E28}
&q_{r}^0(x)=q_{r,\mathrm{rw2}}(x,-1.0),\,x\in[-5.0,10.0],
\end{align}
with the Dirichlet boundary conditions
\begin{align}\label{E29}
q_{r}^{\mathrm{lb}}(t)=q_{r,\mathrm{rw2}}(-5.0,t),\,q_{r}^{\mathrm{ub}}(t)=q_{r,\mathrm{rw2}}(10.0,t),\,t\in[-1.0,1.0],\,r=1,2.
\end{align}

With the aid of Matlab, the sampling points are selected by means of the finite difference method with spatial-temporal region $[-5.0,10.0]\times[-1.0,1.0]$, and the vector RWs \eqref{E22} are discretized into $[2000\times1000]$ data points which contain initial-boundary value condition \eqref{E28} and \eqref{E29}. Then, one can obtain the training dataset by randomly extracting $N_q = 2000$ from original initial-boundary value condition dataset and $N_f= 30000$ collocation points produced via the LHS in spatial-temporal region. Through 20000 Adam iterations and 23068 L-BFGS iterations to optimize loss function $\mathscr{L}(\bar{\Theta})$, the latent vector RWs $\rm\uppercase\expandafter{\romannumeral2}$ $q_r(x,t)$ have been successfully recovered by using the IPINN, and the network achieved relative $\mathbb{L}_2$ error of 3.575306$\rm e^{-3}$ for the RW $q_1(x,t)$ and relative $\mathbb{L}_2$ error of 3.321024$\rm e^{-3}$ for the RW $q_2(x,t)$, and the total number of iterations is 43068.

\begin{figure}[htbp]
\centering
\subfigure[]{
\begin{minipage}[t]{0.48\textwidth}
\centering
\includegraphics[height=6.5cm,width=6cm]{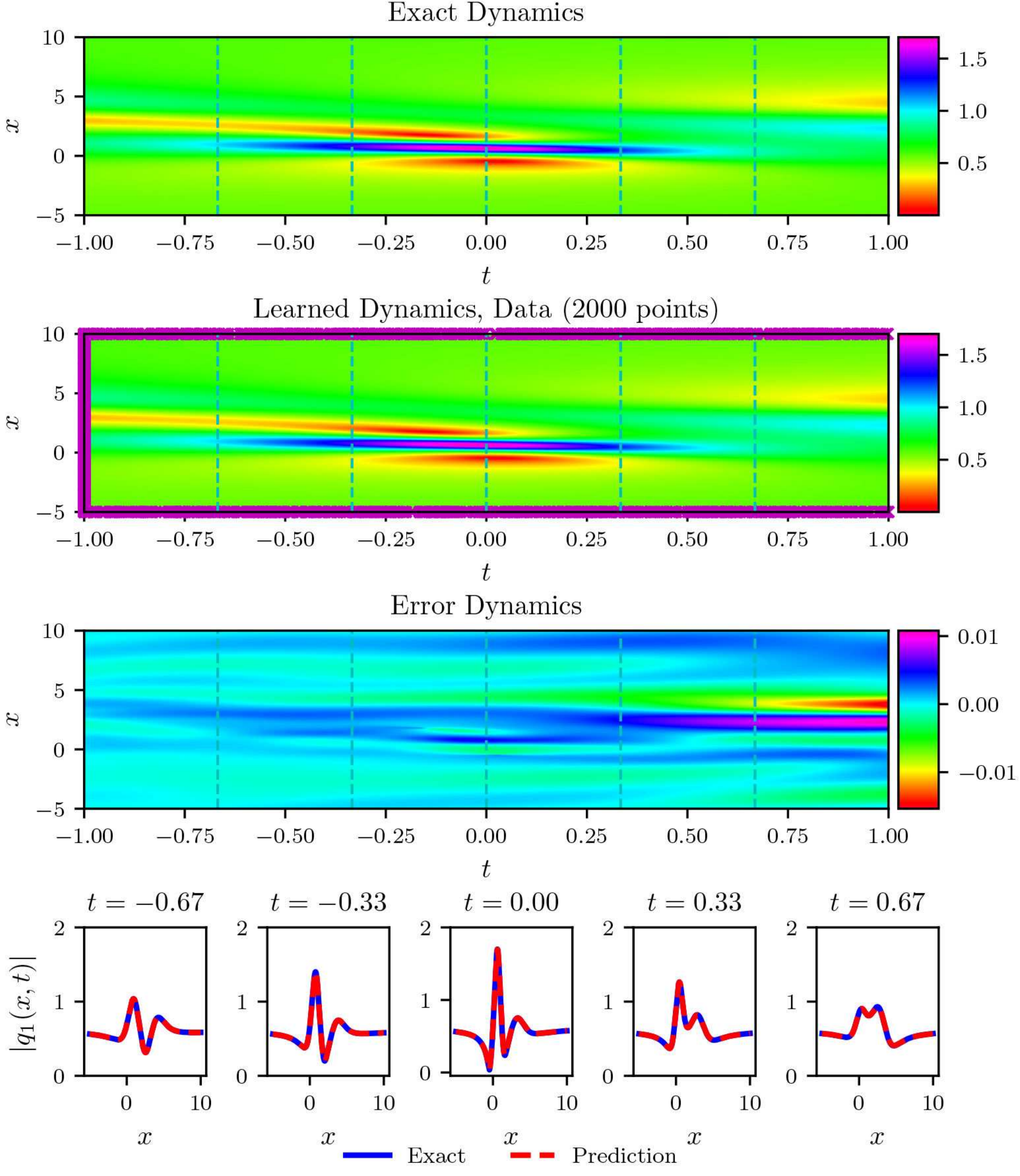}
\end{minipage}
}%
\subfigure[]{
\begin{minipage}[t]{0.48\textwidth}
\centering
\includegraphics[height=6.5cm,width=5cm]{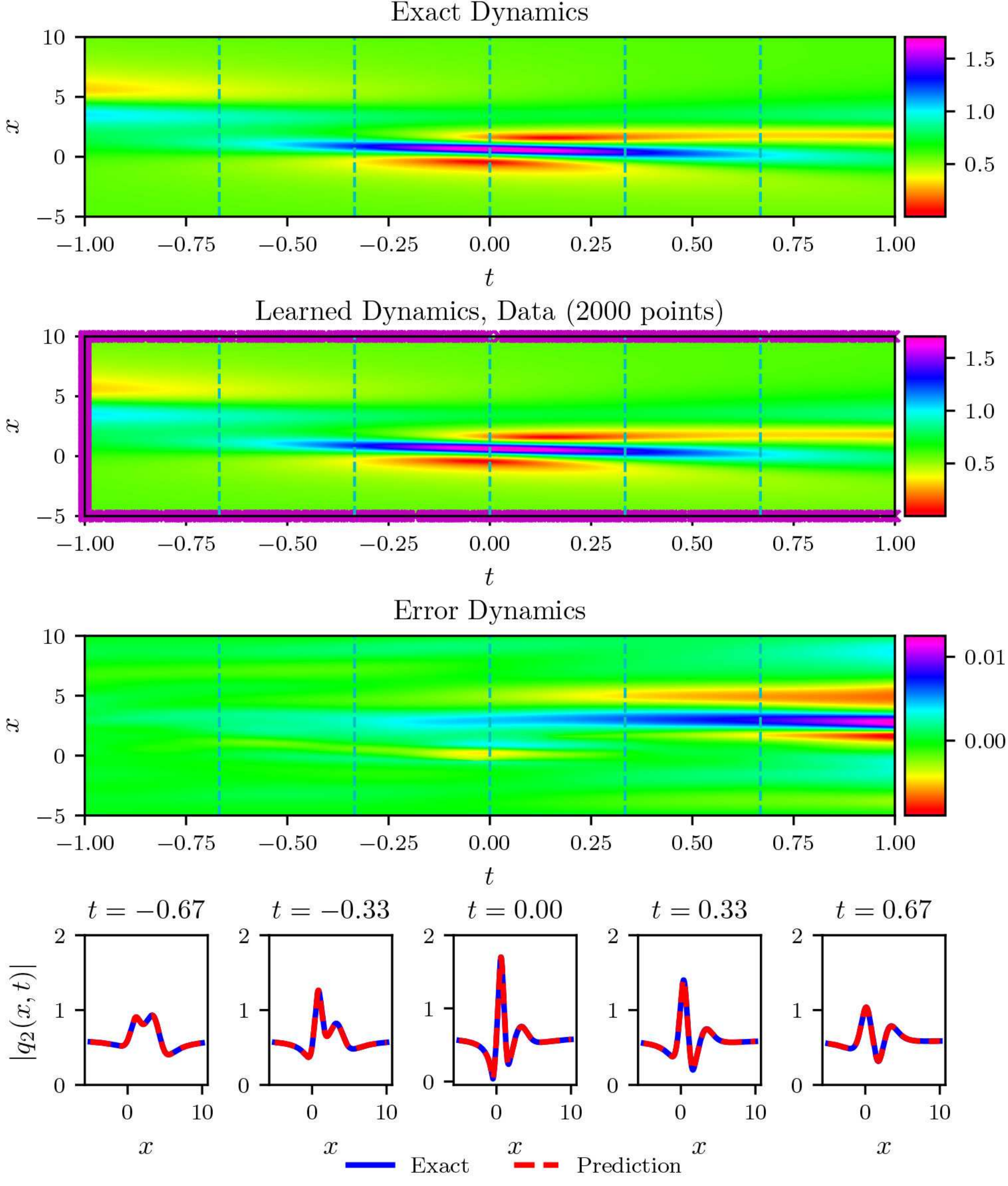}
\end{minipage}%
}%
\centering
\caption{(Color online) The vector RWs $\rm\uppercase\expandafter{\romannumeral2}$ $q_r(x,t)$ $(r=1,2)$ resulted from the IPINN with the randomly chosen initial and boundary points $N_q=2000$ which have been shown by using mediumorchid $``\times"$ in learned dynamics , and $N_f = 30000$ collocation points in the corresponding spatiotemporal region. The exact, learned and error dynamics density plots for the vector RWs $\rm\uppercase\expandafter{\romannumeral2}$ $q_r(x,t)$ with five distinct tested times $t=-0.67, -0.33, 0.00, 0.33$ and 0.67 (darkturquoise dashed lines), and the sectional drawings which contain the learned and explicit vector RWs $\rm\uppercase\expandafter{\romannumeral2}$ $q_r(x,t)$ at the aforementioned five distinct times: (a) The density plots and sectional drawings for the RW $\rm\uppercase\expandafter{\romannumeral2}$ $q_1(x,t)$; (b) The density plots and sectional drawings for the RW $\rm\uppercase\expandafter{\romannumeral2}$ $q_2(x,t)$.}
\label{F14}
\end{figure}

\begin{figure}[htbp]
\centering

\begin{minipage}[t]{0.99\textwidth}
\centering
\includegraphics[height=6cm,width=14cm]{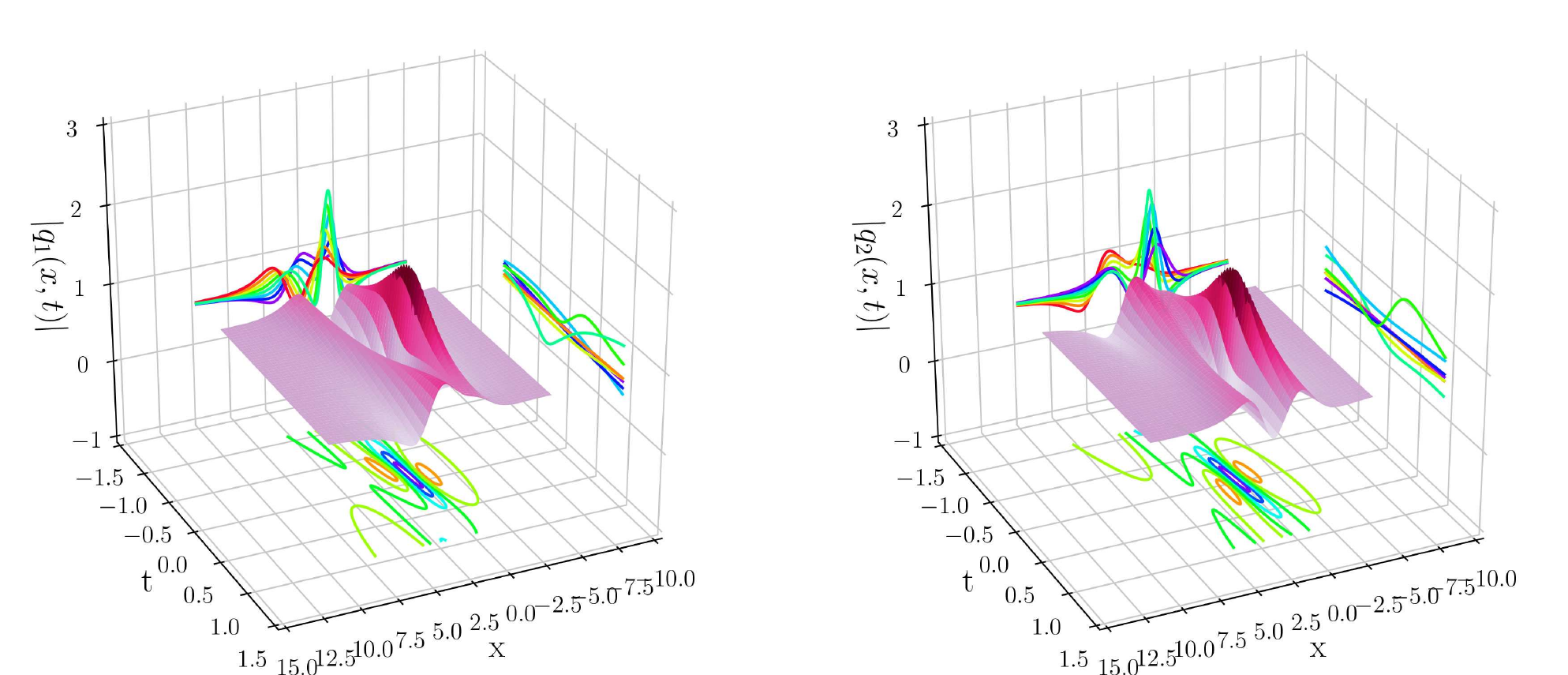}
\end{minipage}
\centering
\caption{(Color online) The three-dimensional plots with contour map on three planes of the predicted vector RWs $\rm\uppercase\expandafter{\romannumeral2}$ $q_r(x,t)$ $(r=1,2)$ based on the IPINN: (Left side panel) The 3D plot for the $q_1(x,t)$; (Right side panel) The 3D plot for the $q_2(x,t)$.}
\label{F15}
\end{figure}

\begin{figure}[htbp]
\centering
\subfigure[]{
\begin{minipage}[t]{0.48\textwidth}
\centering
\includegraphics[height=5cm,width=7cm]{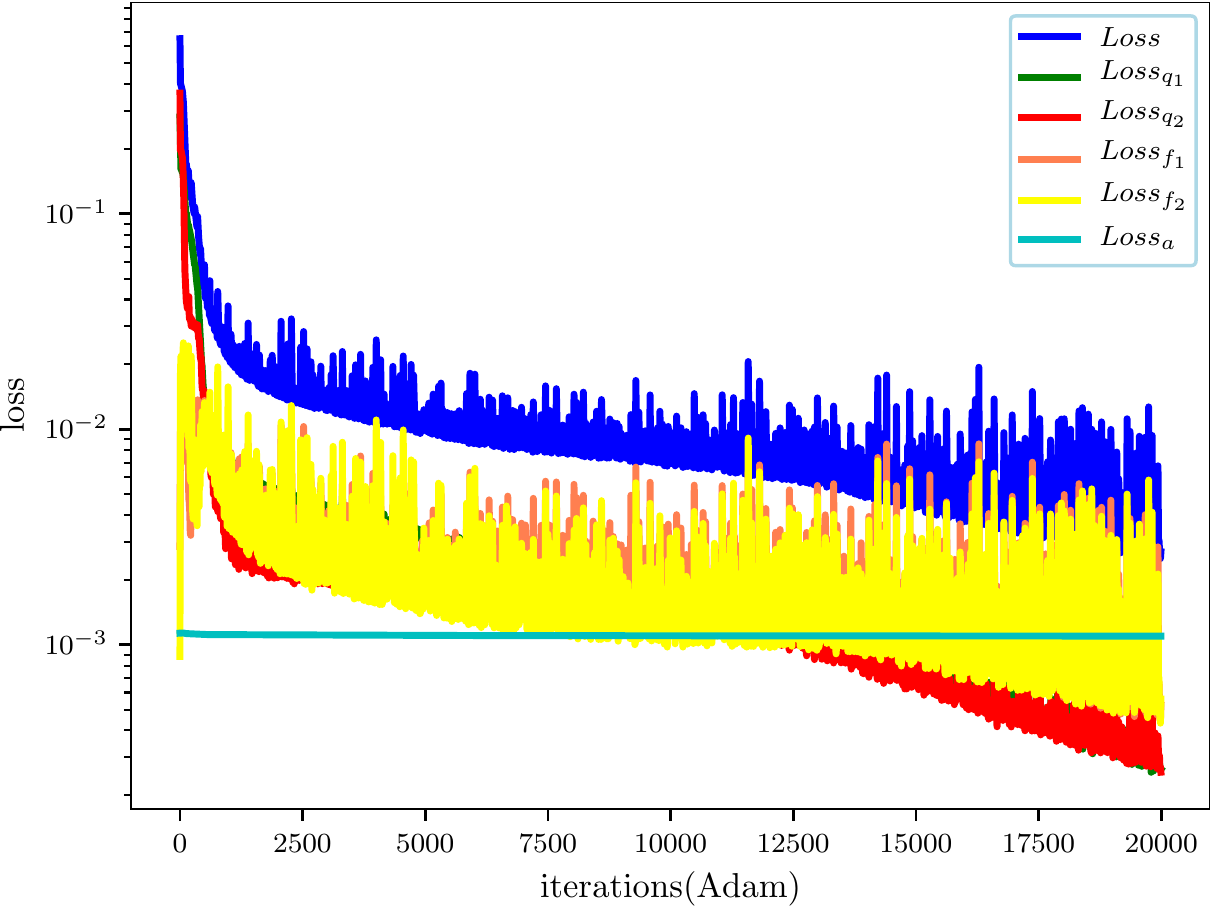}
\end{minipage}
}%
\subfigure[]{
\begin{minipage}[t]{0.48\textwidth}
\centering
\includegraphics[height=5cm,width=7cm]{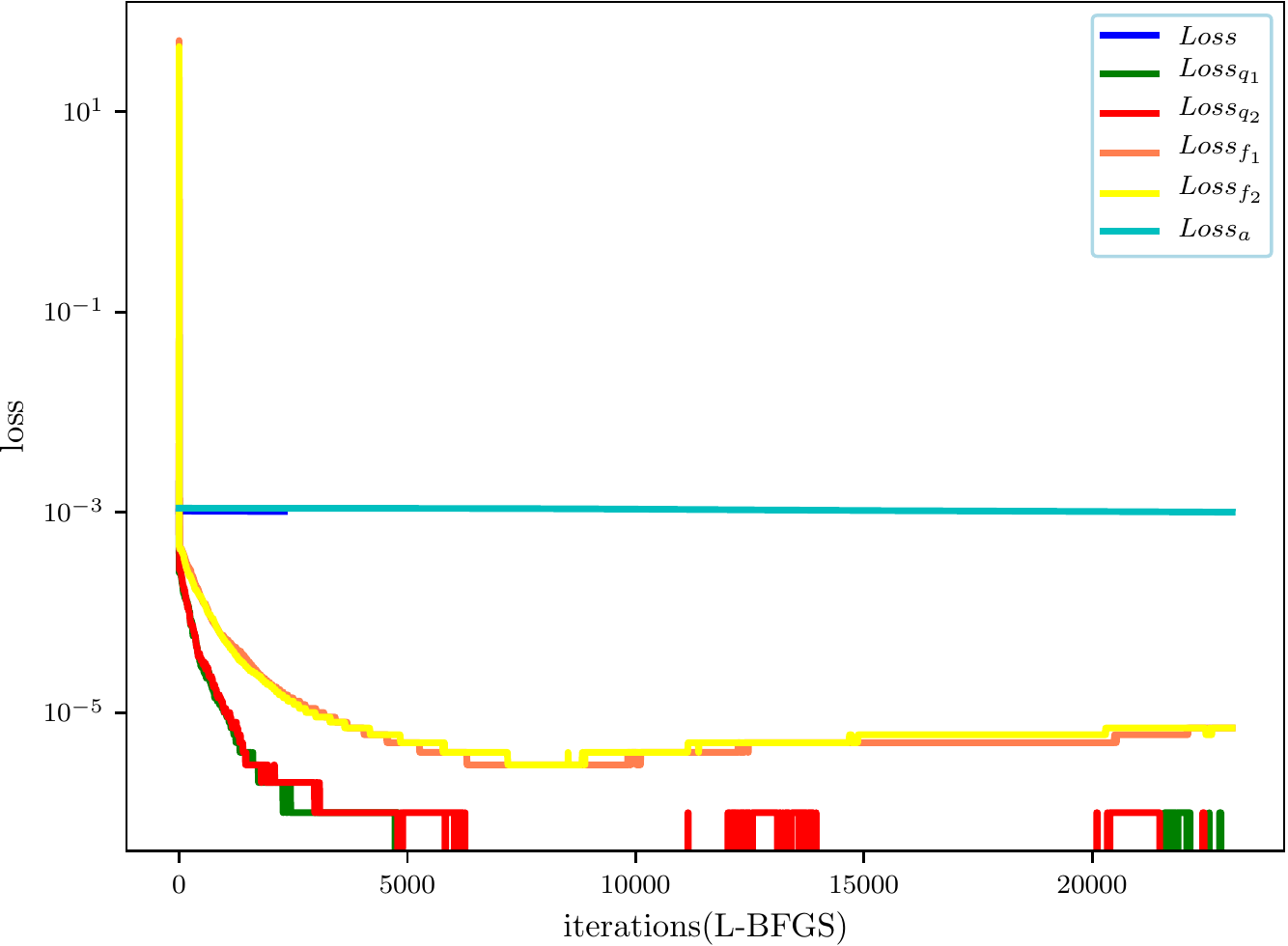}
\end{minipage}%
}%
\centering
\caption{(Color online) The loss function curve figures of the vector RWs $\rm\uppercase\expandafter{\romannumeral2}$ $q_r(x,t)$ $(r=1,2)$ arising from the IPINN with the 20000 steps Adam and 23068 steps L-BFGS optimizations: (a) The loss function curve for the 20000 Adam optimization iterations; (b) The loss function curve for the 23068 L-BFGS optimization iterations.}
\label{F16}
\end{figure}

Figs. \ref{F14} - \ref{F16} exhibit the corresponding training results stemmed from the IPINN for the vector RWs $\rm\uppercase\expandafter{\romannumeral2}$ $q_r(x,t)$ $(r=1,2)$ of the Manakov system with the initial boundary value problem \eqref{E28} and \eqref{E29}. In Fig. \ref{F14}, the exact, learned and error dynamics density plots with corresponding amplitude scale size on the right side have been given out, it is worth mentioning that the $N_q=2000$ training data points involved in the initial-boundary condition are marked by mediumorchid symbol $``\times"$ in the learned density graphs both in (a) and (b) of Fig. \ref{F14}. Meanwhile, the sectional drawings which include the learned and exact vector RWs $\rm\uppercase\expandafter{\romannumeral2}$ $q_r(x,t)$ have been shown at the five distinct times pointed out in the exact, learned and error dynamics density plots by using darkturquoise dashed lines in the bottom panel of Fig. \ref{F14}. Similar to the predicted vector RWs $\rm\uppercase\expandafter{\romannumeral1}$, from the density plots of learning dynamics and section diagrams in Fig. \ref{F14}, one can see that the two predicted component solutions $q_r(x,t)$ are mirror symmetrical about the central axis of panels (a) and (b) of Fig. \ref{F14}. Fig. \ref{F15} displays the three-dimensional plots with contour map on three planes of the predicted vector RWs $\rm\uppercase\expandafter{\romannumeral2}$ $q_r(x,t)$ based on the IPINN. Fig. \ref{F16} exhibits the loss function curve figures of the vector RWs $\rm\uppercase\expandafter{\romannumeral2}$ $q_r(x,t)$ arising from the IPINN with the 20000 steps Adam and 23068 steps L-BFGS optimizations on the loss function $\mathscr{L}(\bar{\Theta})$.

\subsection{Data-driven vector rogue waves \rm\uppercase\expandafter{\romannumeral3}}
In order to obtain the data-driven vector RWs \rm\uppercase\expandafter{\romannumeral3} of the Manakov system with $\lambda_1=1$ and $\lambda_2=2$ by using the 9-layer IPINN with 40 neurons per layer, we are committed to studying the initial value conditions
\begin{align}\label{E30}
&q_{r}^0(x)=q_{r,\mathrm{rw3}}(x,-0.5),\,x\in[-8.0,8.0],
\end{align}
and the Dirichlet boundary conditions
\begin{align}\label{E31}
q_{r}^{\mathrm{lb}}(t)=q_{r,\mathrm{rw3}}(-8.0,t),\,q_{r}^{\mathrm{ub}}(t)=q_{r,\mathrm{rw3}}(8.0,t),\,t\in[-0.5,0.5],\,r=1,2.
\end{align}

We employ the traditional finite difference scheme on even grids in Matalb to simulate Eq. \eqref{E23} which contains the initial data \eqref{E30} and boundary data \eqref{E31} to obtain the original training data. Specifically, divide spatial region $[-8.0, 8.0]$ into 2000 points and temporal region $[-0.5, 0.5]$ into 1000 points, the vector RWs $\rm\uppercase\expandafter{\romannumeral3}$ are discretized into 1000 snapshots accordingly. We generate a smaller training dataset containing initial-boundary data by randomly extracting $N_q=2000$ from original dataset and $N_f = 30000$ collocation points which are generated in corresponding spatial-temporal region by utilizing the LHS method. After that, introducing the dataset of initial and boundary points into the IPINN, and taking 20000 Adam iterations and 26037 L-BFGS iterations to optimize loss function $\mathscr{L}(\bar{\Theta})$, the latent vector RWs $\rm\uppercase\expandafter{\romannumeral2}$ $q_r(x,t)$ have been successfully learned by tuning all learnable parameters of the IPINN, and the network achieved relative $\mathbb{L}_2$ error of 2.197789$\rm e^{-3}$ for the RW $q_1(x,t)$ and relative $\mathbb{L}_2$ error of 3.661877$\rm e^{-3}$ for the RW $q_2(x,t)$, and the total number of iterations is 46037.

\begin{figure}[htbp]
\centering
\subfigure[]{
\begin{minipage}[t]{0.48\textwidth}
\centering
\includegraphics[height=6.5cm,width=6cm]{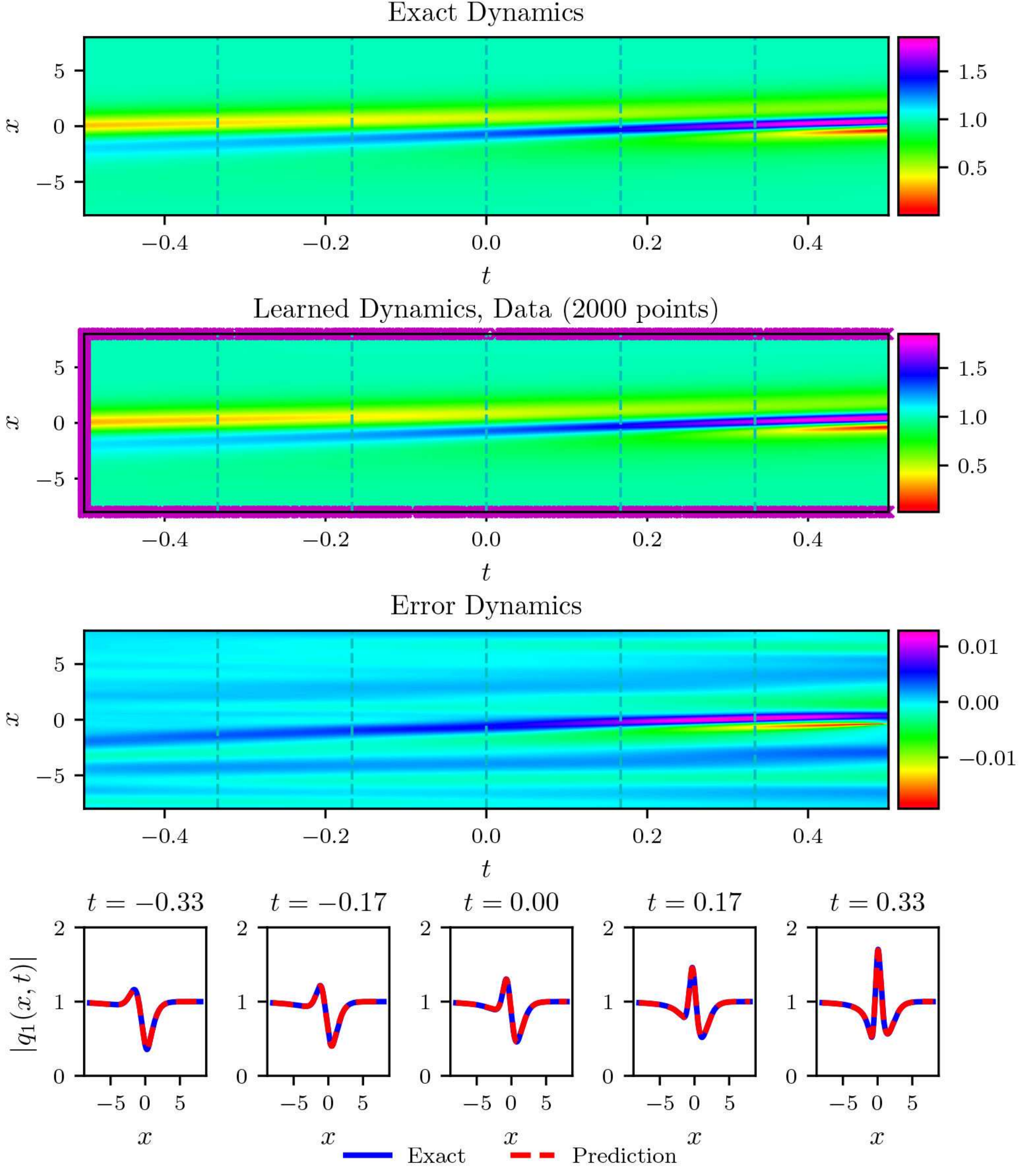}
\end{minipage}
}%
\subfigure[]{
\begin{minipage}[t]{0.48\textwidth}
\centering
\includegraphics[height=6.5cm,width=5cm]{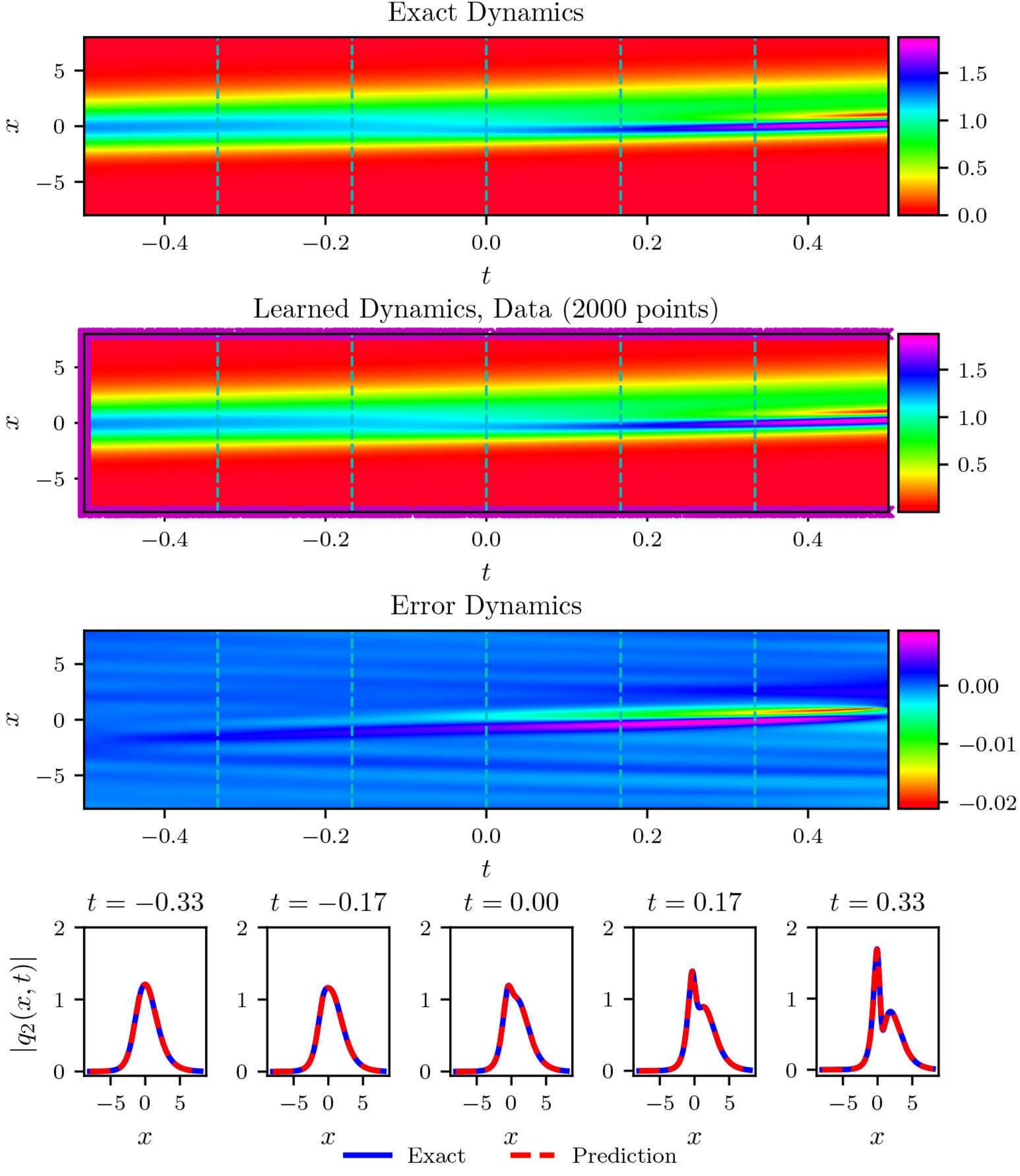}
\end{minipage}%
}%
\centering
\caption{(Color online) The vector RWs $\rm\uppercase\expandafter{\romannumeral3}$ $q_r(x,t)$ $(r=1,2)$ resulted from the IPINN with the randomly chosen initial and boundary points $N_q=2000$ which have been shown by using mediumorchid $``\times"$ in learned dynamics , and $N_f = 30000$ collocation points in the corresponding spatiotemporal region. The exact, learned and error dynamics density plots for the the RWs $\rm\uppercase\expandafter{\romannumeral3}$ $q_r(x,t)$ with five distinct tested times $t=-0.33, -0.17, 0.00, 0.17$ and 0.33 (darkturquoise dashed lines), and the sectional drawings which contain the learned and explicit the RWs $\rm\uppercase\expandafter{\romannumeral3}$ $q_r(x,t)$ at the aforementioned five distinct times: (a) The density plots and sectional drawings for the RW $\rm\uppercase\expandafter{\romannumeral3}$ $q_1(x,t)$; (b) The density plots and sectional drawings for the RW $\rm\uppercase\expandafter{\romannumeral3}$ $q_2(x,t)$.}
\label{F17}
\end{figure}

\begin{figure}[htbp]
\centering

\begin{minipage}[t]{0.99\textwidth}
\centering
\includegraphics[height=6cm,width=14cm]{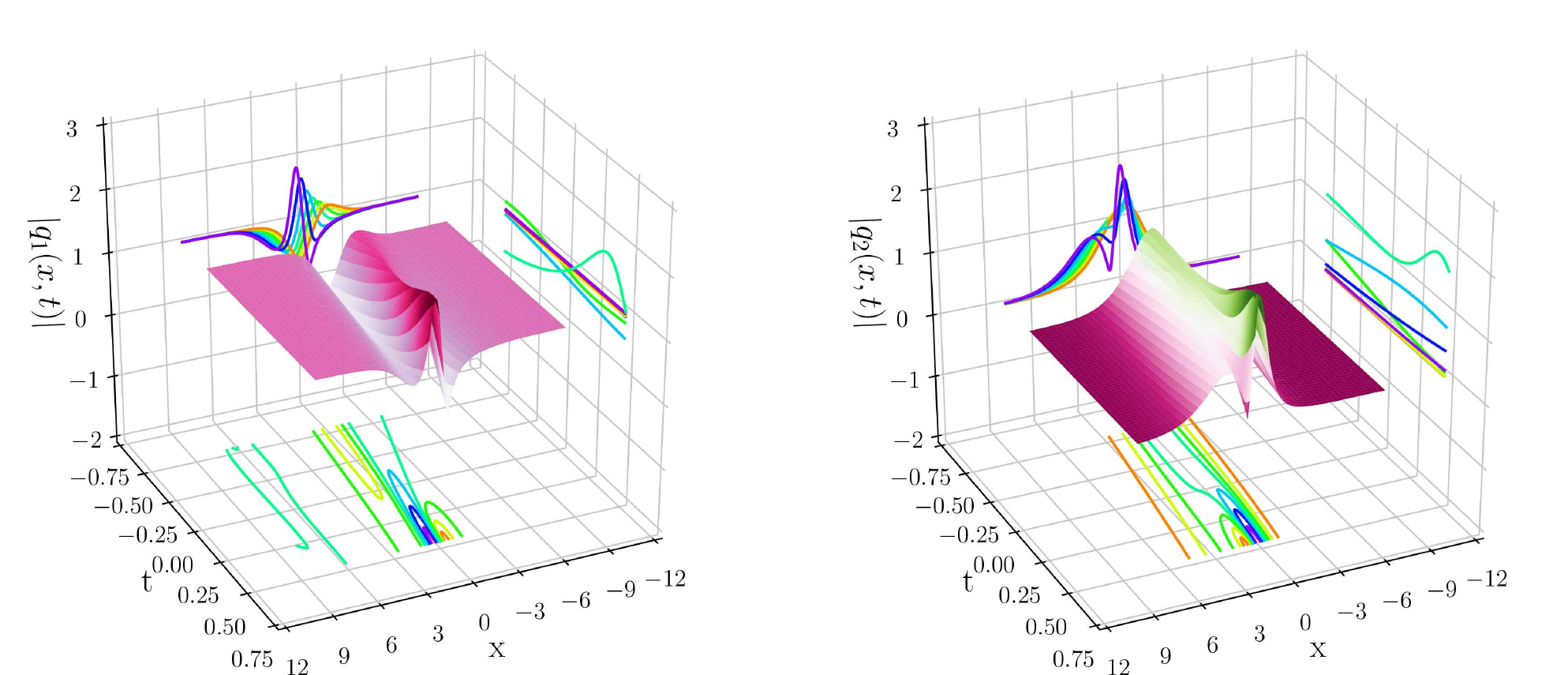}
\end{minipage}
\centering
\caption{(Color online) The three-dimensional plots with contour map on three planes of the predicted vector RWs $\rm\uppercase\expandafter{\romannumeral3}$ $q_r(x,t)$ $(r=1,2)$ based on the IPINN: (Left side panel) The 3D plot for the $q_1(x,t)$; (Right side panel) The 3D plot for the $q_2(x,t)$.}
\label{F18}
\end{figure}

\begin{figure}[htbp]
\centering
\subfigure[]{
\begin{minipage}[t]{0.48\textwidth}
\centering
\includegraphics[height=5cm,width=7cm]{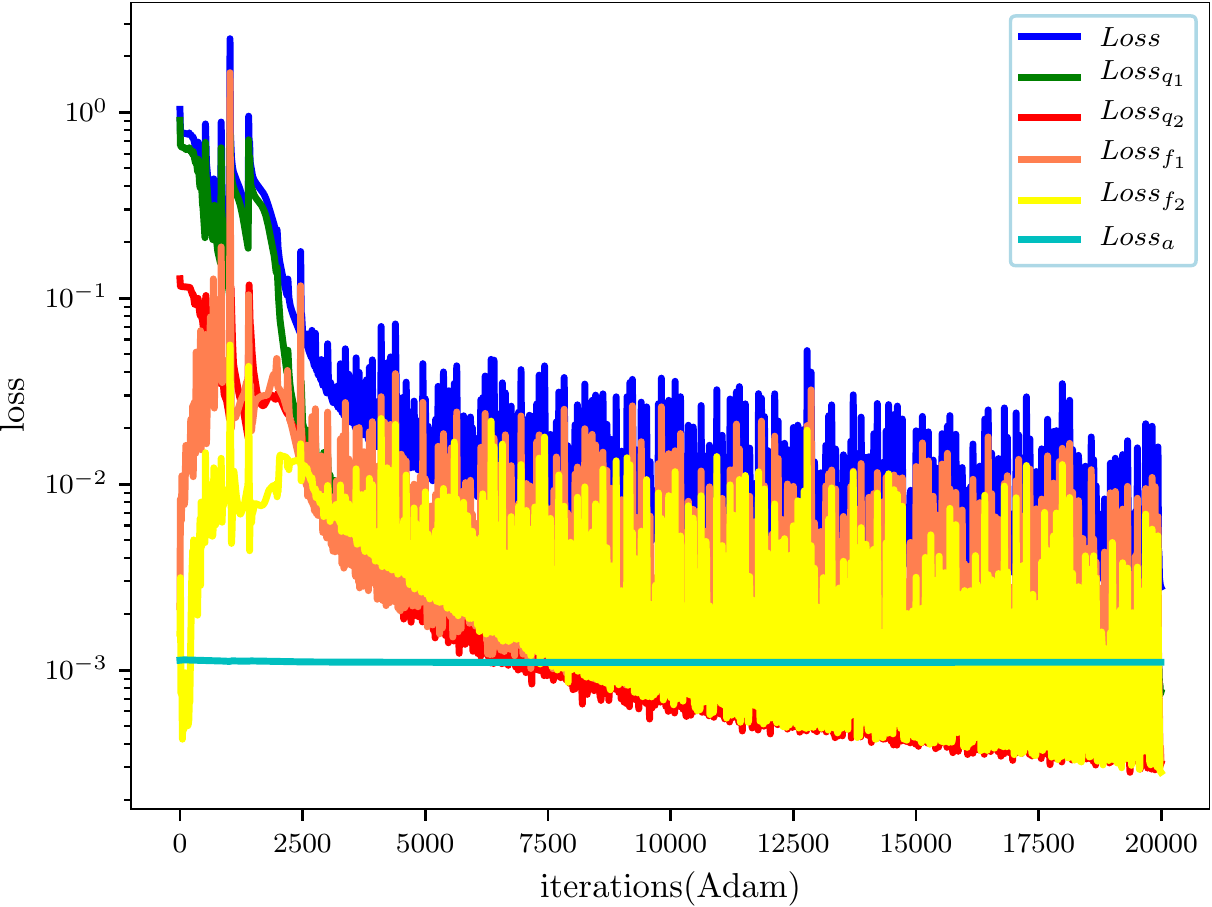}
\end{minipage}
}%
\subfigure[]{
\begin{minipage}[t]{0.48\textwidth}
\centering
\includegraphics[height=5cm,width=7cm]{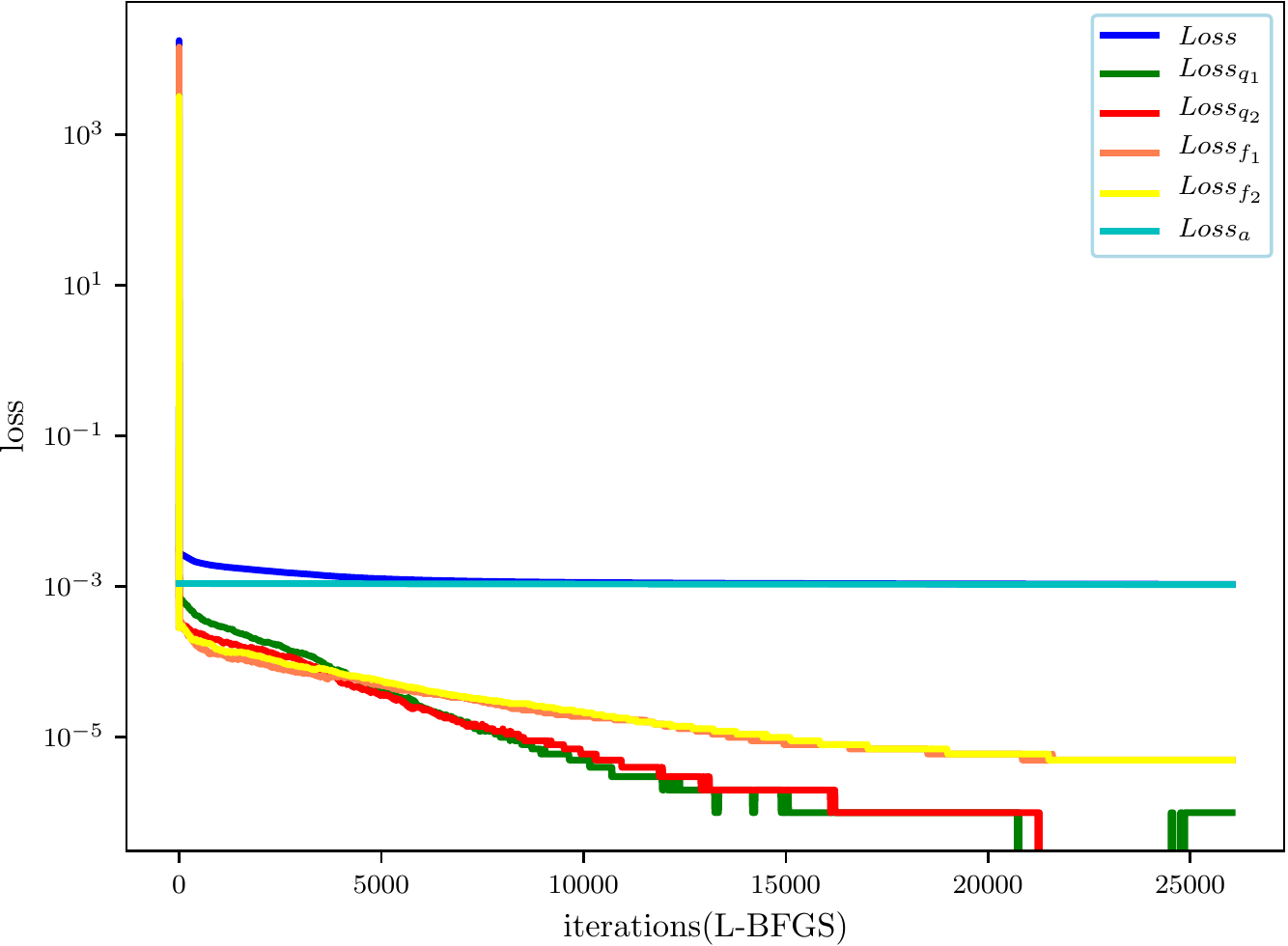}
\end{minipage}%
}%
\centering
\caption{(Color online) The loss function curve figures of the vector RWs $\rm\uppercase\expandafter{\romannumeral3}$ $q_r(x,t)$ $(r=1,2)$ arising from the IPINN with the 20000 steps Adam and 26037 steps L-BFGS optimizations: (a) The loss function curve for the 20000 Adam optimization iterations; (b) The loss function curve for the 26.37 L-BFGS optimization iterations.}
\label{F19}
\end{figure}

Figs. \ref{F17} - \ref{F19} provide the training results arising from the IPINN for the vector RWs $\rm\uppercase\expandafter{\romannumeral3}$ $q_r(x,t)$ $(r=1,2)$ of the Manakov system with the initial boundary value problem \eqref{E30} and \eqref{E31}. In Fig. \ref{F17}, the exact, learned and error dynamics density plots with corresponding amplitude scale size on the right side have been exhibited, it is worth mentioning that the $N_q=2000$ training data points involved in the initial-boundary condition are marked by mediumorchid symbol $``\times"$ in the learned density plots both in (a) and (b) of Fig. \ref{F17}. Meanwhile, the sectional drawings which include the learned and exact vector RWs $\rm\uppercase\expandafter{\romannumeral3}$ $q_r(x,t)$ have been shown at the five distinct times pointed out in the exact, learned and error dynamics density plots by using darkturquoise dashed lines in the bottom panel of Fig. \ref{F17}. Fig. \ref{F18} displays the three-dimensional plots with contour map on three planes of the predicted vector RWs $\rm\uppercase\expandafter{\romannumeral3}$ $q_r(x,t)$ based on the IPINN. From Figs. \ref{F17} and \ref{F18}, with the development of time, one can observe that the first component $q_1(x,t)$ of vector RWs $\rm\uppercase\expandafter{\romannumeral3}$ $q_r(x,t)$ is an interaction solution that dark soliton gradually generates a RW on the side of the dark soliton, while the second component $q_2(x,t)$ is an interaction solution that the bright soliton is gradually emerges a RW on the wave crest of bright soliton. Fig. \ref{F19} exhibits the loss function curve figures of the vector RWs $\rm\uppercase\expandafter{\romannumeral3}$ $q_r(x,t)$ arising from the IPINN with the 20000 steps Adam and 26037 steps L-BFGS optimizations on the loss function $\mathscr{L}(\bar{\Theta})$. In particular, from the (a) of Fig. \ref{F19}, these loss functions present an unstable iteration curve in the first 2500 iterations of Adam optimizer, and the values of the loss functions have an upward trend at this time.

\subsection{Data-driven vector rogue waves \rm\uppercase\expandafter{\romannumeral4}}
Similarly, considering the initial condition and Dirichlet boundary condition of the Manakov system with $\lambda_1=1$ and $\lambda_2=2$ to obtain the data-driven vector RWs \rm\uppercase\expandafter{\romannumeral4} by using the 9-layer IPINN with 40 neurons per layer, the $[L_0, L_1]$ and $[T_0, T_1]$ in Eq. \eqref{E1} are taken as $[-10.0, 5.0]$ and $[-0.5, 0.5]$, respectively. We immediately obtain the initial value conditions
\begin{align}\label{E32}
&q_{r}^0(x)=q_{r,\mathrm{rw4}}(x,-0.5),\,x\in[-10.0,5.0],
\end{align}
and the Dirichlet boundary conditions
\begin{align}\label{E33}
q_{r}^{\mathrm{lb}}(t)=q_{r,\mathrm{rw4}}(-10.0,t),\,q_{r}^{\mathrm{ub}}(t)=q_{r,\mathrm{rw4}}(5.0,t),\,t\in[-0.5,0.5],\,r=1,2.
\end{align}

Similarly, discretizing Eq. \eqref{E24} with the aid of the traditional finite difference scheme on even grids, and we obtain the original training data which contain initial data \eqref{E32} and boundary data \eqref{E33} by dividing separately the spatial region $[-10.0, 5.0]$ into 2000 points and the temporal region $[-0.5, 0.5]$ into 1000 points. Then, one can generate a smaller training dataset that contains partial initial-boundary data by randomly extracting $N_q = 2000$ from original dataset and $N_f = 30000$ collocation points which are produced by the LHS. After that, the latent vector RWs $\rm\uppercase\expandafter{\romannumeral4}$ $q_r(x,t)$ have been successfully learned by tuning all learnable parameters of the IPINN, and the network achieved relative $\mathbb{L}_2$ error of 1.637892$\rm e^{-3}$ for the RW $q_1(x,t)$ and relative $\mathbb{L}_2$ error of 1.784348$\rm e^{-3}$ for the RW $q_2(x,t)$, and the total number of iterations is 51110.

\begin{figure}[htbp]
\centering
\subfigure[]{
\begin{minipage}[t]{0.48\textwidth}
\centering
\includegraphics[height=6.5cm,width=6cm]{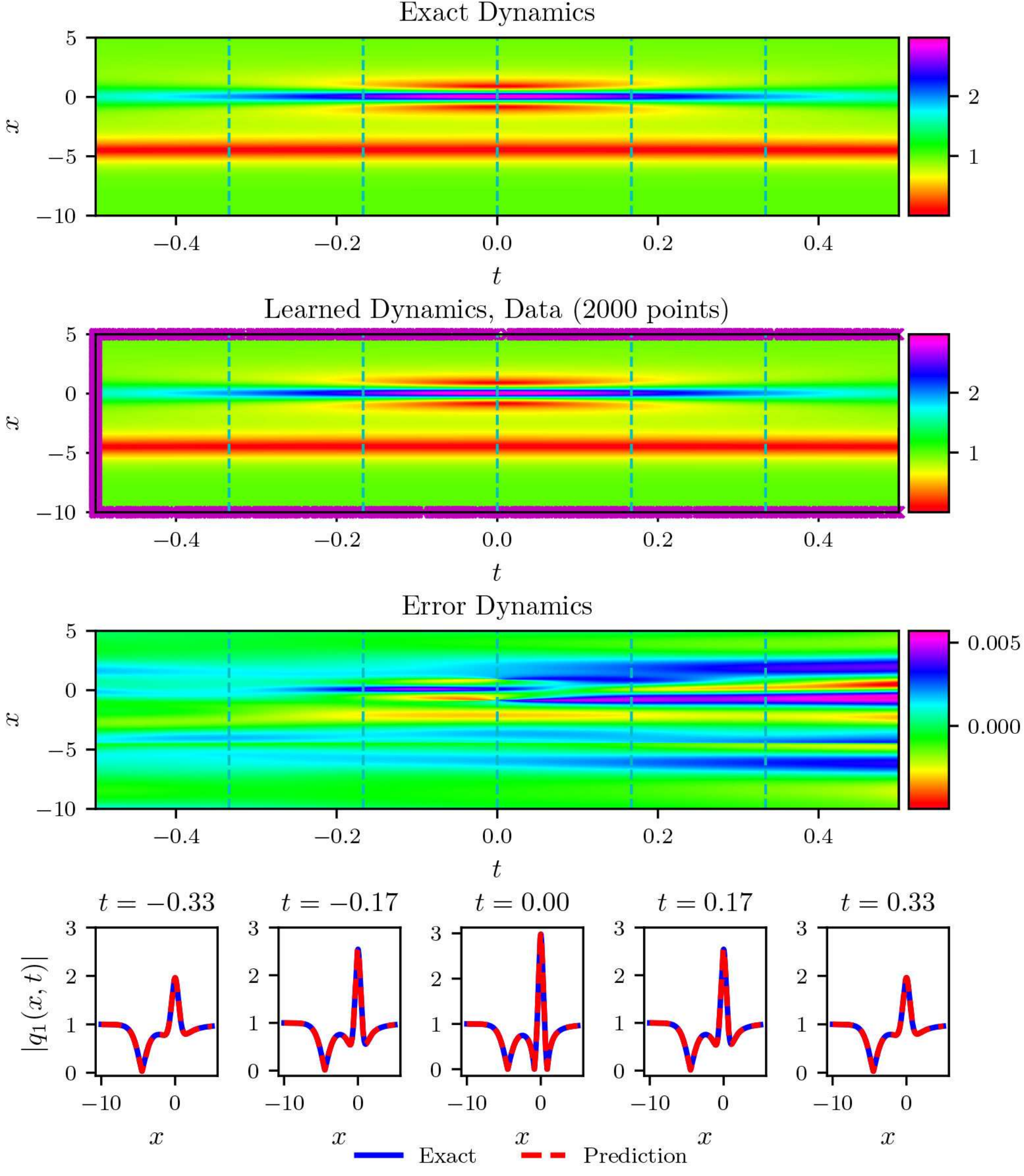}
\end{minipage}
}%
\subfigure[]{
\begin{minipage}[t]{0.48\textwidth}
\centering
\includegraphics[height=6.5cm,width=5cm]{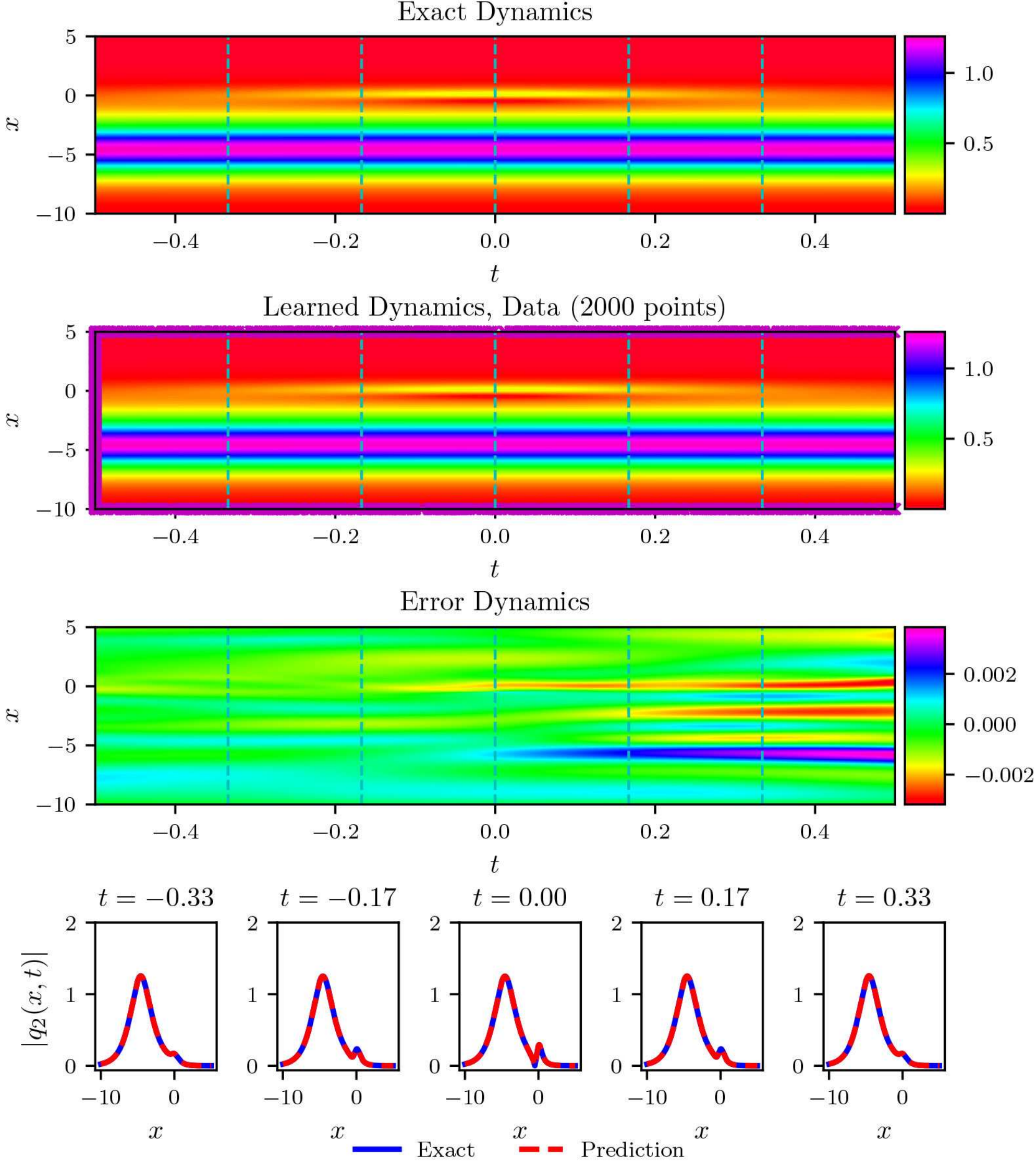}
\end{minipage}%
}%
\centering
\caption{(Color online) The vector RWs $\rm\uppercase\expandafter{\romannumeral4}$ $q_r(x,t)$ $(r=1,2)$ resulted from the IPINN with the randomly chosen initial and boundary points $N_q=2000$ which have been shown by using mediumorchid $``\times"$ in learned dynamics , and $N_f = 30000$ collocation points in the corresponding spatiotemporal region. The exact, learned and error dynamics density plots for the vector RWs $\rm\uppercase\expandafter{\romannumeral4}$ $q_r(x,t)$ with five distinct tested times $t=-0.33, -0.17, 0.00, 0.17$ and 0.33 (darkturquoise dashed lines), and the sectional drawings which contain the learned and explicit vector RWs $\rm\uppercase\expandafter{\romannumeral4}$ $q_r(x,t)$ at the aforementioned five distinct times: (a) The density plots and sectional drawings for the RW $\rm\uppercase\expandafter{\romannumeral4}$ $q_1(x,t)$; (b) The density plots and sectional drawings for the RW $\rm\uppercase\expandafter{\romannumeral4}$ $q_2(x,t)$.}
\label{F20}
\end{figure}

\begin{figure}[htbp]
\centering

\begin{minipage}[t]{0.99\textwidth}
\centering
\includegraphics[height=6cm,width=14cm]{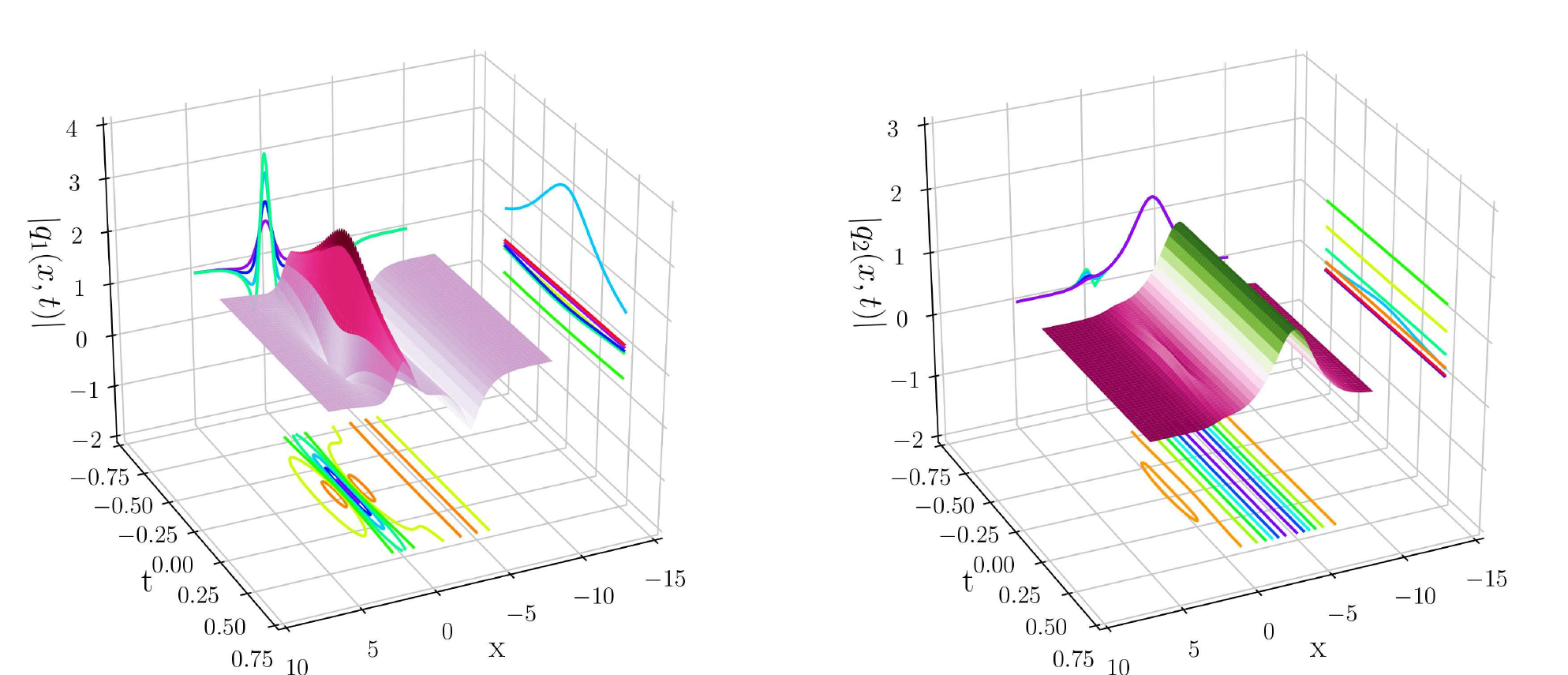}
\end{minipage}
\centering
\caption{(Color online) The three-dimensional plots with contour map on three planes of the predicted vector RWs $\rm\uppercase\expandafter{\romannumeral4}$ $q_r(x,t)$ $(r=1,2)$ based on the IPINN: (Left side panel) The 3D plot for the $q_1(x,t)$; (Right side panel) The 3D plot for the $q_2(x,t)$.}
\label{F21}
\end{figure}

\begin{figure}[htbp]
\centering
\subfigure[]{
\begin{minipage}[t]{0.48\textwidth}
\centering
\includegraphics[height=5cm,width=7cm]{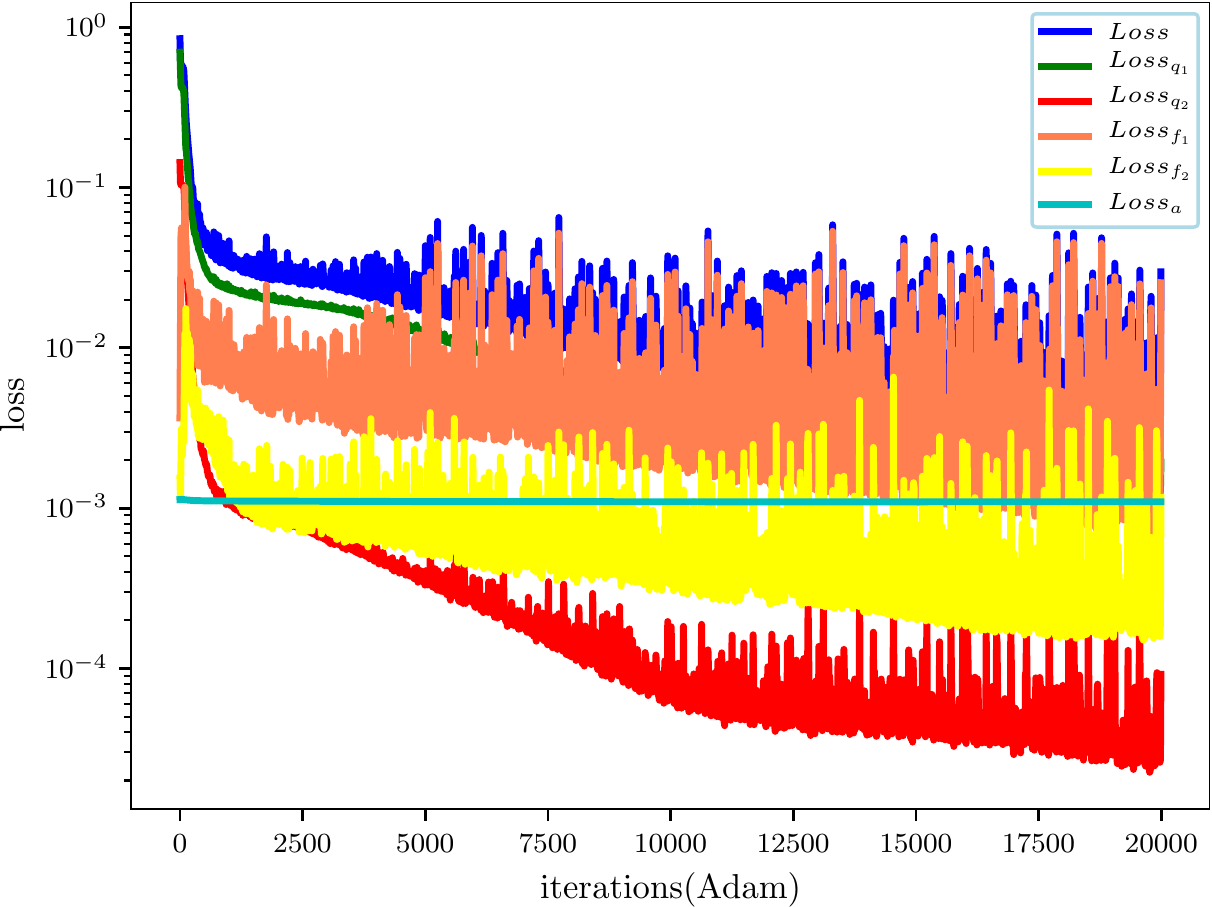}
\end{minipage}
}%
\subfigure[]{
\begin{minipage}[t]{0.48\textwidth}
\centering
\includegraphics[height=5cm,width=7cm]{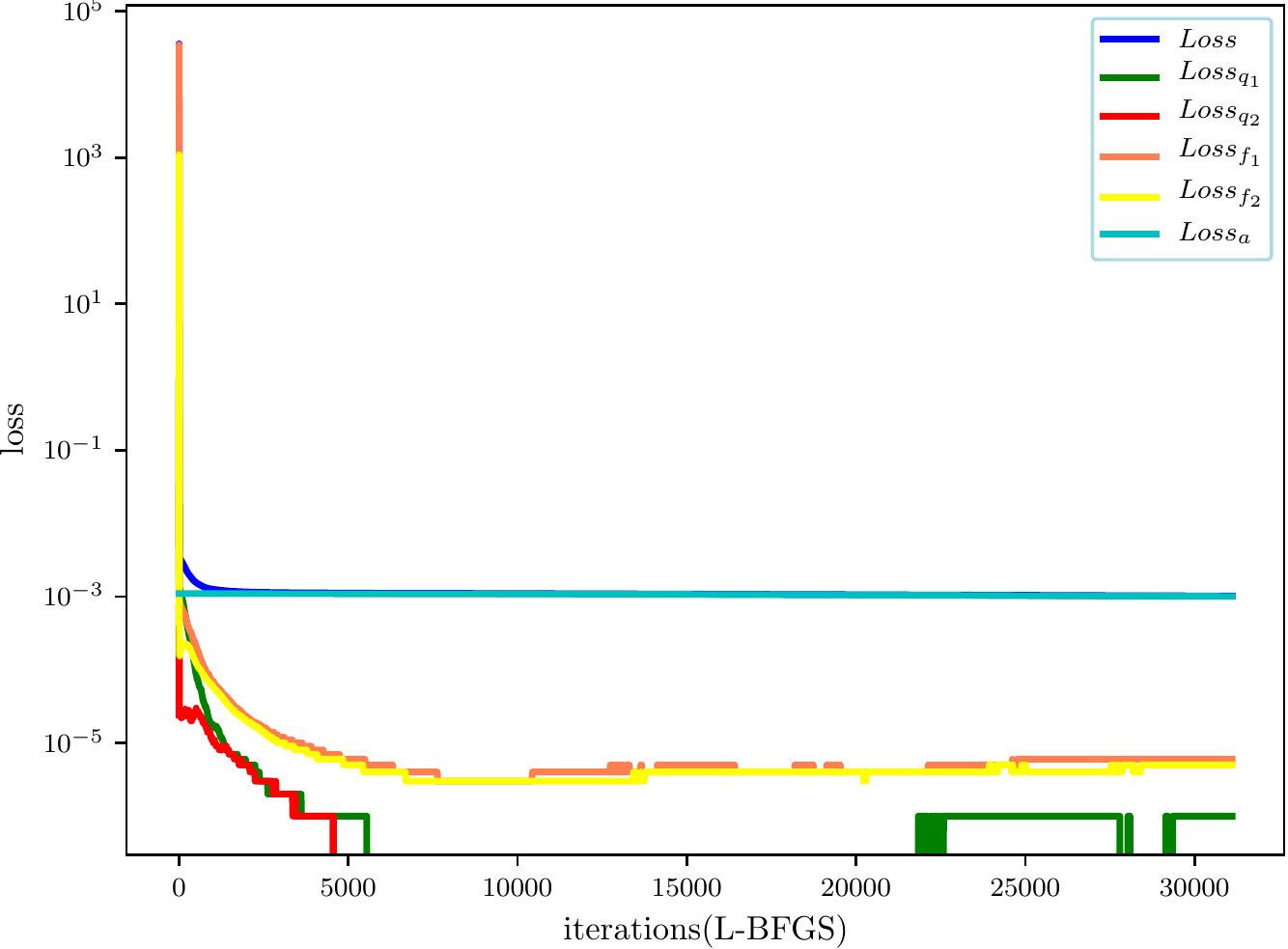}
\end{minipage}%
}%
\centering
\caption{(Color online) The loss function curve figures of the vector RWs $\rm\uppercase\expandafter{\romannumeral4}$ $q_r(x,t)$ $(r=1,2)$ arising from the IPINN with the 20000 steps Adam and 31110 steps L-BFGS optimizations: (a) The loss function curve for the 20000 Adam optimization iterations; (b) The loss function curve for the 31110 L-BFGS optimization iterations.}
\label{F22}
\end{figure}

Figs. \ref{F20} - \ref{F22} provide the training results arising from the IPINN for the vector RWs $\rm\uppercase\expandafter{\romannumeral4}$ $q_r(x,t)$ $(r=1,2)$ of the Manakov system with the initial boundary value problem \eqref{E32} and \eqref{E33}. In Fig. \ref{F20}, the exact, learned and error dynamics density plots with corresponding amplitude scale size on the right side have been exhibited, it is worth mentioning that the $N_q=2000$ training data points involved in the initial-boundary condition are marked by mediumorchid symbol $``\times"$ in the learned density plots both in (a) and (b) of Fig. \ref{F20}. Meanwhile, the sectional drawings which include the learned and exact vector RWs $\rm\uppercase\expandafter{\romannumeral4}$ $q_r(x,t)$ have been shown at the five distinct times pointed out in the exact, learned and error dynamics density plots by using darkturquoise dashed lines in the bottom panel of Fig. \ref{F20}. Fig. \ref{F21} displays the three-dimensional plots with contour map on three planes of the predicted vector RWs $\rm\uppercase\expandafter{\romannumeral4}$ $q_r(x,t)$ based on the IPINN. From Figs. \ref{F20} and \ref{F21}, with the development of time, one can observe that the first component $q_1(x,t)$ of vector RWs $\rm\uppercase\expandafter{\romannumeral4}$ $q_r(x,t)$ is the interaction solution composed of dark soliton and RW, while the second component $q_2(x,t)$ is the interaction solution composed of bright soliton and RW whose the amplitude is much lower than that of bright soliton. Different from the case where the RW in the second component of vector RW $\rm\uppercase\expandafter{\romannumeral3}$ emerges on the wave crest, RWs of this part both appear on one side of the bright-dark soliton. Fig. \ref{F22} exhibits the loss function curve figures of the vector RWs $\rm\uppercase\expandafter{\romannumeral4}$ $q_r(x,t)$ arising from the IPINN with the 20000 steps Adam and 31110 steps L-BFGS optimizations on the loss function $\mathscr{L}(\bar{\Theta})$.

\subsection{Data-driven vector rogue waves \rm\uppercase\expandafter{\romannumeral5}}
Next, considering the initial conditions $q_{r}^0(x)$ and Dirichlet boundary conditions $q_{r}^{\mathrm{lb}}(t)$ and $q_{r}^{\mathrm{ub}}(t)$ of Eq. \eqref{E1} arising from the vector RWs \rm\uppercase\expandafter{\romannumeral5} Eq. \eqref{E25} for obtaining the data-driven vector RWs \rm\uppercase\expandafter{\romannumeral5} by using the 9-layer IPINN with 40 neurons per layer, we set $[L_0,L_1]$ and $[T_0,T_1]$ in Eq. \eqref{E1} as $[-8.0,6.0]$ and $[-0.25,0.25]$, respectively. After that, the corresponding initial conditions can be written as belows
\begin{align}\label{E34}
&q_{r}^0(x)=q_{r,\mathrm{rw5}}(x,-0.25),\,x\in[-8.0,6.0],
\end{align}
and the Dirichlet boundary conditions
\begin{align}\label{E35}
q_{r}^{\mathrm{lb}}(t)=q_{r,\mathrm{rw5}}(-8.0,t),\,q_{r}^{\mathrm{ub}}(t)=q_{r,\mathrm{rw5}}(6.0,t),\,t\in[-0.25,0.25],\,r=1,2.
\end{align}

Here, employing the same data discretization method in section 4.1, and  producing the training data which consists of initial data \eqref{E34} and boundary data \eqref{E35} by dividing the spatial region $[-8.0,6.0]$ into 2000 points and the temporal region $[-0.25,0.25]$ into 1000 points. We generate a smaller training dataset that containing initial-boundary data by randomly extracting $N_q=2000$ from original dataset and $N_f=30000$ collocation points which are generated by the LHS method. After giving the dataset of initial and boundary points, the latent data-driven vector RWs \rm\uppercase\expandafter{\romannumeral5} have been successfully learned by tuning all learnable parameters of the IPINN and utilizing 20000 Adam iterations and 14602 L-BFGS iterations to regulate the loss function \eqref{E6}, and the network achieved relative $\mathbb{L}_2$ error of 6.325696$\rm e^{-3}$ for the RW $q_1(x,t)$ and relative $\mathbb{L}_2$ error of 7.654479$\rm e^{-3}$ for the RW $q_2(x,t)$, and the total number of iterations is 34602.

\begin{figure}[htbp]
\centering
\subfigure[]{
\begin{minipage}[t]{0.48\textwidth}
\centering
\includegraphics[height=6.5cm,width=6cm]{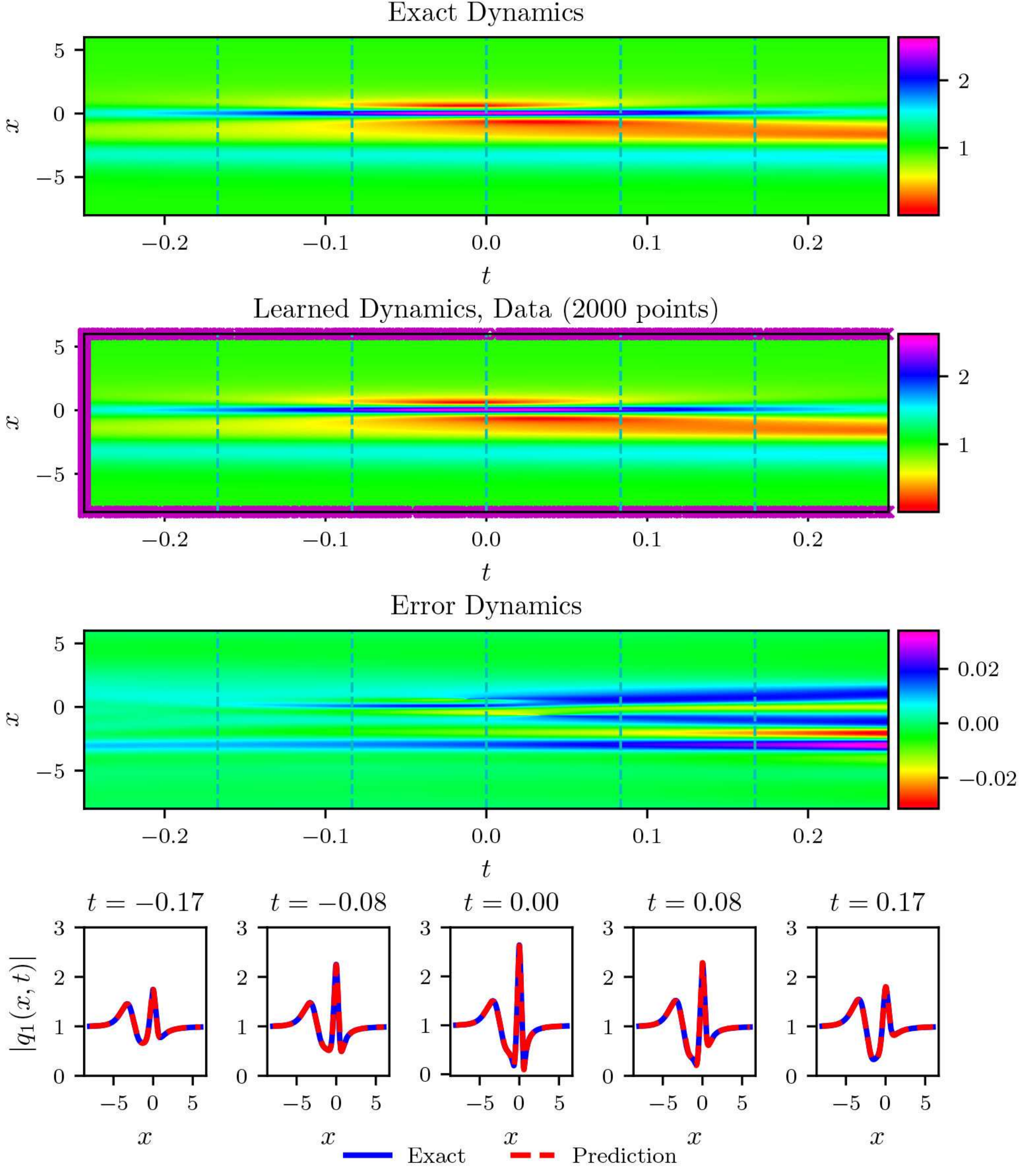}
\end{minipage}
}%
\subfigure[]{
\begin{minipage}[t]{0.48\textwidth}
\centering
\includegraphics[height=6.5cm,width=5cm]{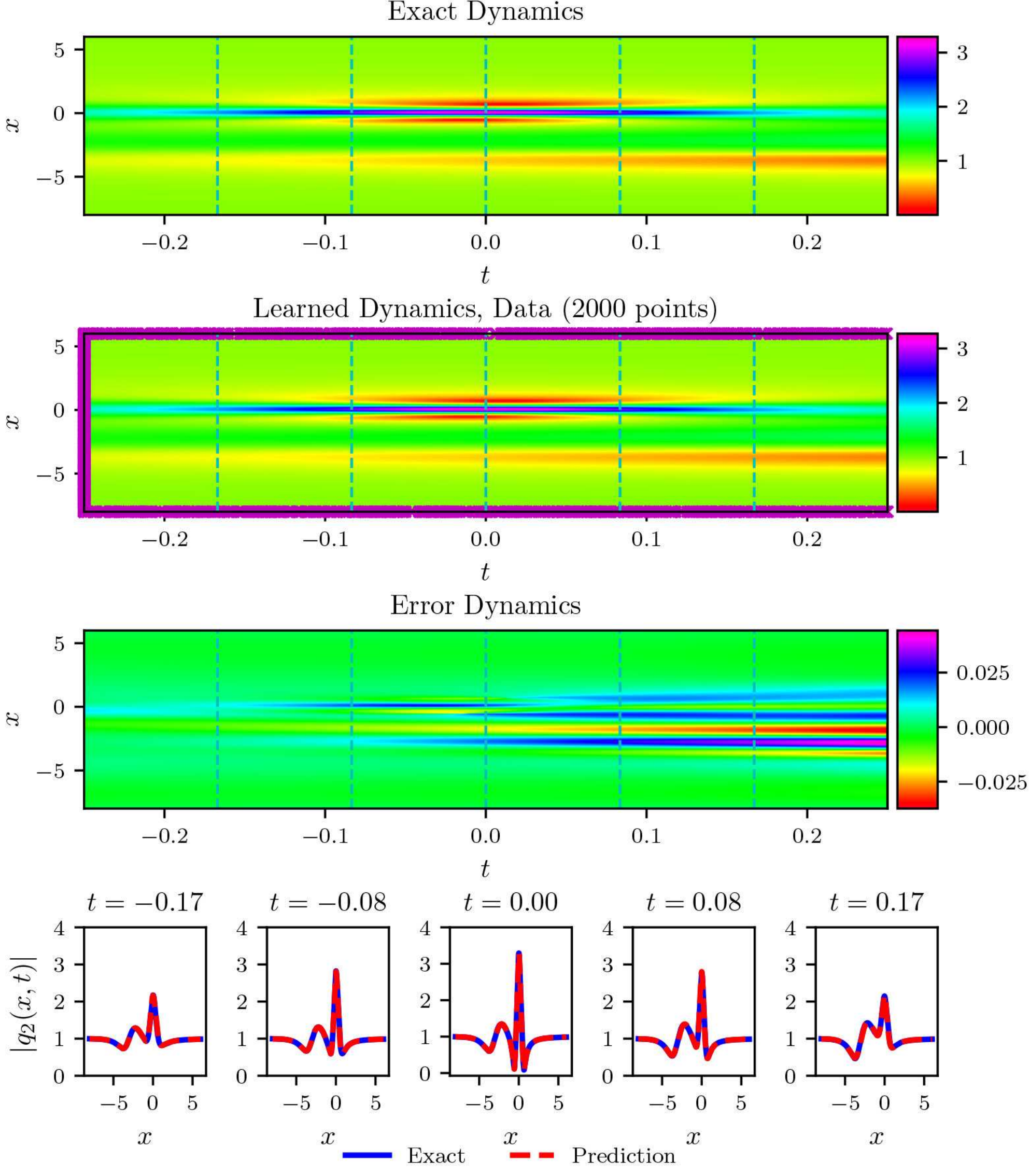}
\end{minipage}%
}%
\centering
\caption{(Color online) The vector RWs $\rm\uppercase\expandafter{\romannumeral5}$ $q_r(x,t)$ $(r=1,2)$ resulted from the IPINN with the randomly chosen initial and boundary points $N_q=2000$ which have been shown by using mediumorchid $``\times"$ in learned dynamics , and $N_f = 30000$ collocation points in the corresponding spatiotemporal region. The exact, learned and error dynamics density plots for the vector RWs $\rm\uppercase\expandafter{\romannumeral5}$ $q_r(x,t)$ with five distinct tested times $t=-0.17, -0.08, 0.00, 0.08$ and 0.17 (darkturquoise dashed lines), and the sectional drawings which contain the learned and explicit vector RW $\rm\uppercase\expandafter{\romannumeral5}$ $q_r(x,t)$ at the aforementioned five distinct times: (a) The density plots and sectional drawings for the RW $\rm\uppercase\expandafter{\romannumeral5}$ $q_1(x,t)$; (b) The density plots and sectional drawings for the RW $\rm\uppercase\expandafter{\romannumeral5}$ $q_2(x,t)$.}
\label{F23}
\end{figure}

\begin{figure}[htbp]
\centering

\begin{minipage}[t]{0.99\textwidth}
\centering
\includegraphics[height=6cm,width=14cm]{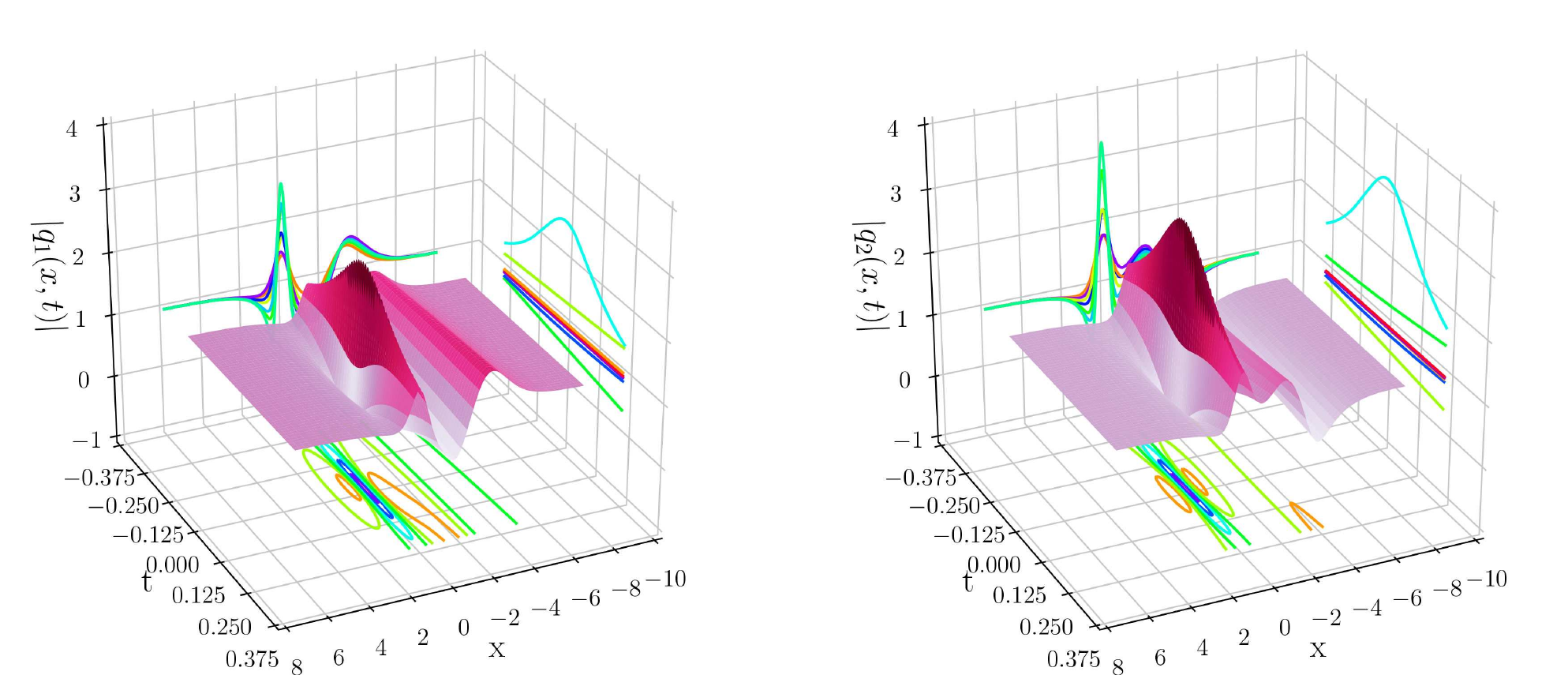}
\end{minipage}
\centering
\caption{(Color online) The three-dimensional plots with contour map on three planes of the predicted vector RWs $\rm\uppercase\expandafter{\romannumeral5}$ $q_r(x,t)$ $(r=1,2)$ based on the IPINN: (Left side panel) The 3D plot for the $q_1(x,t)$; (Right side panel) The 3D plot for the $q_2(x,t)$.}
\label{F24}
\end{figure}

\begin{figure}[htbp]
\centering
\subfigure[]{
\begin{minipage}[t]{0.48\textwidth}
\centering
\includegraphics[height=5cm,width=7cm]{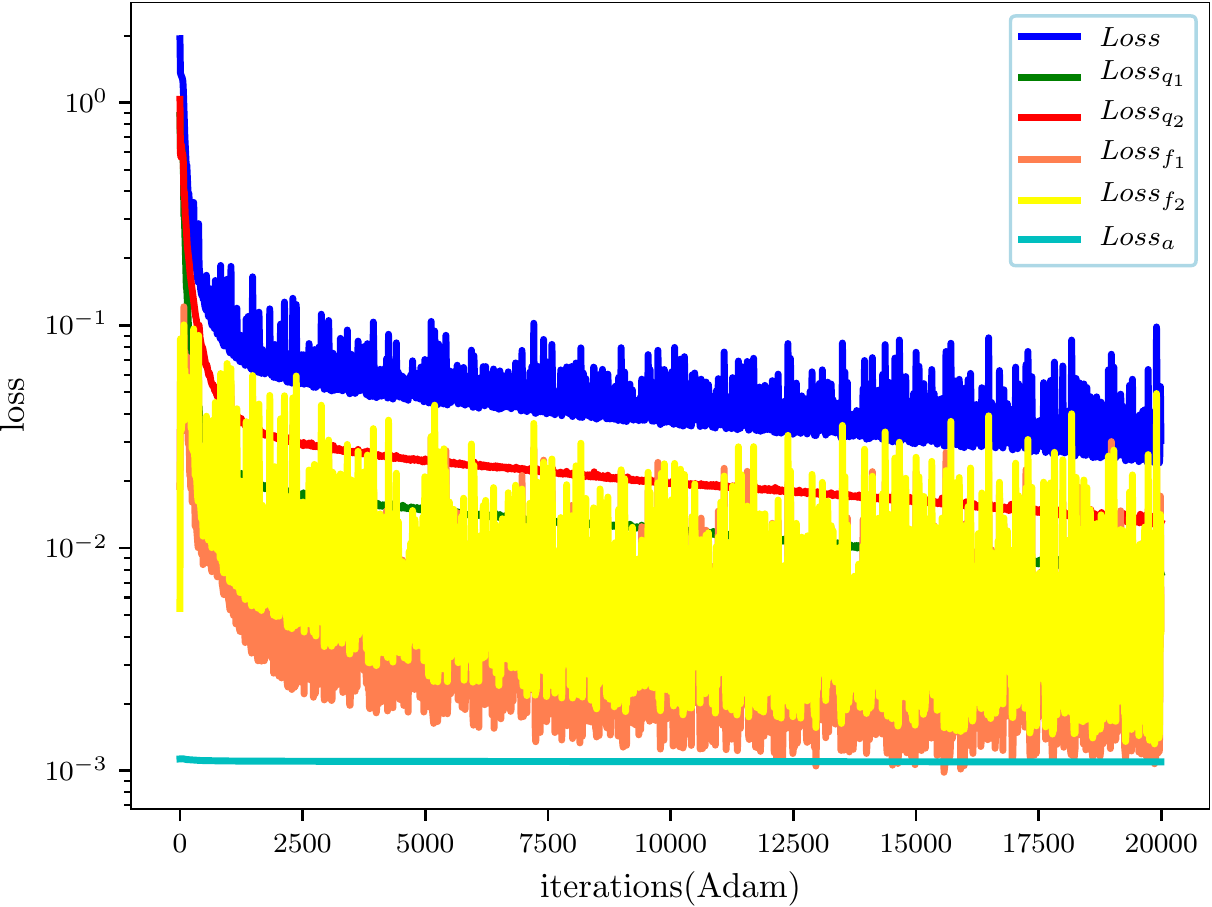}
\end{minipage}
}%
\subfigure[]{
\begin{minipage}[t]{0.48\textwidth}
\centering
\includegraphics[height=5cm,width=7cm]{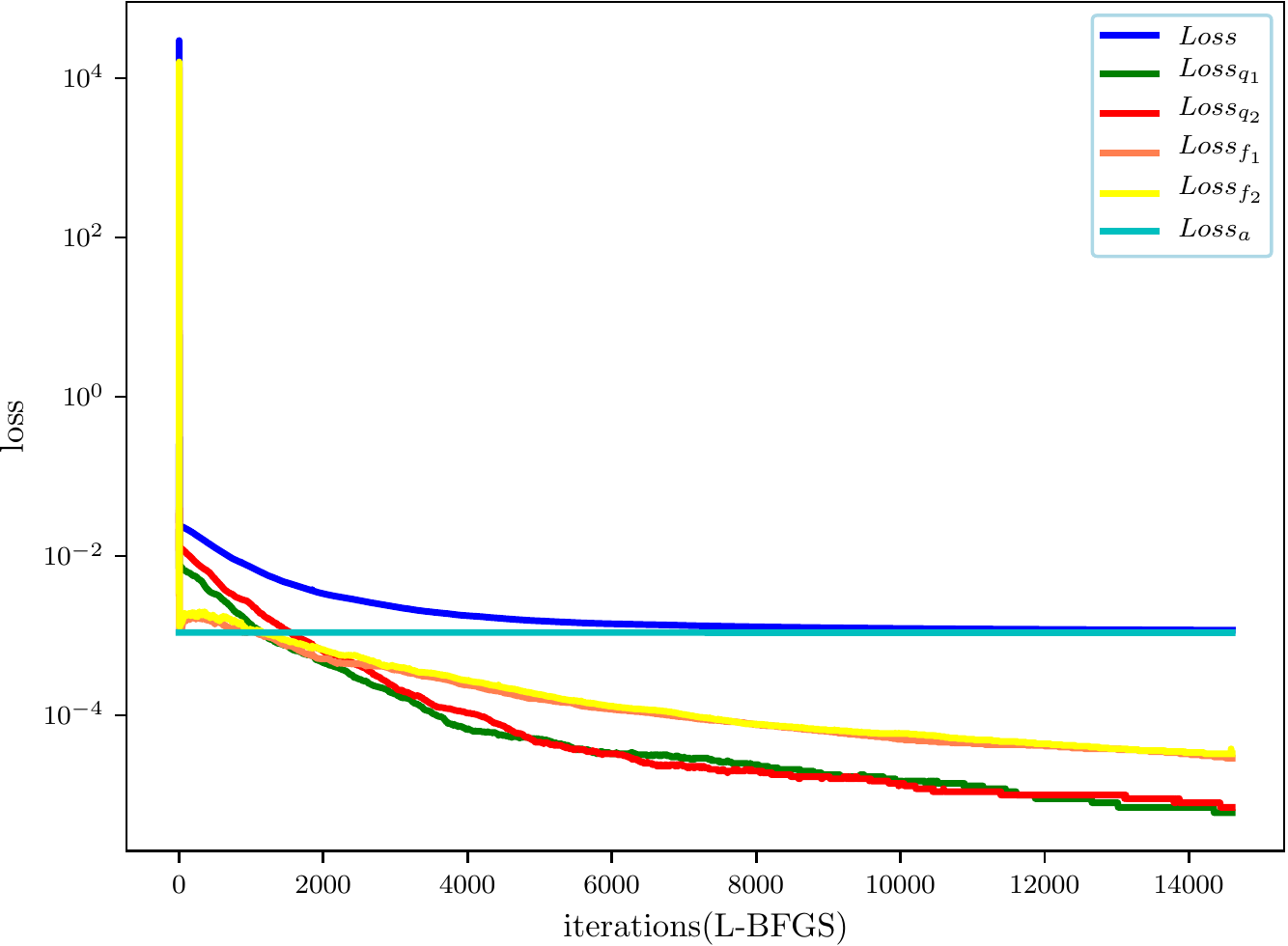}
\end{minipage}%
}%
\centering
\caption{(Color online) The loss function curve figures of the vector RWs $\rm\uppercase\expandafter{\romannumeral5}$ $q_r(x,t)$ $(r=1,2)$ arising from the IPINN with the 20000 steps Adam and 14602 steps L-BFGS optimizations: (a) The loss function curve for the 20000 Adam optimization iterations; (b) The loss function curve for the 14602 L-BFGS optimization iterations.}
\label{F25}
\end{figure}

Figs. \ref{F23} - \ref{F25} display the deep learning results come from the IPINN for the vector RWs $\rm\uppercase\expandafter{\romannumeral5}$ $q_r(x,t)$ $(r=1,2)$ of the Manakov system with the initial-boundary value problem \eqref{E34} and \eqref{E35}. In Fig. \ref{F23}, the exact, learned and error dynamics density plots with corresponding amplitude scale size on the right side have been exhibited, it is worth mentioning that the $N_q=2000$ training data points involved in the initial-boundary condition are marked by mediumorchid symbol $``\times"$ in the learned density plots both in (a) and (b) of Fig. \ref{F23}. Meanwhile, the sectional drawings which include the predictive and exact vector RWs $\rm\uppercase\expandafter{\romannumeral5}$ $q_r(x,t)$ have been shown at the five distinct times pointed out in the exact, learned and error dynamics density plots by using darkturquoise dashed lines in the bottom panel of Fig. \ref{F23}. Fig. \ref{F24} displays the three-dimensional plots with contour map on three planes of the predicted vector RWs $\rm\uppercase\expandafter{\romannumeral5}$ $q_r(x,t)$ based on the IPINN. Fig. \ref{F25} exhibits the loss function curve figures of the vector RWs $\rm\uppercase\expandafter{\romannumeral5}$ $q_r(x,t)$ arising from the IPINN with the 20000 steps Adam and 14602 steps L-BFGS optimizations on the loss function $\mathscr{L}(\bar{\Theta})$.

In this section, we construct a variety of vector RWs of the Manakov system with $\lambda_1=1$ and $\lambda_2=2$ by using the IPINN, in which contain simple vector RWs in subsection 4.1 and complex vector RWs in subsections 4.2 - 4.5. Although we can accurately simulate both simple and complex vector RWs, we can only restore their dynamic behavior for complex vector RWs in a short time interval, and a large number of training experiments show that the IPINN has poor dynamic behavior training effect in a long time region. However, for the vector RWs $\rm\uppercase\expandafter{\romannumeral5}$ formed by the interaction between breathers and RWs, we know that the periodic oscillation can be better displayed in the long-time region. Therefore, how to make the IPINN algorithm train localized waves with good results in the long-time region is an unsolved problem. Finally, we provide a summary of all the aforementioned training results in following Tab. \ref{Tab-RW}.

\begin{table}[htbp]
  \caption{Relative $\mathbb{L}_2$ errors of three different RWs in IPINN model}
  \label{Tab-RW}
  \centering
  \scalebox{0.70}{
  \begin{tabular}{l|c|c|c|c|c}
  \toprule
  \diagbox{\textbf{Component}}{\textbf{RW Types}} & Vector RWs $\rm\uppercase\expandafter{\romannumeral1}$ & Vector RWs $\rm\uppercase\expandafter{\romannumeral2}$ & Vector RWs $\rm\uppercase\expandafter{\romannumeral3}$ & Vector RWs $\rm\uppercase\expandafter{\romannumeral4}$ & Vector RWs $\rm\uppercase\expandafter{\romannumeral5}$\\
  \hline
  Component $q_1$   & 7.505391$\rm e$-03 & 3.575306$\rm e$-03 & 2.197789$\rm e$-03 & 1.637892$\rm e$-03 & 6.325696$\rm e$-03 \\
  \hline
  Component $q_2$   & 7.988676$\rm e$-03 & 3.321024$\rm e$-03 & 3.661877$\rm e$-03 & 1.784348$\rm e$-03 & 7.654479$\rm e$-03 \\
  \bottomrule
  \end{tabular}}
\end{table}

\section{Data-driven parameters discovery of Manakov system}
In this section, we focus on the data-driven parameters discovery of Manakov system with unknown parameters \eqref{E1} by using IPINN model. At this time, the physics-informed part becomes the following form
\begin{align}\label{E-pi2}
\begin{split}
&f_u:=-v_t+\lambda_1u_{xx}+\lambda_2u(u^2+v^2+m^2+n^2),\\
&f_v:=u_t+\lambda_1v_{xx}+\lambda_2v(u^2+v^2+m^2+n^2),\\
&f_m:=-n_t+\lambda_1m_{xx}+\lambda_2m(u^2+v^2+m^2+n^2),\\
&f_n:=m_t+\lambda_1n_{xx}+\lambda_2n(u^2+v^2+m^2+n^2),
\end{split}
\end{align}
in which $\lambda_1$ and $\lambda_2$ are parameters to be learned. The unknown parameters $\lambda_1$ and $\lambda_2$ are initialized to $\lambda_1=\lambda_2=0$ in IPINN. Then we define the relative error of unknown parameters is
\begin{align}\label{E36}
RE=\frac{|\hat{\lambda}_{\alpha}-\lambda_{\alpha}|}{\lambda_{\alpha}}\times100\%,\,\alpha=1, 2,
\end{align}
where the $\hat{\lambda}_{\alpha}$ and $\lambda_{\alpha}$ represent predicted value and true value, respectively. All noise interference in this section is added to the randomly chosen small data set, the specific form is as follows
\begin{align}\label{E37}
\begin{split}
Data_{-}train = &Data_{-}train + noise*np.std(Data_{-}train)*np.random.randn\\
&(Data_{-}train.shape[0], Data_{-}train.shape[1]),
\end{split}
\end{align}
where $noise$ and $Data_{-}train$ indicate the noise intensity and a small randomly chosen training data set, respectively. The $np.std(\cdot)$ returns the standard deviation of an array element, and $np.random.randn(\cdot,\cdot)$ returns a set of samples with a standard normal distribution.

For learning the parameters $\lambda_1$ and $\lambda_2$ in Eq. \eqref{E1} with the aid of the IPINN with neuron-wise locally adaptive activation function, and considering the initial conditions and Dirichlet boundary conditions of Eq. \eqref{E1} arising from the vector RWs $\rm\uppercase\expandafter{\romannumeral1}$ \eqref{E21} by using the 9-layer IPINN with 40 neurons per layer, we set the spatiotemporal region $(x,t)\in[-3,6]\times[-0.5,0.5]$. Thus the corresponding initial conditions can be written as following
\begin{align}\label{E38}
\begin{split}
&q_{r}^0(x)=q_{r,\mathrm{rw1}}(x,-0.5),\,x\in[-3.0,6.0],
\end{split}
\end{align}
and the Dirichlet boundary conditions become
\begin{align}\label{E39}
q_{r}^{\mathrm{lb}}(t)=q_{r,\mathrm{rw1}}(-3.0,t),\,q_{r}^{\mathrm{ub}}(t)=q_{r,\mathrm{rw1}}(6.0,t),\,t\in[-0.5,0.5],\,r=1,2.
\end{align}

Then we use the same data discretization method in section 4, and generate the original data set by dividing the spatial region $[-3.0,6.0]$ into 2000 points and temporal region $[-0.5,0.5]$ into 1500 points. We obtain a smaller training dataset that containing initial-boundary data \eqref{E38} and \eqref{E39} by randomly extracting $N_q=3000$ from original dataset and $N_f=30000$ collocation points which are yielded by the LHS method. After that, the latent data-driven unknown parameters $\lambda_1$ and $\lambda_2$ have been successfully learned by tuning all learnable parameters of the IPINN and utilizing Adam iterations and L-BFGS iterations to regulate the loss function $\mathscr{L}(\bar{\Theta})$.

\begin{figure}[htbp]
\centering
\subfigure[]{
\begin{minipage}[t]{0.48\textwidth}
\centering
\includegraphics[height=4.5cm,width=6.5cm]{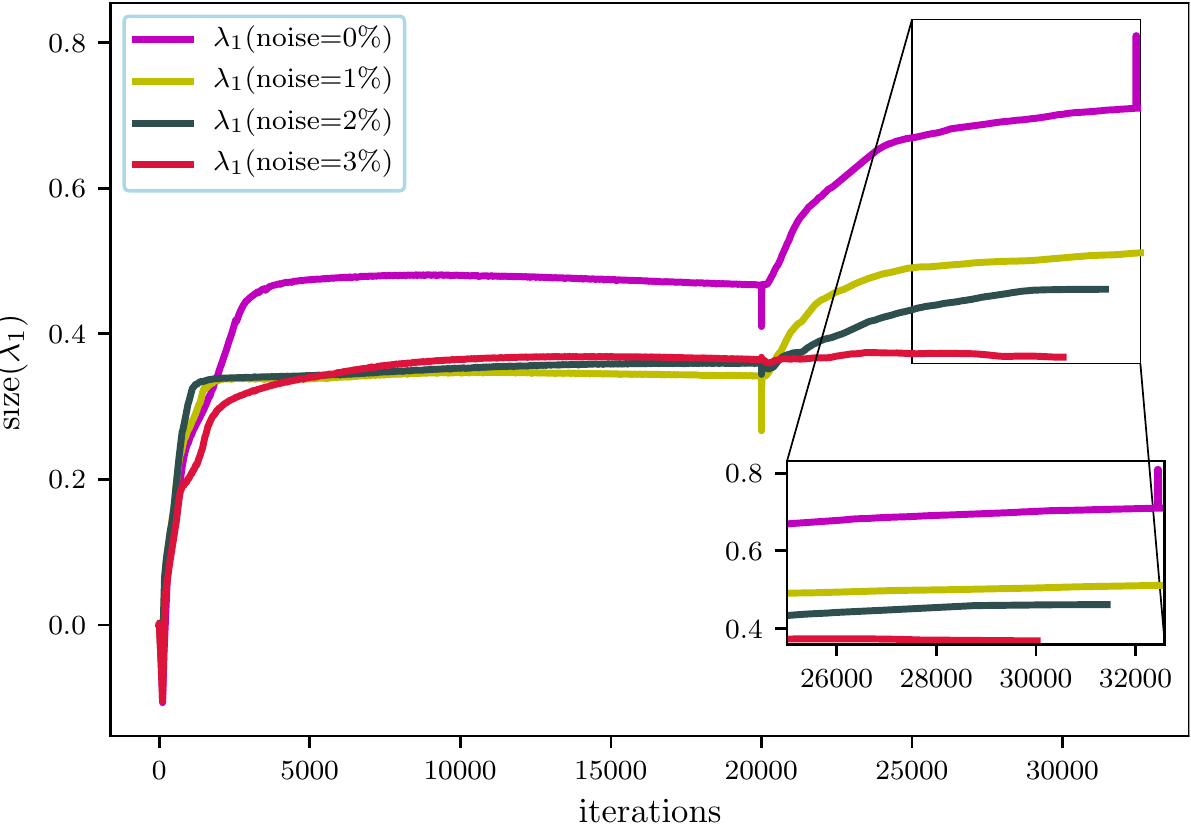}
\end{minipage}
}%
\subfigure[]{
\begin{minipage}[t]{0.48\textwidth}
\centering
\includegraphics[height=4.5cm,width=6.5cm]{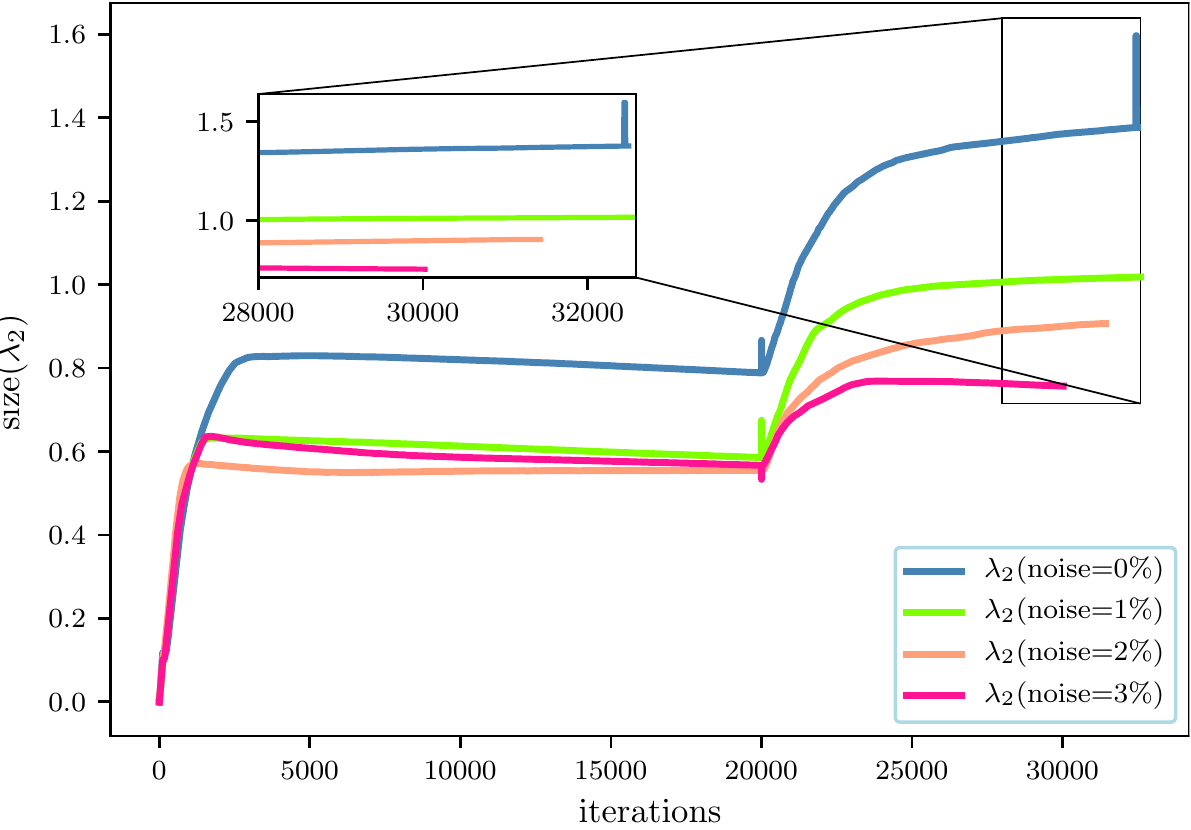}
\end{minipage}%
}\\%
\subfigure[]{
\begin{minipage}[t]{0.48\textwidth}
\centering
\includegraphics[height=4.5cm,width=6.5cm]{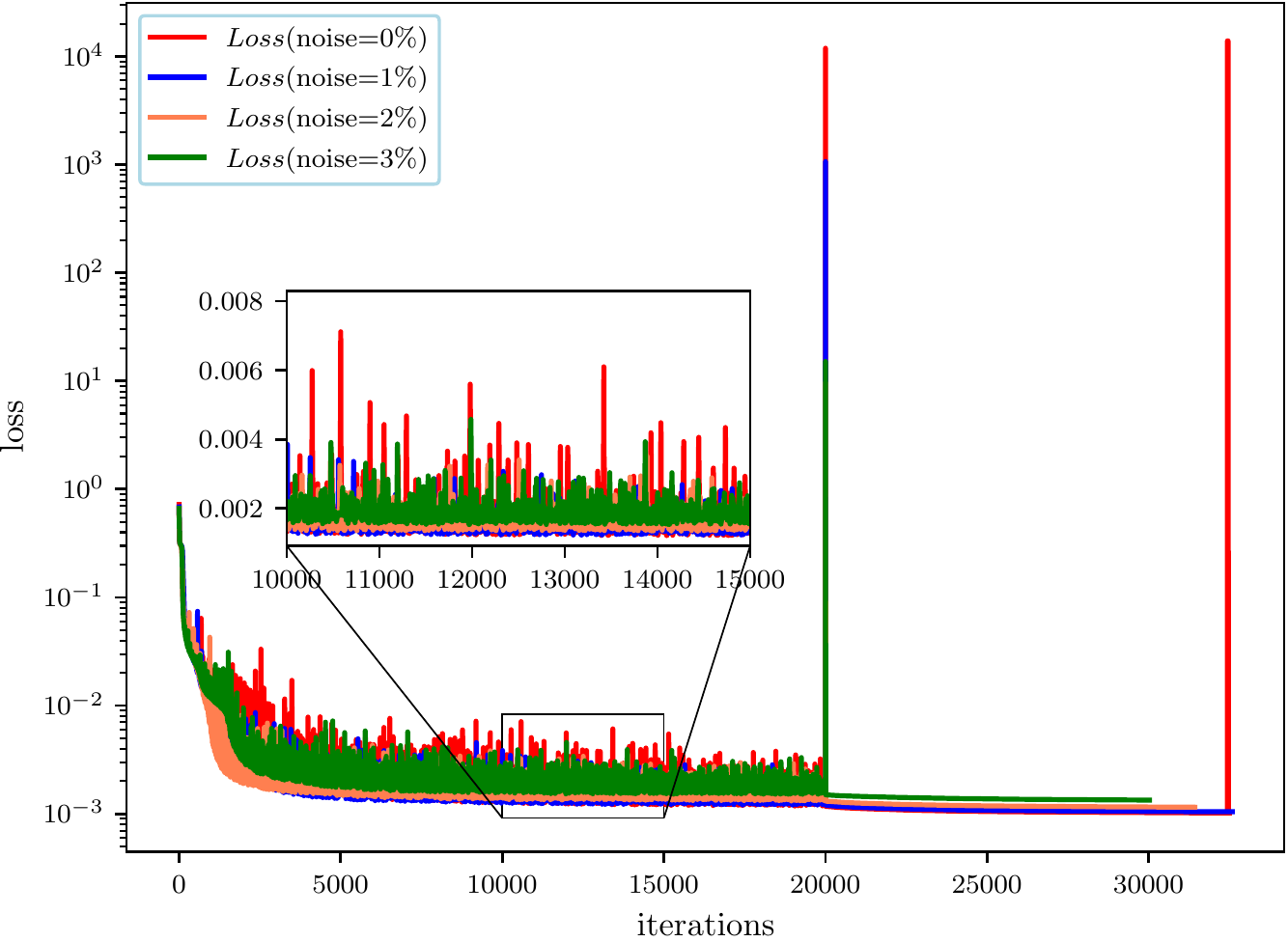}
\end{minipage}
}%
\subfigure[]{
\begin{minipage}[t]{0.48\textwidth}
\centering
\includegraphics[height=4.5cm,width=6.5cm]{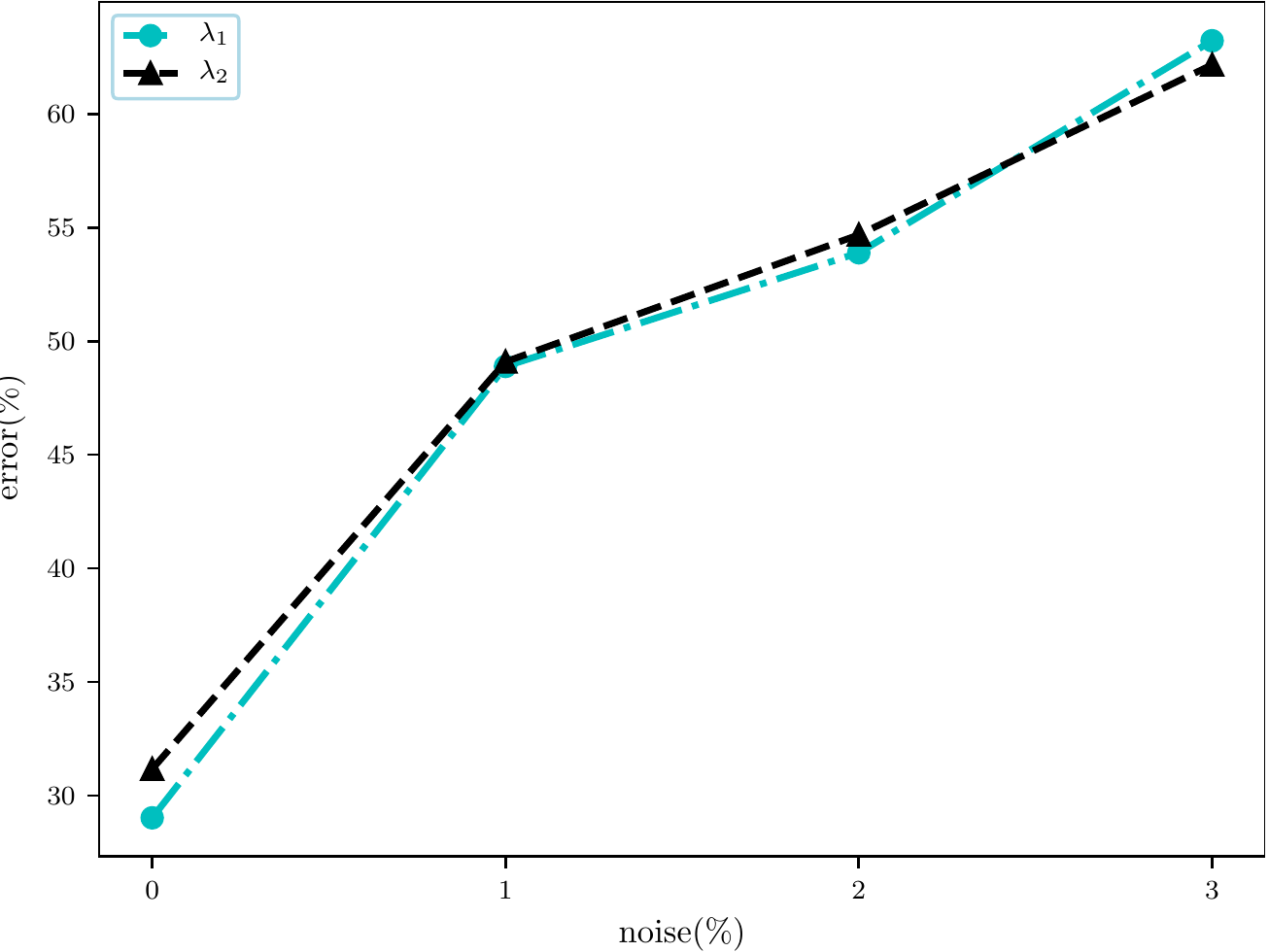}
\end{minipage}%
}%
\centering
\caption{(Color online) The training results of parameter discover by means of the IPINN with 20000 Adam iterations and various number of L-BFGS iterations: (a)-(b) the variation curves of unknown coefficients $\lambda_1$ and $\lambda_2$ with different noise intensity; (c) the variation curve of loss function with different noise intensity; (d) unknown coefficients $\lambda_1$ and $\lambda_2$ error variation plots under different interference noise.}
\label{F26}
\end{figure}

Fig. \ref{F26} displays the training results of unknown parameters under the above initial boundary value conditions by using the IPINN with 20000 Adam iterations and diverse number of L-BFGS iterations. Fig. \ref{F26} (a)-(b) exhibit the variation curves of unknown coefficients $\lambda_1$ and $\lambda_2$ with different noise intensity. The panel (c) of Fig. \ref{F26} shows the variation curve of loss function with different noise intensity. While the error variation plots of unknown coefficients $\lambda_1$ and $\lambda_2$  under different interference noise are revealed in Fig. \ref{F26} (d). However, we find that the learning effect of unknown parameters is not ideal, from training result by using clean initial-boundary data (noise=$0\%$) in IPINN, the relative error of $\lambda_1$ reaches about $15\%$, while the relative error of $\lambda_2$ exceeds $30\%$ from Fig. \ref{F26}. Once the noise intensity is increased, the relative error also increases in direct proportion from Fig. \ref{F26} (d). Thus we urgently need to find IPINN algorithm which can accurately and effectively learn unknown parameters. Fortunately, we find that the PINN with with neuron-wise locally adaptive activation function and $L^2$ norm parameter regularization shows amazing effect in studying the inverse problem of Yajima-Oikawa system \cite{PuYO2021}. Therefore, we introduce $L^2$ norm parameter regularization into IPINN to study the parameter discovery problem of Manakov system.

From Ref. \cite{PuYO2021}, we construct a new loss function $\widetilde{\mathscr{L}}(\bar{\Theta})$ with $L^2$ norm weight decay, as following
\begin{align}\label{E40}
&\widetilde{\mathscr{L}}(\bar{\Theta})=Loss_{PR}=\mathscr{L}(\bar{\Theta})+\frac{\beta}{2}\|\textbf{W}\|^2_2,
\end{align}
where $\mathscr{L}(\bar{\Theta})$ and $\textbf{W}$ have been defined in Eq. \eqref{E6} and Eq. \eqref{E-buchong}.

\begin{figure}[htbp]
\centering
\subfigure[]{
\begin{minipage}[t]{0.48\textwidth}
\centering
\includegraphics[height=4.5cm,width=6.5cm]{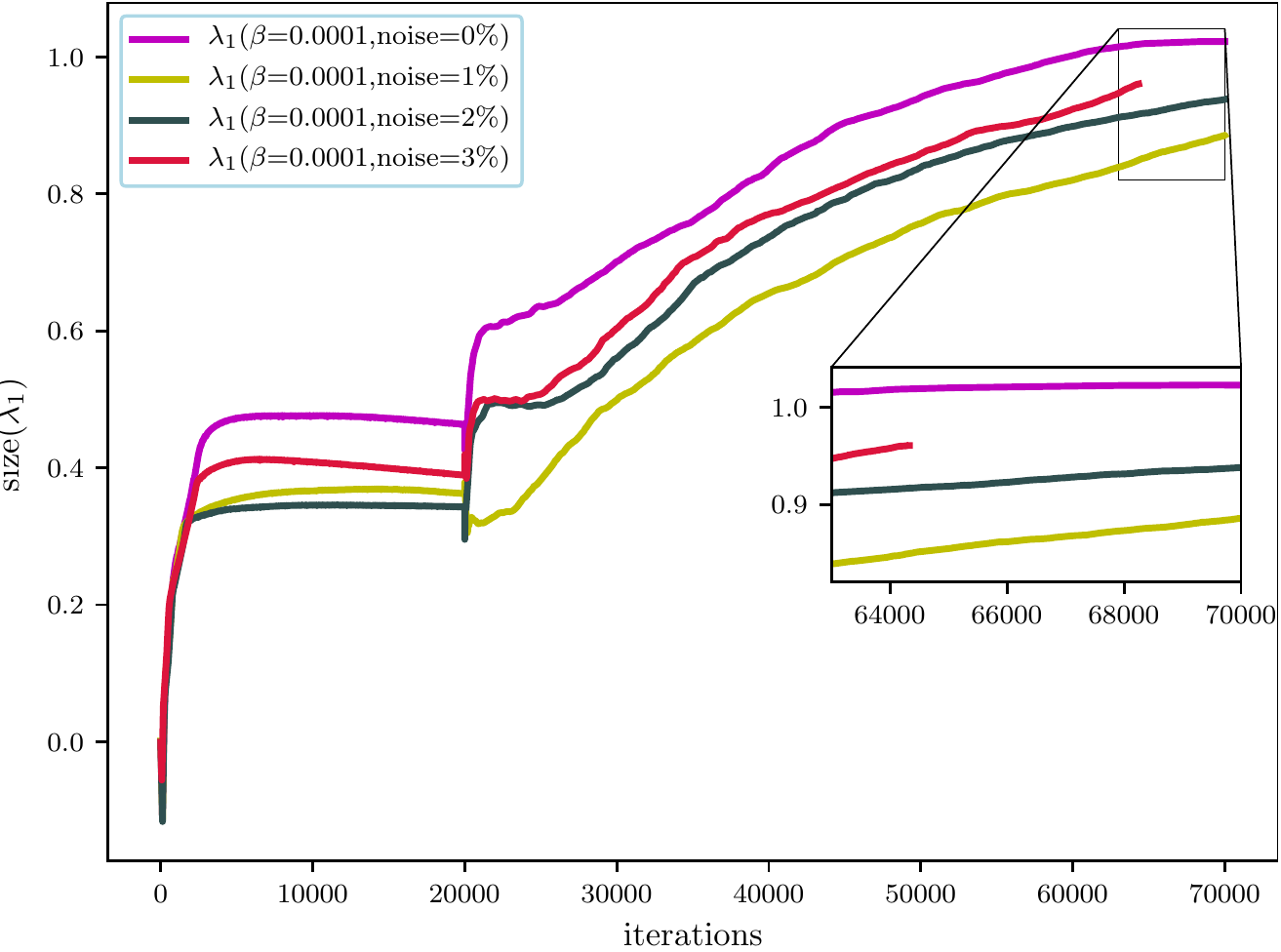}
\end{minipage}
}%
\subfigure[]{
\begin{minipage}[t]{0.48\textwidth}
\centering
\includegraphics[height=4.5cm,width=6.5cm]{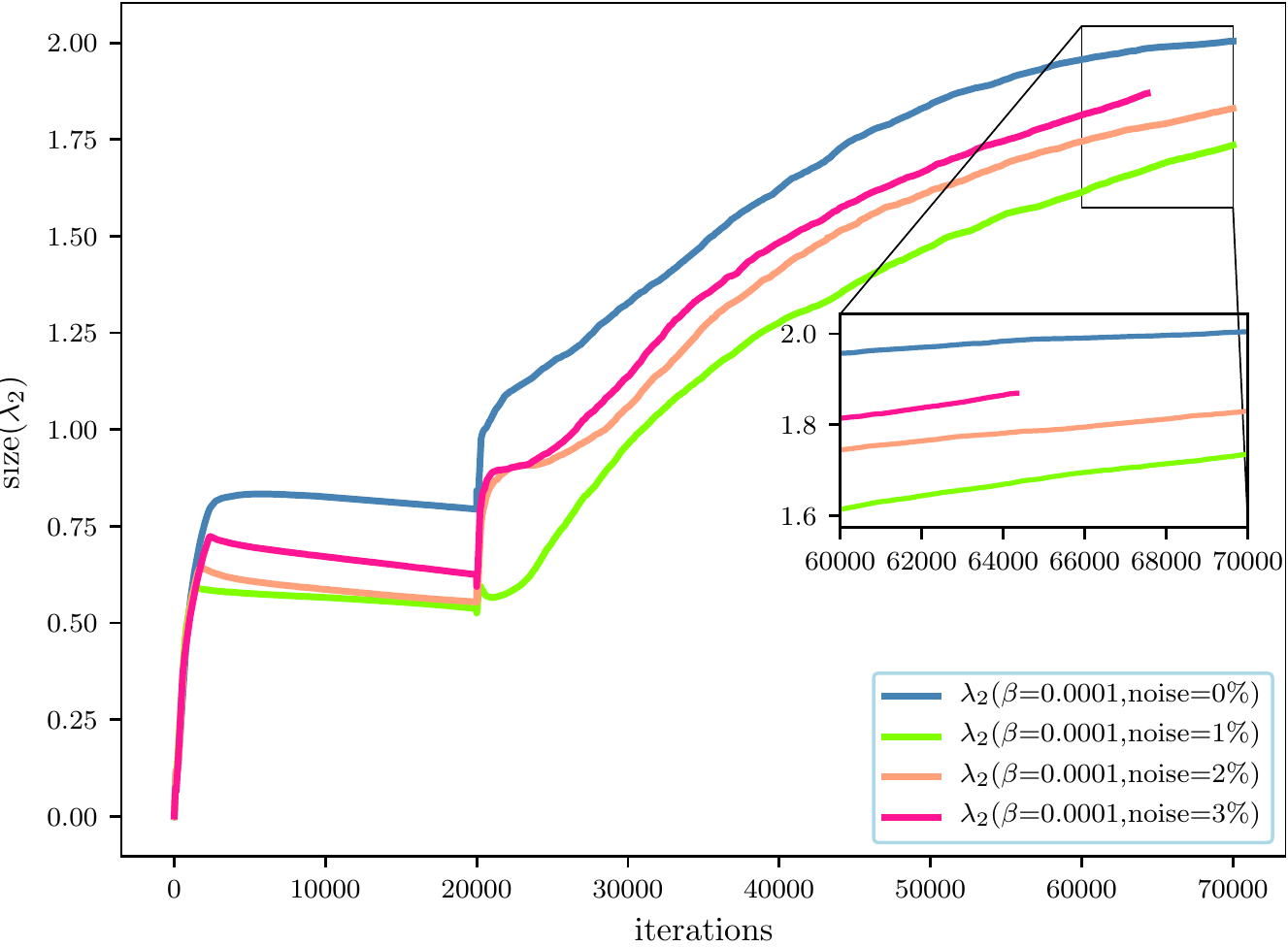}
\end{minipage}%
}\\%
\subfigure[]{
\begin{minipage}[t]{0.48\textwidth}
\centering
\includegraphics[height=4.5cm,width=6.5cm]{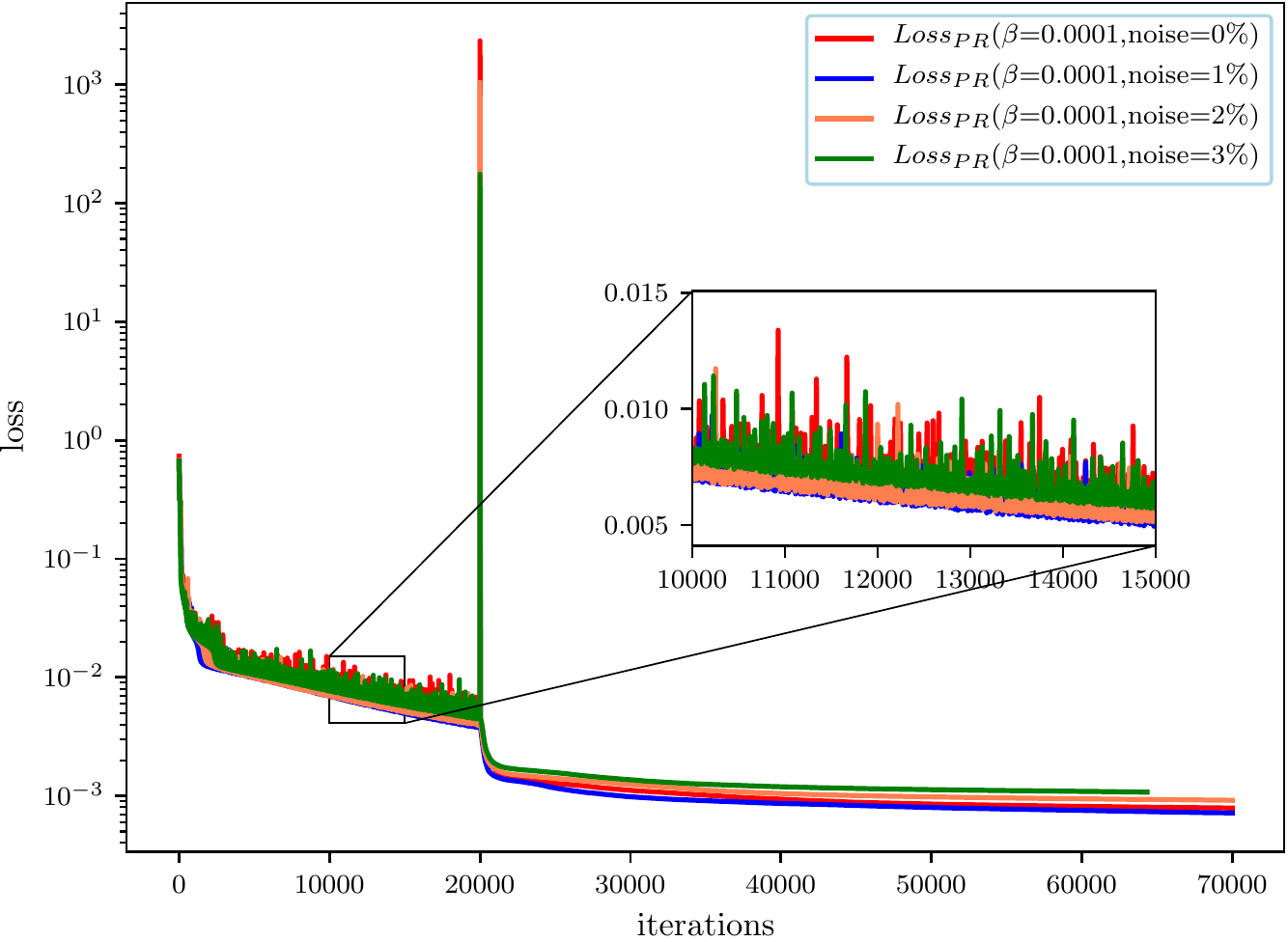}
\end{minipage}
}%
\subfigure[]{
\begin{minipage}[t]{0.48\textwidth}
\centering
\includegraphics[height=4.5cm,width=6.5cm]{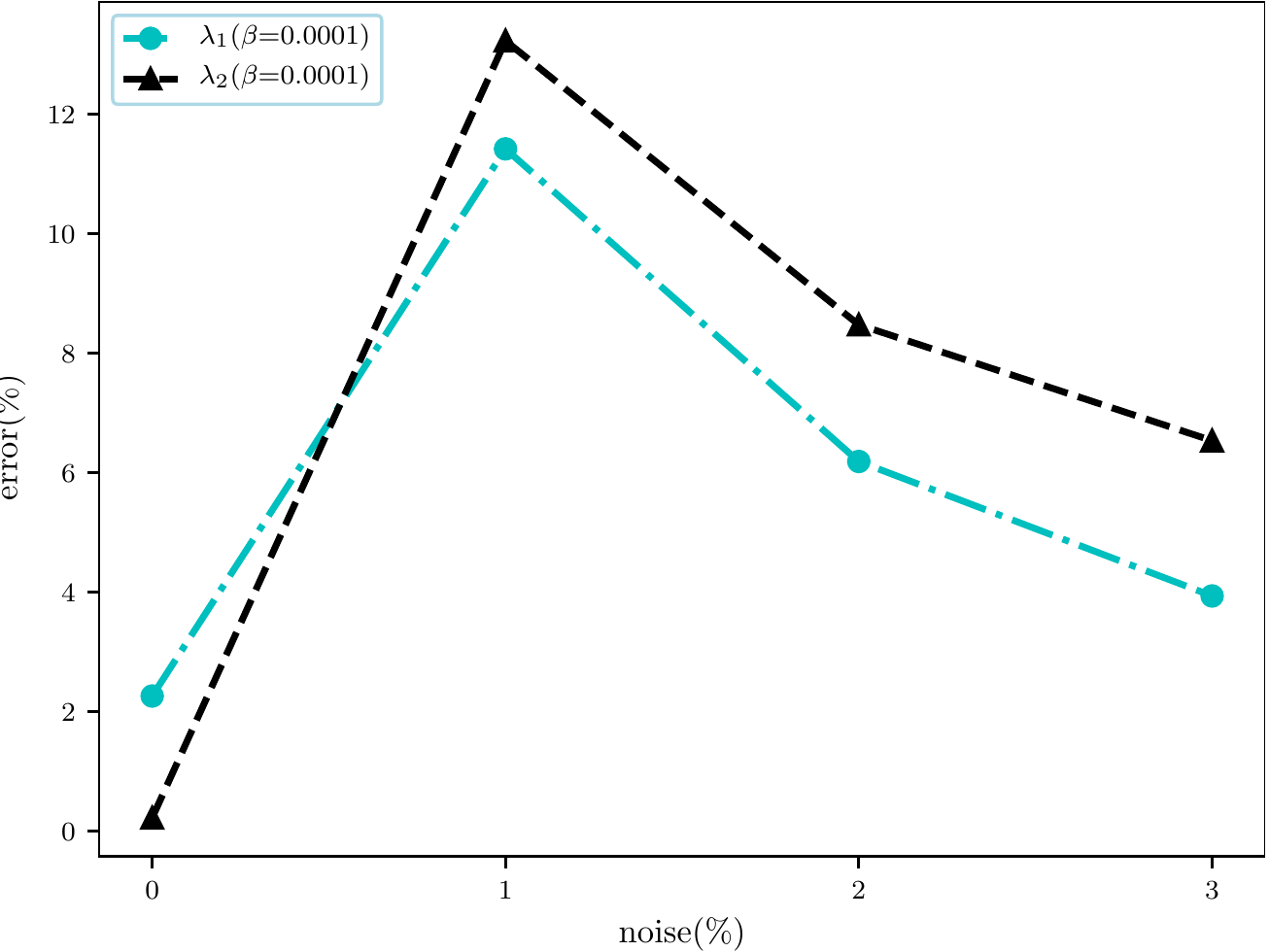}
\end{minipage}%
}%
\centering
\caption{(Color online) The training results of parameter discover by means of the IPINN with $\beta=0.0001$ weight decay: (a)-(b) the variation curves of unknown coefficients $\lambda_1$ and $\lambda_2$ with different noise intensity; (c) the variation curve of loss function with different noise intensity; (d) unknown coefficients $\lambda_1$ and $\lambda_2$ error variation plots under different interference noise.}
\label{F27}
\end{figure}

Next, we utilize the same initial-boundary value data points as IPINN without parametric regularization, after 20000 Adam iterations and various number of L-BFGS iterations, then corresponding numerical results are exhibited in Fig. \ref{F27} by using IPINN with $\beta=0.0001$ weight decay. Fig. \ref{F27} (a) and (b) display the variation curves of unknown coefficients $\lambda_1$ and $\lambda_2$ with different noise intensity, Fig. \ref{F27} (c) depicts the variation curve of loss function with different noise intensity, and the noise intensity and relative error plots are shown in Fig. \ref{F27} (d). The specific numerical results show that $\lambda_1$ obtains the minimum relative error of 2.263904$\%$ as the noise intensity is $0\%$, and maximum relative error 11.419546$\%$ as the noise intensity is 1$\%$, whereas $\lambda_2$ obtains the minimum relative error of 0.225925$\%$ as the noise intensity is 0$\%$, and maximum relative error 13.228714$\%$ as the noise intensity is 1$\%$. Compared with the numerical results obtained by IPINN without parameter regularization in Fig. \ref{F26}, the relative error of parameters learned by IPINN with $L^2$ norm parameter regularization is small as a whole, especially when the noise intensity is 0$\%$, the parameters training effect is the best. In other words, the training effect has been very good in the case of clean data. However, when using the data with noise interference for training, the relative error of the prediction parameters is larger than that of the training results of clean data. Therefore, we need to reset the weight decay coefficient to improve the training effect of using the data with noise interference.

\begin{figure}[htbp]
\centering
\subfigure[]{
\begin{minipage}[t]{0.48\textwidth}
\centering
\includegraphics[height=4.5cm,width=6.5cm]{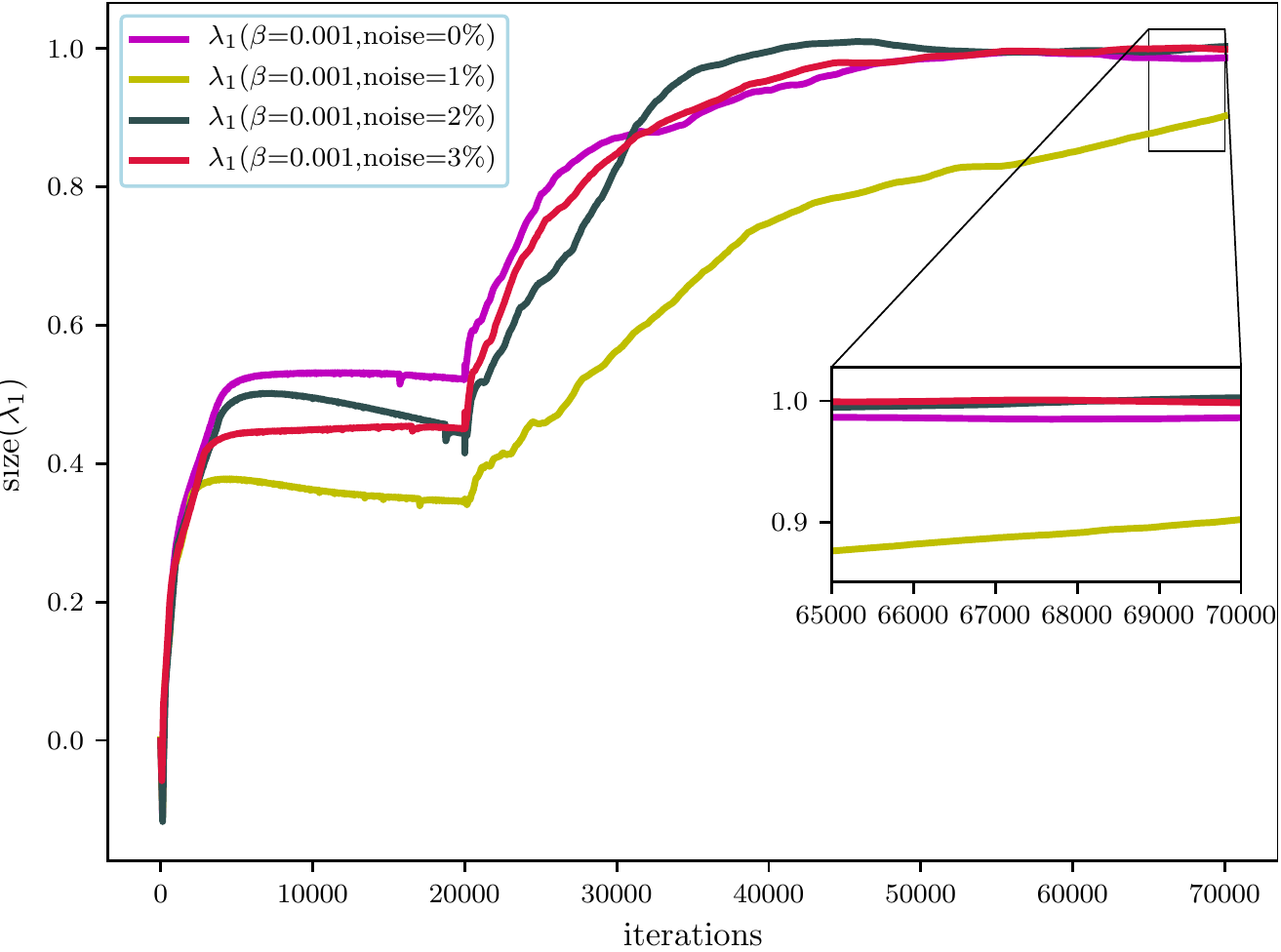}
\end{minipage}
}%
\subfigure[]{
\begin{minipage}[t]{0.48\textwidth}
\centering
\includegraphics[height=4.5cm,width=6.5cm]{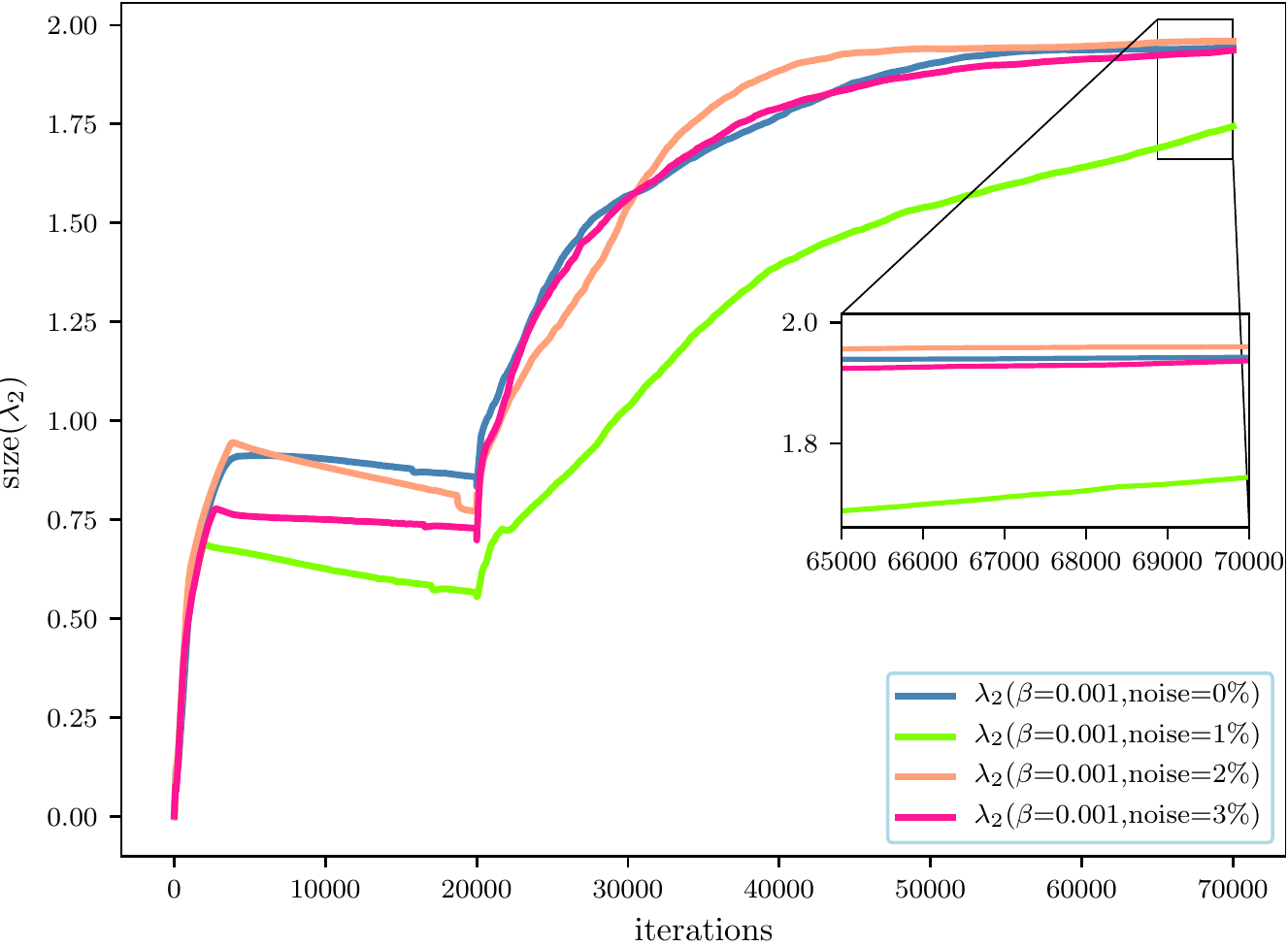}
\end{minipage}%
}\\%
\subfigure[]{
\begin{minipage}[t]{0.48\textwidth}
\centering
\includegraphics[height=4.5cm,width=6.5cm]{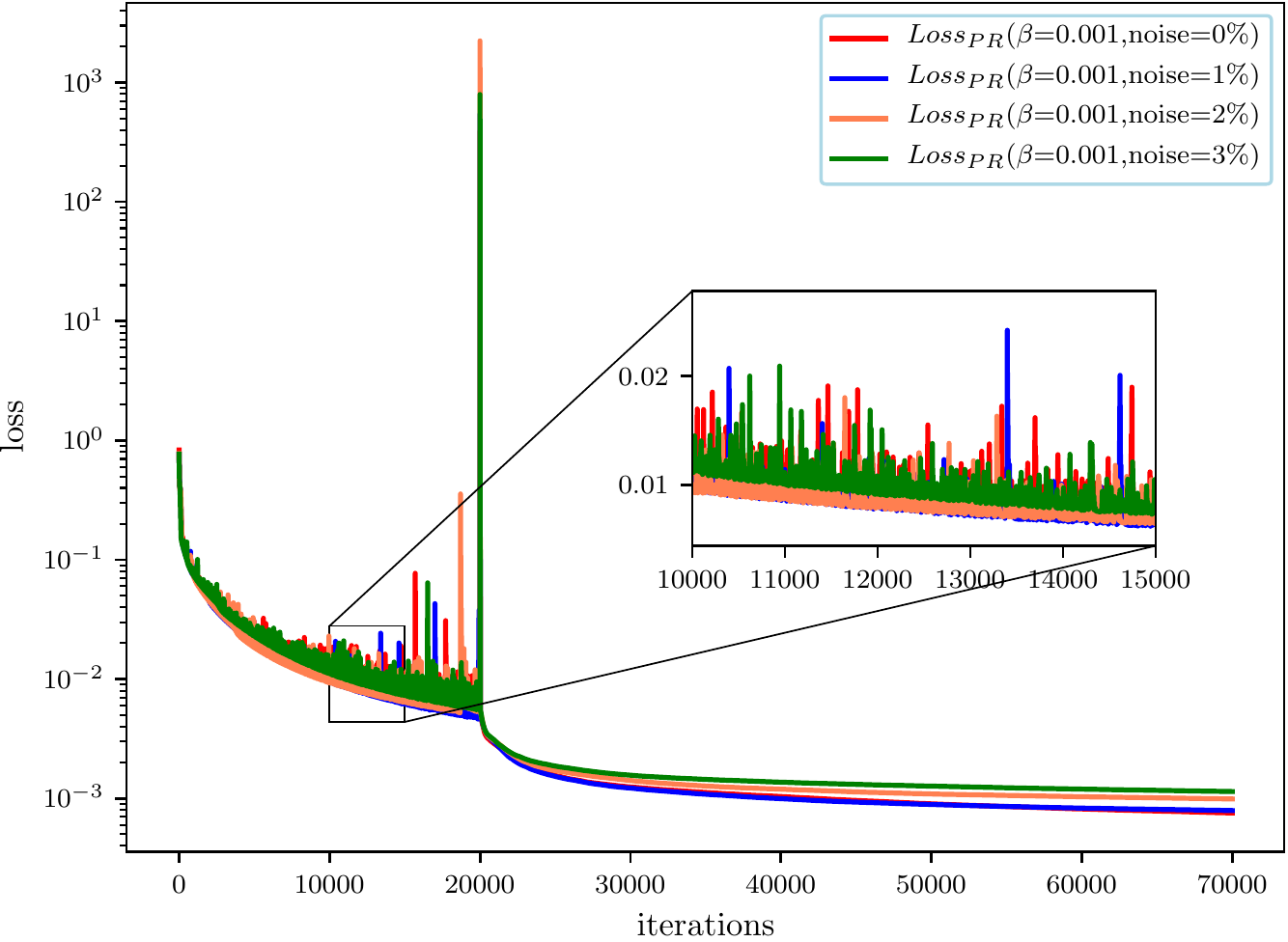}
\end{minipage}
}%
\subfigure[]{
\begin{minipage}[t]{0.48\textwidth}
\centering
\includegraphics[height=4.5cm,width=6.5cm]{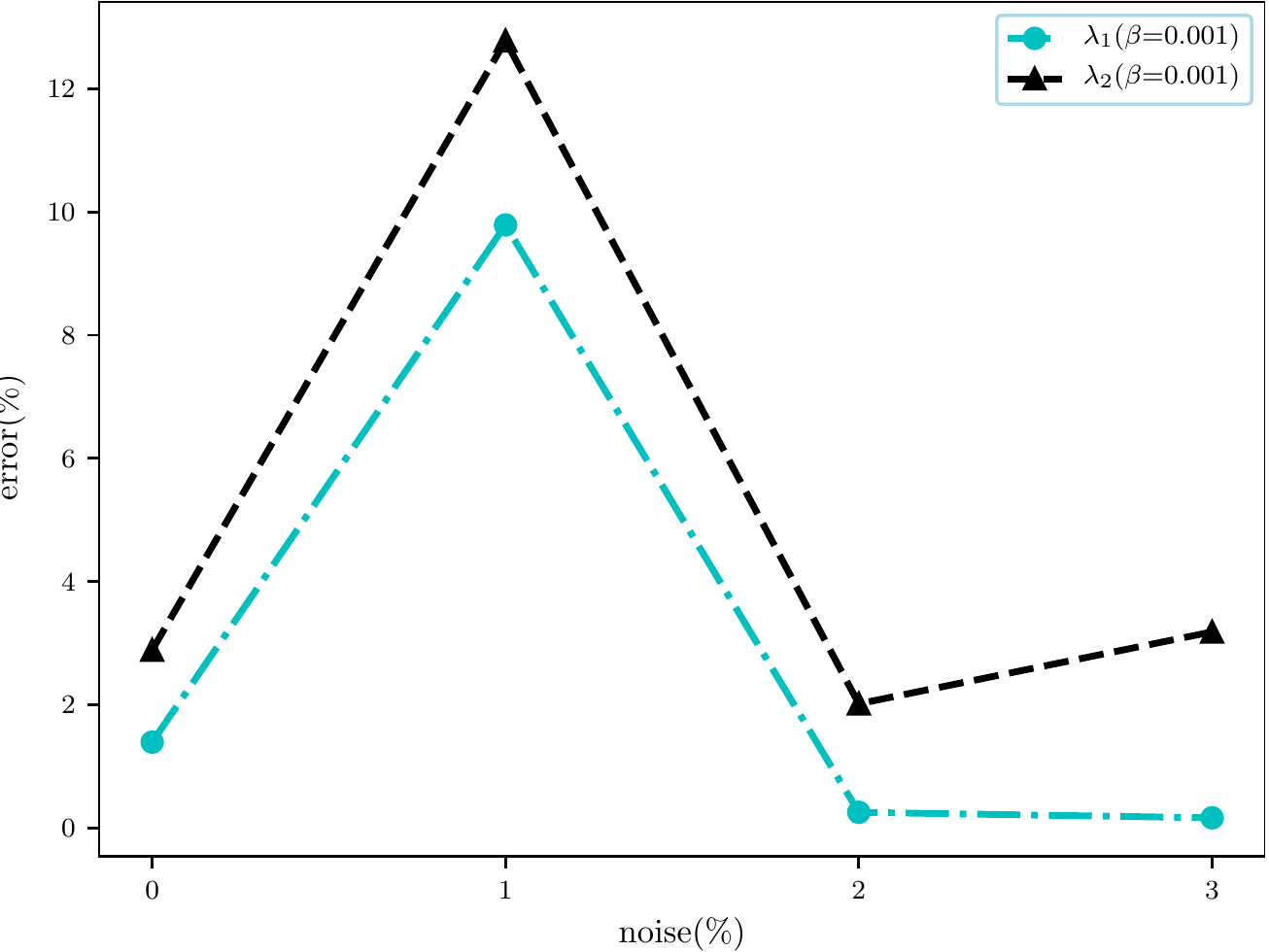}
\end{minipage}%
}%
\centering
\caption{(Color online) The training results of parameter discover by means of the IPINN with $\beta=0.001$ weight decay: (a)-(b) the variation curves of unknown coefficients $\lambda_1$ and $\lambda_2$ with different noise intensity; (c) the variation curve of loss function with different noise intensity; (d) unknown coefficients $\lambda_1$ and $\lambda_2$ error variation plots under different interference noise.}
\label{F28}
\end{figure}

Now we attempt to increase the weight decay coefficient by an order of magnitude to further obtain a better training effect in the case of noise data, so we set $\beta=0.001$ in IPINN with $L^2$ norm penalty, and employ same initial-boundary value data points as IPINN without parametric regularization, then corresponding training results are depicted in Fig. \ref{F28} by using 20000 Adam iterations and various number of L-BFGS iterations in IPINN with parametric regularization. From Fig. \ref{F28}, one can find that the training effect with data of various noise is much better than that with clean data, except that the training effect is poor when the noise intensity is 1$\%$. Although the training effect is not ideal when using individual noise intensity data, the generalization training shows that IPINN with coefficient $\beta=0.001$ of weight decay has excellent noise resistance. Of course, we can also adjust coefficient of the weight decay to adapt to the various problems at hand.

So to summarise, In this section, the parameter discovery problem of Manakov system is studied firstly through IPINN, and it is found that the training effect is not good when using clean data and noisy data. Therefore, we introduce the parameter regularization strategy into IPINN, and study the parameters discovery problem of Manakov system by means of IPINN with two different parameter regularization coefficients. Specifically, we find that when the weight decay coefficient $\beta=0.0001$, The training effect is excellent when using clean data, while the numerical results are not good when using noisy data. However, when the weight decay coefficient $\beta=0.001$, not only the training effect using noisy data is better than that using clean data, but also the relative error of parameters is small as a whole. Finally, we provide a summary of all the aforementioned training results in following Tab. \ref{Tab-PR}.
\begin{table}[htbp]
  \caption{Comparison of correct Manakov system and identified Manakov system obtained by means of the IPINN with different noise intensities and weight hyper-parameters $\beta$.}
  \label{Tab-PR}
  \centering
  \scalebox{0.45}{
  \begin{tabular}{l|c|c|c}
  \toprule
  \diagbox{\scriptsize{\textbf{Manakov system}}}{\scriptsize{\textbf{hyper-parameters}}} & $\beta=0$ & $\beta=0.0001$ & $\beta=0.001$\\
  \hline
  \scriptsize{Correct Manakov system}   & \makecell[c]{$\mathrm{i}q_{1t}+q_{1xx}+2(|q_1|^2+|q_2|^2)q_1=0$ \\ $\mathrm{i}q_{2t}+q_{2xx}+2(|q_1|^2+|q_2|^2)q_2=0$ \\ $\lambda_1$ error: 0$\%$ \\ $\lambda_2$ error: 0$\%$} & \makecell[c]{$\mathrm{i}q_{1t}+q_{1xx}+2(|q_1|^2+|q_2|^2)q_1=0$ \\ $\mathrm{i}q_{2t}+q_{2xx}+2(|q_1|^2+|q_2|^2)q_2=0$ \\ $\lambda_1$ error: 0$\%$ \\ $\lambda_2$ error: 0$\%$} & \makecell[c]{$\mathrm{i}q_{1t}+q_{1xx}+2(|q_1|^2+|q_2|^2)q_1=0$ \\ $\mathrm{i}q_{2t}+q_{2xx}+2(|q_1|^2+|q_2|^2)q_2=0$ \\ $\lambda_1$ error: 0$\%$ \\ $\lambda_2$ error: 0$\%$}\\
  \hline
  \scriptsize{Identified Manakov system (clean data)}   & \makecell[c]{$\mathrm{i}q_{1t}+0.709768q_{1xx}+1.376634(|q_1|^2+|q_2|^2)q_1=0$ \\ $\mathrm{i}q_{2t}+0.709768q_{2xx}+1.376634(|q_1|^2+|q_2|^2)q_2=0$ \\ $\lambda_1$ error: 29.023170$\%$ \\ $\lambda_2$ error: 31.168324$\%$} & \makecell[c]{$\mathrm{i}q_{1t}+1.022639q_{1xx}+2.004519(|q_1|^2+|q_2|^2)q_1=0$ \\ $\mathrm{i}q_{2t}+1.022639q_{2xx}+2.004519(|q_1|^2+|q_2|^2)q_2=0$ \\ $\lambda_1$ error: 2.263904$\%$ \\ $\lambda_2$ error: 0.225925$\%$} & \makecell[c]{$\mathrm{i}q_{1t}+0.986093q_{1xx}+1.942293(|q_1|^2+|q_2|^2)q_1=0$ \\ $\mathrm{i}q_{2t}+0.986093q_{2xx}+1.942293(|q_1|^2+|q_2|^2)q_2=0$ \\ $\lambda_1$ error: 1.390725$\%$ \\ $\lambda_2$ error: 2.885365$\%$}\\
  \hline
  \scriptsize{Identified Manakov system (1$\%$ noise)}   & \makecell[c]{$\mathrm{i}q_{1t}+0.511133q_{1xx}+1.018215(|q_1|^2+|q_2|^2)q_1=0$ \\ $\mathrm{i}q_{2t}+0.511133q_{2xx}+1.018215(|q_1|^2+|q_2|^2)q_2=0$ \\ $\lambda_1$ error: 48.886692$\%$ \\ $\lambda_2$ error: 49.089260$\%$} & \makecell[c]{$\mathrm{i}q_{1t}+0.885805q_{1xx}+1.735426(|q_1|^2+|q_2|^2)q_1=0$ \\ $\mathrm{i}q_{2t}+0.885805q_{2xx}+1.735426(|q_1|^2+|q_2|^2)q_2=0$ \\ $\lambda_1$ error: 11.419546$\%$ \\ $\lambda_2$ error: 13.228714$\%$} & \makecell[c]{$\mathrm{i}q_{1t}+0.902120q_{1xx}+1.744415(|q_1|^2+|q_2|^2)q_1=0$ \\ $\mathrm{i}q_{2t}+0.902120q_{2xx}+1.744415(|q_1|^2+|q_2|^2)q_2=0$ \\ $\lambda_1$ error: 9.788042$\%$ \\ $\lambda_2$ error: 12.779254$\%$}\\
  \hline
  \scriptsize{Identified Manakov system (2$\%$ noise)}   & \makecell[c]{$\mathrm{i}q_{1t}+0.461134q_{1xx}+0.906474(|q_1|^2+|q_2|^2)q_1=0$ \\ $\mathrm{i}q_{2t}+0.461134q_{2xx}+0.906474(|q_1|^2+|q_2|^2)q_2=0$ \\ $\lambda_1$ error: 53.886570$\%$ \\ $\lambda_2$ error: 54.676281$\%$} & \makecell[c]{$\mathrm{i}q_{1t}+0.938114q_{1xx}+1.830483(|q_1|^2+|q_2|^2)q_1=0$ \\ $\mathrm{i}q_{2t}+0.938114q_{2xx}+1.830483(|q_1|^2+|q_2|^2)q_2=0$ \\ $\lambda_1$ error: 6.188589$\%$ \\ $\lambda_2$ error: 8.475840$\%$} & \makecell[c]{$\mathrm{i}q_{1t}+1.002510q_{1xx}+1.959728(|q_1|^2+|q_2|^2)q_1=0$ \\ $\mathrm{i}q_{2t}+1.002510q_{2xx}+1.959728(|q_1|^2+|q_2|^2)q_2=0$ \\ $\lambda_1$ error: 0.251031$\%$ \\ $\lambda_2$ error: 2.013606$\%$}\\
  \hline
  \scriptsize{Identified Manakov system (3$\%$ noise)}   & \makecell[c]{$\mathrm{i}q_{1t}+0.367840q_{1xx}+0.756801(|q_1|^2+|q_2|^2)q_1=0$ \\ $\mathrm{i}q_{2t}+0.367840q_{2xx}+0.756801(|q_1|^2+|q_2|^2)q_2=0$ \\ $\lambda_1$ error: 63.216011$\%$ \\ $\lambda_2$ error: 62.159954$\%$} & \makecell[c]{$\mathrm{i}q_{1t}+0.960615q_{1xx}+1.869262(|q_1|^2+|q_2|^2)q_1=0$ \\ $\mathrm{i}q_{2t}+0.960615q_{2xx}+1.869262(|q_1|^2+|q_2|^2)q_2=0$ \\ $\lambda_1$ error: 3.938526$\%$ \\ $\lambda_2$ error: 6.536925$\%$} & \makecell[c]{$\mathrm{i}q_{1t}+0.998383q_{1xx}+1.936382(|q_1|^2+|q_2|^2)q_1=0$ \\ $\mathrm{i}q_{2t}+0.998383q_{2xx}+1.936382(|q_1|^2+|q_2|^2)q_2=0$ \\ $\lambda_1$ error: 0.161713$\%$ \\ $\lambda_2$ error: 3.180903$\%$}\\
  \bottomrule
  \end{tabular}}
\end{table}

\section{Conclusions and discussions}

In this paper, we propose an IPINN framework with neuron-wise locally adaptive activation function and slope recovery term for solving data-driven localized waves and recovering unknown parameters of Manakov system for the first time, which provides a very important theoretical basis and training experience for investigating data-driven localized waves and parameters discover of other two-component or even multi-component coupled nonlinear systems. Since the Manakov system possesses two component solutions $q_r(x,t)$ $(r=1,2)$, the number of initial-boundary value conditions is twice that of the (1 + 1) dimensional single NLS \cite{Pu2021}, thus the IPINN framework yield four outputs and four nonlinear equation constraints instead of two outputs and two nonlinear equation constraints of the (1 + 1) dimensional single NLS, so we extended and improved PINN algorithm by increasing the number of output components and physics constraints. Then we are committed to study the data-driven vector localized waves, which contain vector solitons, vector breathers and various vector RWs, as well as parameters discovery for the Manakov system by employing the IPINN approach with small sample data set. The abundant numerical results show that the IPINN model can effectively and accurately recover the different dynamical behaviors of localized waves for Manakov system fairly. However, we firstly study the parameters discover of Manakov system with unknown parameters by means of the IPINN, and find that the training effect is very poor whether using clean data or noisy data. Therefore, in order to train the parameters more accurately, we introduce the $L^2$ norm parameter regularization with adjustable weight coefficients $\beta$ into the routine IPINN, and find that once using the appropriate weight coefficients, the unknown parameters can be recovered accurately and effectively whether using clean data or noise data.

Compared with classical PINN \cite{Pu2021}, the PINN method with adaptive activation function and slope recovery term loss function has better training effect and better convergence speed for training derivative nonlinear Schr\"{o}dinger equation by means of a large number of training experiments and vivid plots \cite{PuJ2021,PuJC2021}. Therefore, starting from the aforementioned PINN approach, we establish an IPINN for investigating data-driven vector localized waves and parameters discovery of Manakov system. Furthermore, in the original PINN model, we only employ the L-BFGS optimization algorithm to optimize the loss function, but we utilize not only the L-BFGS optimizer, but also the Adam optimizer to optimize the loss function in the novel IPINN framework. Moreover, we believe that we can easily construct the IPINN of more component coupled nonlinear systems to obtain vector localized waves, and we can also build corresponding IPINN of the high-dimensional multi-component coupled nonlinear systems for studying high-dimensional vector localized waves. These are also the contents that we will further study in the future.

The IPINN method showcases a series of results of various interesting problems in the interdisciplinary field of applied mathematics and computational science, and opens a new path for using NN to recover unknown solutions and correspondingly discover the unknown parametric equations in mathematical physics and deep learning. This also provides an significant theoretical basis and experimental reference for solving some previously unsolvable big data spatial-temporal problems and high-dimensional science. Furthermore, the problem of solving multi-component coupled nonlinear systems occupies an important position in many scientific fields, this paper also provides a very powerful deep learning NN framework for these disciplines to make more professional research. In future research, we will focus on finding and proposing more efficient NN algorithms to study a variety of nonlinear systems, in which how to improve PINN model with integrable system theory is a significant problem that needs further research.

\section*{Declaration of competing interest}
The authors declare that they have no known competing financial interests or personal relationships that could have appeared to influence the work reported in this paper.

\section*{Acknowledgements}
\hspace{0.3cm}
The authors gratefully acknowledge the support of the National Natural Science Foundation of China (No. 12175069) and Science and Technology Commission of Shanghai Municipality (No. 21JC1402500 and No. 18dz2271000).

\end{document}